\documentclass[bibtex]{aa}  

\usepackage{graphicx}
\usepackage{caption}
\usepackage{txfonts}
\usepackage{natbib}
\usepackage{booktabs,multirow}
\DeclareUnicodeCharacter{00A0}{~}
\usepackage[flushleft]{threeparttable}
\usepackage{color}
\usepackage{tabularx}
\usepackage[normalem]{ulem}
\begin{document} 
   \title{Determining the effects of clumping and porosity on the chemistry in a non-uniform AGB outflow
   }
	
   \author{M. Van de Sande \inst{\ref{inst1}} \and
           J.O. Sundqvist \inst{\ref{inst1}} \and
           T.J. Millar \inst{\ref{inst2}} \and
           D. Keller \inst{\ref{inst1}} \and
           W. Homan \inst{\ref{inst1}} \and
           A. de Koter \inst{\ref{inst3},\ref{inst1}} \and
           L. Decin \inst{\ref{inst1}} \and
           F. De Ceuster \inst{\ref{inst4}}
          }

   \institute{Department of Physics and Astronomy, Institute of Astronomy, KU Leuven, Celestijnenlaan 200D, 3001 Leuven, Belgium \label{inst1} \\ \email{marie.vandesande@kuleuven.be}
         \and Astrophysics Research Centre, School of Mathematics and Physics, Queen’s University Belfast, University Road, Belfast BT7 1NN, UK  \label{inst2} 
         \and Astronomical Institute Anton Pannekoek, University of Amsterdam, Science Park 904, PO Box 94249, 1090 GE Amsterdam, The Netherlands  \label{inst3} 
         \and Department of Physics and Astronomy, University College London, Gower Street, London WC1E 6BT, UK \label{inst4}
             }
             
   \date{Received ; accepted }

\abstract
{ In the inner regions of AGB outflows, several molecules have been detected with abundances much higher than those predicted from thermodynamic equilibrium chemical models. The presence of the majority of these species can be explained by shock-induced non-equilibrium chemical models, where shocks caused by the pulsating star take the chemistry out of equilibrium in the inner region. 
Moreover, a non-uniform density structure has been detected in several AGB outflows. Both large-scale structures, such as spirals and disks, and small-scale density inhomogeneities or clumps have been observed. These structures may also have a considerable impact on the circumstellar chemistry. A detailed parameter study on the quantitative effects of a non-homogeneous outflow has so far not been performed.
}
{ We examine the effects of a non-uniform density distribution within an AGB outflow on its chemistry by considering a stochastic, clumpy density structure.
}
{ We implement a \emph{porosity formalism} for treating the increased leakage of light associated with radiation transport through a clumpy, porous medium. We then use this method to examine the effects from the altered UV radiation field penetration on the chemistry, accounting also for the increased reaction rates of two-body processes in the overdense clumps. The specific clumpiness is determined by three parameters: the characteristic length scale of the clumps at the stellar surface, the clump volume filling factor, and the inter-clump density contrast. In this paper, the clumps are assumed to have a spatially constant volume filling factor, which implies that they expand as they move outward in the wind. 
}
{ We present a parameter study of the effect of clumping and porosity on the chemistry throughout the outflow. Both the higher density within the clumps and the increased UV radiation field penetration have an important impact on the chemistry, as they both alter the chemical pathways throughout the outflow. The increased amount of UV radiation in the inner region leads to photodissociation of parent species, releasing the otherwise deficient elements. We find an increased abundance in the inner region of all species not expected to be present assuming thermodynamic equilibrium chemistry, such as HCN in O-rich outflows, H$_2$O in C-rich outflows, and NH$_3$ in both.
}
{ A non-uniform density distribution directly influences the chemistry throughout the AGB outflow, both through the density structure itself and through its effect on the UV radiation field. 
Species not expected to be present in the inner region of the outflow assuming {thermodynamic} equilibrium chemistry are now formed in this region, including species that are not formed in greater abundance by shock-induced non-equilibrium chemistry models.
Outflows whose clumps have a large overdensity and that are very porous to the interstellar UV radiation field yield abundances comparable to those observed in O-rich and C-rich outflows for most of the unexpected species investigated. The inner wind abundances of H$_2$O in C-rich outflows and of NH$_3$ in O-rich and C-rich outflows are however underpredicted.
}


   \maketitle

\section{Introduction}				\label{sect:Introduction}

The asymptotic giant branch (AGB) phase of stars with an initial mass up to 8 M$_\odot$ is characterised by strong mass loss via a stellar outflow or wind. The outflow removes the outer stellar layers and creates an extended circumstellar envelope (CSE). Through their outflows, AGB stars eject nuclear reaction products in the form of gas and dust into the interstellar medium and hence play an important role in the chemical enrichment of the Universe {\citep[][and references therein]{Habing2003}}.
The AGB outflow can be split up into three dynamically and chemically different regions: the inner region close to the central star in which the material experiences shocks caused by stellar pulsations and in which the first chemical processing of photospheric gas takes place, the intermediate region where dust condenses from the gas phase and material is accelerated outward, and the outer region where the material has reached its terminal velocity and the chemistry is dominated by photodissociation caused by the penetration of interstellar UV photons.
The molecular composition of the CSE is determined by the elemental carbon-to-oxygen abundance ratio of the AGB star because of the large binding energy of CO. Oxygen-rich stars have C/O < 1, and therefore an oxygen-rich chemistry throughout their outflows. Similarly, carbon-rich stars have C/O > 1 and a carbon-rich chemistry.

The initial molecular composition of the outflow is established by the physical circumstances in its innermost region. {Thermodynamic} equilibrium (TE) can be assumed at the stellar surface, which leads to a certain molecular composition with a clear O- or C-rich signature \citep{Willacy1998}.
However, several molecular species have been detected that are expected, under the assumption of TE chemistry, either to be absent in the inner wind or to have a smaller abundance than observed in the outer regions. 
In this paper, we will refer to these species as \emph{unexpected} species for convenience and brevity.
In O-rich outflows, there are detections of several unexpected C-rich molecules such as HCN, CS, and CN \citep{Omont1993,Bujarrabal1994,Justtanont1996}. Similarly in C-rich outflows, e.g. warm H$_2$O and CH$_3$CN have been detected close to the central star \citep{Decin2010b,Neufeld2011,Agundez2015}. 
In both O- and C-rich outflows, NH$_3$ has been detected with abundances larger than predicted by TE \citep{Keady1993,Menten1995,Menten2010,Wong2017}.
To reproduce the unexpectedly high observed abundances of these molecules in the outer wind, the chemical models of \citet{Willacy1997} and \citet{Willacy1998} relied on the injection of molecules bearing the deficient element.
A proposed solution to these discrepancies in abundance is to take into account the non-equilibrium character of the chemistry in the innermost region of the wind. In this region, the chemistry is taken out of TE due to shocks caused by the pulsating AGB star (e.g. \citeauthor{Duari1999}~\citeyear{Duari1999}, \citeauthor{Cherchneff2006}~\citeyear{Cherchneff2006}, \citeauthor{Gobrecht2016}~\citeyear{Gobrecht2016}).
These shock-induced non-equilibrium chemistry models are able to reproduce most observed abundances, but fail to account for e.g. the observed NH$_3$ abundance in the O-rich star IK Tau \citep{Gobrecht2016}.

Clumpy structures, and non-uniformity in general, have been widely observed in AGB outflows. 
Both large-scale structures, such as spirals (e.g. \citeauthor{Mauron2006}~\citeyear{Mauron2006}, \citeauthor{Maercker2016}~\citeyear{Maercker2016}) and disks (e.g. \citeauthor{Kervella2014}~\citeyear{Kervella2014}), and small-scale inhomogeneities or clumps (e.g. \citeauthor{Khouri2016}~\citeyear{Khouri2016}, \citeauthor{Agundez2017}~\citeyear{Agundez2017}), have been detected. 
As an alternative to the shock-induced non-equilibrium models, \citet{Agundez2010} have proposed the penetration of interstellar UV photons into the inner wind due to the clumpiness of the outflow. 
By allowing UV photons to reach the otherwise shielded inner region of the outflow, the CO and N$_2$ bonds can be broken and the deficient element (C in O-rich outflows, O in C-rich outflows, N in both) is released close to the star. 
\citet{Agundez2010} assumed that a certain fraction of UV photons is able to reach the inner wind uninhibitedly, but did not take the clumpy density distribution of the outflow into account.
Their model results in abundances of unexpected molecules well above TE for both O- and C-rich outflows, and produces NH$_3$ in both.

In this paper, we include a clumpy density distribution in our spherically symmetric chemical model by implementing a \emph{porosity formalism}.
The porosity formalism provides us with a mathematical framework that not only describes the effect of {a clumpy} outflow on its optical depth (and thereby on the penetration of UV photons{, i.e. porosity}), but also allows us to take the {local {overdensities} (i.e. clumping)} into account in the chemical model.
We explore the parameter space of the formalism for both O- and C-rich outflows with various mass-loss rates, and study its effects on the abundances of the unexpected species.
We consider two scenarios:  (i) a simplified outflow where all material is located in mass conserving clumps, (ii) a perhaps more realistic outflow consisting of both a clump component and an inter-clump component.

In Sect. \ref{sect:Model}, we describe the chemical model and reaction network used.
We elaborate on the analytical description of {the clumping and porosity properties} of the outflow and {their} implementation into the model in Sect. \ref{sect:Porosity}.
The results for both O- and C-rich outflows are presented in Sect. \ref{sect:Results}.
Discussion and {conclusions} follow in Sect. \ref{sect:Discussion} and Sect. \ref{sect:Conclusion}.

\section{Chemical model of the outflow}                                        \label{sect:Model}

The chemical reaction network used is the most recent release of the UMIST Database for Astrochemistry (UDfA), \textsc{Rate12} \citep{McElroy2013}, a gas-phase only chemical network containing 6173 reactions involving 467 species. 
Our chemical model is based on the UDfA CSE model. Both the chemical network and CSE model are publicly available\footnote{\url{http://udfa.ajmarkwick.net/index.php?mode=downloads}}.
The one-dimensional model describes a uniformly expanding, spherically symmetric CSE with a constant mass-loss rate. 
The number density profile $n(r)$ hence falls as $1/r^2$, with $r$ the distance from the centre of the star.
The velocity of the material is {assumed to be} constant throughout the outflow. The effect of CO self-shielding is taken into account during the calculation using a single-band approximation \citep{Morris1983}.
H$_2$ is assumed to be completely self-shielded, so that $n_{\mathrm{H}_2}(r) \approx n(r)$.
A more detailed description of the model can be found in \citet{Millar2000}, \citet{Cordiner2009}, and \citet{McElroy2013}.

\begin{table}
	\caption{Physical parameters of the CSE model. } 
	\centering
	\resizebox{\columnwidth}{!}{%
    \begin{tabular}{l l}
    \hline \hline 
    \noalign{\smallskip}
    Mass-loss rate $\dot{M}$        &    $10^{-5}, 10^{-6}, 10^{-7}\ \mathrm{M}_\odot\ \mathrm{yr}^{-1}$ \\
    Stellar temperature $T_*$        & 2000 K \\
    Stellar radius $R_*$             & 5 $\times 10^{13}$ cm \\
    Outflow velocity $\nu_\infty$   & 15 km s$^{-1}$  \\
    Exponent temperature power-law $\epsilon$                       & 0.7 \\
    \hline
    \end{tabular}
    }
    \label{table:Model-PhysPar}    
\end{table}

\begin{table}
	\caption{Parent species and their initial abundances relative to H$_2$ for the C- and O-rich CSE.	Taken from \citet{Agundez2010}.
} 
	\centering
    \begin{threeparttable}
    {\small
    \centering
    \begin{tabular}{l r c c l r c}
    \hline \hline 
    \noalign{\smallskip}
    \multicolumn{3}{c}{Carbon-rich} && \multicolumn{3}{c}{Oxygen-rich}  \\  
    \cline{1-3} \cline{5-7} 
    \noalign{\smallskip}
    Species & Abundance & Ref. & & Species & Abundance & Ref. \\
    \hline
    \noalign{\smallskip}
    He            & 0.17                &        &    & He        & 0.17                &     \\
    CO            & $8\times10^{-4}$    & (1)    &    & CO        & $3\times10^{-4}$    & (1) \\
    N$_2$         & $4\times10^{-5}$    & (2)    &    & N$_2$     & $4\times10^{-5}$    & (2) \\
    C$_2$H$_2$    & $8\times10^{-5}$    & (3)    &    & H$_2$O    & $3\times10^{-4}$    & (4) \\
    HCN           & $2\times10^{-5}$    & (3)    &    & CO$_2$    & $3\times10^{-7}$    & (5) \\
    SiO           & $1.2\times10^{-7}$  & (3)    &    & SiO       & $1.7\times10^{-7}$  & (6) \\
    SiS           & $1\times10^{-6}$    & (3)    &    & SiS       & $2.7\times10^{-7}$  & (7) \\
    CS            & $5\times10^{-7}$    & (3)    &    & SO        & $1\times10^{-6}$    & (8) \\
    SiC$_2$       & $5\times10^{-8}$    & (3)    &    & H$_2$S    & $7\times10^{-8}$    & (9) \\
    HCP           & $2.5\times10^{-8}$  & (3)    &    & PO        & $9\times10^{-8}$    & (10) \\
    \hline 
    \end{tabular}
    \begin{tablenotes}
    \footnotesize
    \item {{References.}} (1) \citet{Teyssier2006}; (2) TE abundance \citep{Agundez2010}; (3) \citet{Agundez2009};
    (4) \citet{Maercker2008}; (5) \citet{Tsuji1997}; (6) \citet{Schoier2006}; (7) \citet{Schoier2007}; (8) \citet{Bujarrabal1994};
    (9) \citet{Ziurys2007}; (10) \citet{Tenenbaum2007}.
    \end{tablenotes}
    }
    \end{threeparttable}
    
    \label{table:Model-Parents}    
\end{table}

The initial abundance of all species is set to zero, except for the parent species. These are defined as having formed just above the photosphere and are injected into the CSE at the start of our model at $r = 2\ R_* = 1 \times 10^{14}$ cm.
The adjustments made to the code by introducing the porosity formalism are described in detail in Sect. \ref{subsect:Porosity:Implementation}.
We have also changed the assumed temperature structure throughout the outflow to a power-law,
\begin{equation}
T(r) = T_* \left( \frac{R_*}{r} \right)^{\epsilon},
\end{equation}
with $T_*$ and $R_*$ the stellar temperature and radius, and $\epsilon$ the exponent characterising the power-law.
The temperature profile has a lower limit of 10 K, as to prevent unrealistically low temperatures in the outer CSE \citep{Cordiner2009}.

In order to compare our results to those of \citet{Agundez2010}, we have adopted the same physical parameters (Table \ref{table:Model-PhysPar}) and the same parent species and their initial abundances for the O- and C-rich CSE (Table \ref{table:Model-Parents}). 
The constant outflow velocity $\nu_\infty$ is the main physical difference with the model of \citet{Agundez2010}, who have assumed an expansion velocity of 5 km s$^{-1}$ for $r \leq 5\ R_*$ {and a velocity of 15 km s$^{-1}$ beyond}.
The \textsc{Rate12} network does not include $^{13}$CO and its self-shielding, in contrast to the chemical network used by \citet{Agundez2010}.

\section{Porosity formalism}                        \label{sect:Porosity}

By assuming that the non-uniformity of the outflow is caused by a statistical ensemble of small-scale density inhomogeneities or clumps, we are able to use a so-called \emph{porosity formalism} \citep{Owocki2004,Owocki2006,Sundqvist2012,Sundqvist2014} to describe the effect of a clumpy outflow on its chemistry.
{See also \citet{Feldmeier2003} and \citet{Oskinova2004,Oskinova2006} for an alternative formulation regarding radiative transfer effects in an inhomogeneous medium.}

Within {the porosity} formalism, we assume that the outflow is a stochastic two-component medium consisting of overdense clumps and a rarefied inter-clump medium.
The clumps are assumed to conserve their mass throughout the outflow.
The fraction of the total volume occupied by the clumps is set by $f_\mathrm{vol}$, the clump volume filling factor.
The mean (laterally averaged) density of the outflow can then be written as
\begin{equation}			\label{Eq:Density-2C}
    \langle \rho \rangle = (1 - f_\mathrm{vol})\ \rho_\mathrm{ic} + f_\mathrm{vol}\ \rho_\mathrm{cl} = \frac{\dot{M}}{4\pi r^2 \nu_\infty},
\end{equation}
where the subscripts `cl' and `ic' refer to the clump and inter-clump density components respectively. The last equality assumes mass-conservation as compared to a corresponding smooth outflow with density $\rho_\mathrm{sm} = \langle \rho \rangle$. For ease of notation, we drop the radial dependency of all expressions.
The density contrast between the inter-clump component and the mean density is set by the parameter {\citep{Sundqvist2014}}
\begin{equation}			\label{Eq:Fic}
	f_\mathrm{ic} \equiv \frac{\rho_\mathrm{ic}}{\langle\rho\rangle}.
\end{equation}
By splitting the outflow into two separate components, their specific over- or under-densities compared to the corresponding smooth outflow can be taken into account in our chemical model.
Note that within our statistical formalism, no assumptions on the specific locations of individual clumps need to be made.

Generally, if clumps become optically thick, this leads both to a local self-shielding of opacity within the clumps and to an increased escape of radiation through the porous channels in between the clumps.
The overall net effect {of a clumpy} outflow is therefore a reduction in its mean opacity and hence its optical depth for a given wind mass. {A clumpy outflow} hence affects the penetration of interstellar UV photons into the circumstellar material, which is important to the chemistry as it induces photodissociation reactions. 
The extinction of the interstellar UV radiation field in the outflow is given by $e^{-\tau}$, with $\tau$ the optical depth.
For a smooth outflow, the radial optical depth is given by
\begin{equation}			\label{Eq:Tau-Smooth}
	\tau = \int_r^\infty \kappa\ \rho_\mathrm{sm}\ \mathrm{d}r = \kappa \int_r^\infty \rho_\mathrm{sm}\ \mathrm{d}r,
\end{equation}
where the latter equality assumes that the opacity $\kappa$ of the outflow is spatially constant. We have here that $\kappa \approx$ {1200} cm$^2$ g$^{-1}$, its calculation can be found in Appendix \ref{Sect:App:Opacity}.
Non-radial rays are accounted for according to Eq. 4 of \cite{Morris1983}.
By introducing the radial optical depth down to the stellar surface for a smooth wind as a reference optical depth $\tau_*$, given by
\begin{equation}
	\tau_* = \frac{\kappa \dot{M}}{4\pi \nu_\infty R_*},
\end{equation}
we can write that $\tau = \tau_* R_*/r$. 
The optical depth of a smooth outflow is therefore proportional to $\dot{M}$. 
For the physical parameters assumed in this paper, we have that $\tau_* \approx$ {800} for $\dot{M} = 10^{-5}\ \mathrm{M}_\odot\ \mathrm{yr}^{-1}$, $\tau_* \approx$ {80} for $\dot{M} = 10^{-6}\ \mathrm{M}_\odot\ \mathrm{yr}^{-1}$ and $\tau_* \approx$ {6} for $\dot{M} = 10^{-7}\ \mathrm{M}_\odot\ \mathrm{yr}^{-1}$. 
The mass-loss rate of a smooth outflow hence strongly influences its optical depth.

The increased interstellar UV radiation field in a clumpy outflow is now taken into account by changing the optical depth $\tau$ to an \emph{effective} optical depth $\tau_\mathrm{eff}$.
Eq. \ref{Eq:Tau-Smooth} then changes to
\begin{equation}			\label{Eq:Tau-Clump}
	\tau_\mathrm{eff} =  \int_r^\infty \kappa_\mathrm{eff}\ \langle\rho\rangle \ \mathrm{d}r,
\end{equation}
with $\tau_\mathrm{eff}$ and $\kappa_\mathrm{eff}$ the effective optical depth and opacity of the clumpy outflow.

Within our statistical porosity model, this effective opacity $\kappa_\mathrm{eff}$ depends on the interaction probability $P$ of a photon with a clump, and thus depends on the optical depth of an individual clump $\tau_\mathrm{cl}$, given by
\begin{equation}         \label{Eq:TauClump-General}
    \tau_\mathrm{cl} = \kappa \  \rho_\mathrm{cl}  \ l,
\end{equation}
with $l$ the characteristic length scale of the clumps. 
The effective extinction $\kappa_\mathrm{eff}\ \langle \rho \rangle$ (per unit length) is then given by
\begin{equation}			\label{Eq:KappaEff-1}
	\kappa_\mathrm{eff}\ \langle \rho \rangle = P\left(\tau_\mathrm{cl}\right)\ n_\mathrm{cl}\ A_\mathrm{cl} = \frac{P(\tau_\mathrm{cl})}{h},
\end{equation}
with $n_\mathrm{cl}$ the clump number density and $A_\mathrm{cl}$ the clump's geometric cross section {\citep[e.g. ][]{Feldmeier2003}}, and where the last equality introduces the \emph{porosity length} $h$ \citep{Owocki2004,Sundqvist2012}. This paper assumes clumps that are spherical or randomly oriented with an arbitrary shape, which then leads to a statistically isotropic effective opacity {\citep[see][]{Sundqvist2012}}.
The porosity length represents the local mean free path between two clumps.
Since the total volume associated with one clump is $V_\mathrm{tot} = 1/n_\mathrm{cl}$, we can also use $V_\mathrm{cl} = l^3$, $A_\mathrm{cl} = l^2$, and $f_\mathrm{vol} = V_\mathrm{cl}/V_\mathrm{tot}$ to rewrite the porosity length as
\begin{equation}			\label{Eq:H-lfvol}
	h = \frac{l}{f_\mathrm{vol}}.
\end{equation}

The probability that a photon gets absorbed by a given clump is $P\left(\tau_\mathrm{cl}\right) = 1 - e^{-\tau_\mathrm{cl}}$. 
But the relation between the effective and smooth opacity in principal depends also on the properties of the clumpy outflow. 
Assuming here an exponential distribution of clumps (i.e. a Markovian mixture, see \citeauthor{Sundqvist2012}~\citeyear{Sundqvist2012}),
\begin{equation}
	f\left(\tau_\mathrm{cl}\right) = \frac{e^{-\tau/\tau_\mathrm{cl}}}{\tau_\mathrm{cl}}
\end{equation}
with 
\begin{equation}
	\langle	\tau \rangle	 = \int_0^\infty \tau f\left( \tau \right) \mathrm{d}\tau = \tau_\mathrm{cl},
\end{equation}
and an outflow where all material is located inside of clumps {(a so-called one-component model, see Sect. \ref{subsect:Porosity:OneComp})}, gives upon integration the bridging law for the effective opacity of the outflow \citep{Owocki2006,Sundqvist2012}
\begin{equation}            \label{Eq:BridgingLaw-1C}
    \frac{\kappa_\mathrm{eff}}{\kappa} = \frac{1}{1+\tau_\mathrm{cl}}.
\end{equation}
By using this expression in the calculation of the effective optical depth (Eq. \ref{Eq:Tau-Clump}), we can now take the effect of porosity upon the radiation field into account for a one-component outflow.
Note that as $\tau_\mathrm{cl}$ goes to zero in Eq. \ref{Eq:BridgingLaw-1C}, the smooth opacity and wind optical depths are recovered. For the other limit of $\tau_\mathrm{cl} \gg 1$, we obtain instead $\kappa_\mathrm{eff} \langle\rho\rangle = 1/h$. That is, for such fully opaque clumps the effective extinction is solely affected by the porosity length, and is independent of $\kappa$. 
For this case, $h$ also represents the actual local mean free path of photons.

This paper further assumes that the clump volume filling factor $f_\mathrm{vol}$ is spatially constant.
Since we also assume mass conserving clumps, this leads to clumps expanding according to
\begin{equation}        \label{Eq:L-Rdependence}
    l(r) =  l_* \left(\frac{r}{R_*}\right)^{2/3},
\end{equation}
with $l_*$ the characteristic length scale near the stellar surface.
Following Eq. \ref{Eq:H-lfvol}, $h$ then has the same radial dependence
\begin{equation}        \label{Eq:H-Rdependence}
    h(r) =  h_* \left(\frac{r}{R_*}\right)^{2/3},
\end{equation}
with $h_* = l_*/f_\mathrm{vol}$ the porosity length at the stellar surface.

Two cases are considered in this paper: {the previously introduced} simplified case {of a one-component model} where all material is located inside clumps and the inter-clump component is effectively void (Sect. \ref{subsect:Porosity:OneComp}), and a case where material is located {in} both the clump and inter-clump components{, the two-component model} (Sect. \ref{subsect:Porosity:TwoComp}).

\subsection{One-component outflow}                \label{subsect:Porosity:OneComp}

\begin{figure}
\centering
\includegraphics[width=1.\columnwidth]{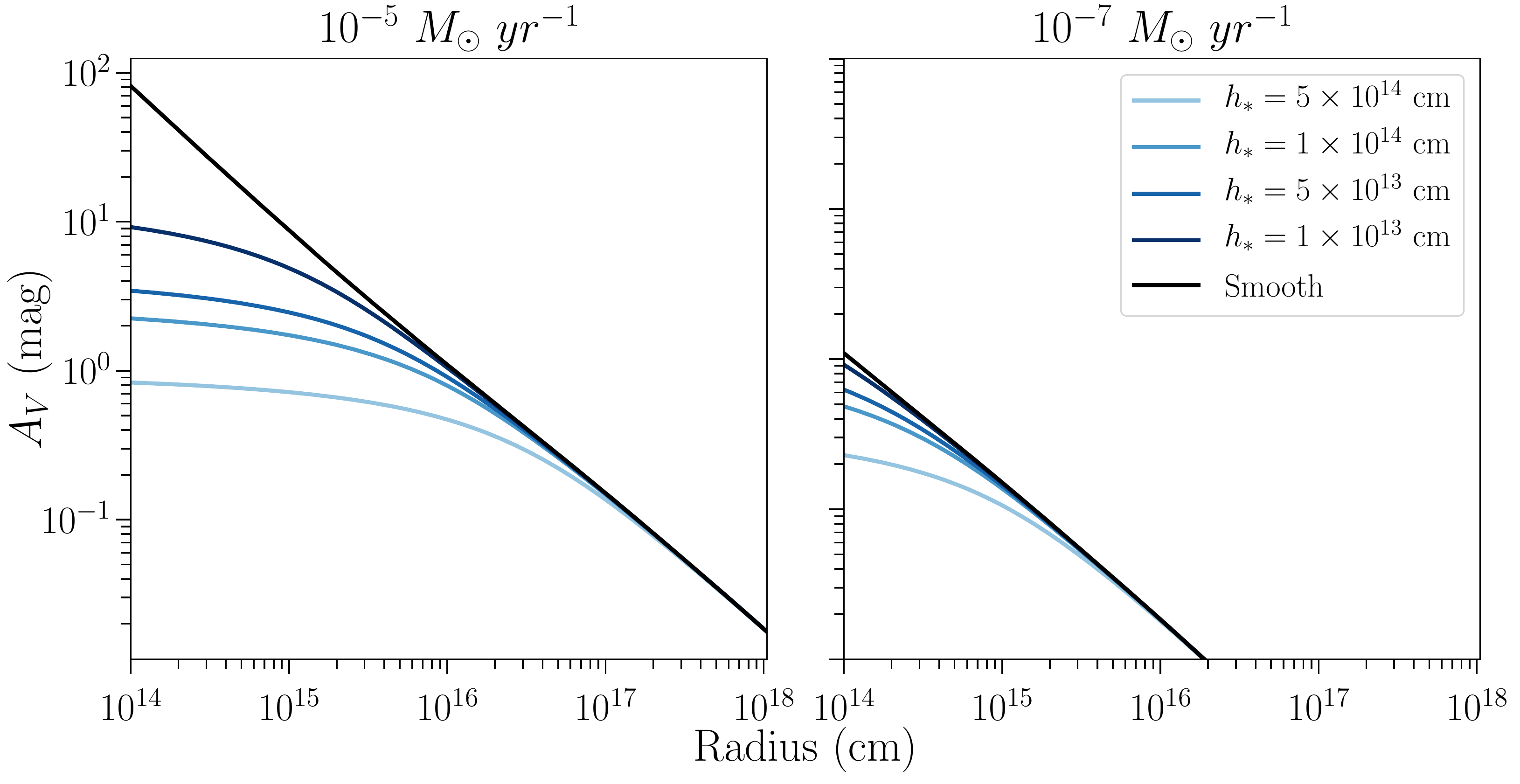}
\caption{Visual extinction $A_V$ throughout a one-component outflow with $\dot{M} = 10^{-5}\ \mathrm{M}_\odot\ \mathrm{yr}^{-1}$ (left panel) and $\dot{M} = 10^{-7}\ \mathrm{M}_\odot\ \mathrm{yr}^{-1}$ (right panel).
Results for a smooth, uniform outflow are shown in black. Results for different values of $h_*$ are shown in blue. 
For reference, $1\ R_* = 5 \times 10^{13}$ cm.
Fig. \ref{fig:Porosity-RF-1C-varH} shows the corresponding decrease in interstellar UV radiation field.
}
\label{fig:Porosity-AV-1C-varH}
\end{figure}

In a one-component outflow, all material is located inside clumps and the inter-clump component is {hence} effectively void. Such an extreme outflow gives us an upper limit to the effect of porosity on the chemistry within the outflow, both regarding its effect on the density contrast between a clumpy outflow and a smooth outflow and the increase in UV radiation field throughout the clumpy outflow.

As the inter-clump component is effectively void, Eq. \ref{Eq:Density-2C} reduces to
\begin{equation}        \label{Eq:Density-1C}
    \langle \rho \rangle =  f_\mathrm{vol}\ \rho_\mathrm{cl}.
\end{equation}
{Using Eqs. \ref{Eq:H-lfvol}, \ref{Eq:H-Rdependence}, and \ref{Eq:Density-1C}, the clump optical depth (Eq. \ref{Eq:TauClump-General}) can be written as}
\begin{equation}			\label{Eq:TauClump-1C}
\begin{split}
	\tau	_\mathrm{cl} &= \kappa\ \langle\rho\rangle\ h \\
					 &= \tau_*\ R_*^{1/3}\ h_*\ r^{-4/3},
\end{split}
\end{equation}
Using Eq. \ref{Eq:BridgingLaw-1C}, the effective optical depth throughout a one-component clumpy outflow is then given by
\begin{equation}        		\label{Eq:IntegralTauEff-1C}   
\begin{split}
    \tau_\mathrm{eff} &= \kappa \int_r^\infty  \langle\rho\rangle\ \frac{1}{1+\tau_\mathrm{cl}} \mathrm{d}r \\
					&= \tau_*\ R_* \int_r^\infty \frac{1}{r^2 +\tau_*\ R_*^{1/3}\ h_*\ r^{2/3}}\ \mathrm{d}r,
\end{split}
\end{equation}
{where Eq. \ref{Eq:TauClump-1C} has been used for $\tau_\mathrm{cl}$.
}
If $h_* = 0$ (i.e. $\tau_\mathrm{cl} = 0$), this indeed reduces to the optical depth of a smooth wind $\tau = \tau_*\ R_*/r$. 
The analytic solution of this integral can be found in Appendix \ref{subsect:App:TauEff:1C}.
Note here that the scaling of the effective optical depth of a one-component medium is influenced only by the porosity length $h_*$.

For larger values of $h_*$, both the clump optical depth and the mean free path in between the clumps increase (Fig. \ref{fig:Porosity-AV-1C-varH}). 
As the clump optical depth increases, the ratio of the effective opacity of the clumpy outflow to that of the smooth outflow $\kappa_\mathrm{eff}/\kappa$ decreases. 
A larger fraction of the interstellar radiation field is therefore able to reach the inner and intermediate regions of the outflow: the visual extinction is smaller in these regions compared to that of a smooth outflow.
In addition, as the mean free path in between the clumps increases, the onset of the decrease in visual extinction (compared to the visual extinction of a smooth outflow, associated with the position where $\tau_\mathrm{cl}$ is no longer negligible) is located further out in the outflow (Fig. \ref{fig:Porosity-AV-1C-varH}).
{Since $\tau_\mathrm{cl}$ scales with $\tau_*$, the clump optical depth decreases with decreasing mass-loss rate. Following Eq. \ref{Eq:BridgingLaw-1C}, this leads to a less pronounced effect of porosity on the effective opacity $\kappa_\mathrm{eff}$ and optical depth $\tau_\mathrm{eff}$ for lower mass-loss rates.
}

\subsection{Two-component outflow}                \label{subsect:Porosity:TwoComp}

\begin{figure}
\centering
\includegraphics[width=1.\columnwidth]{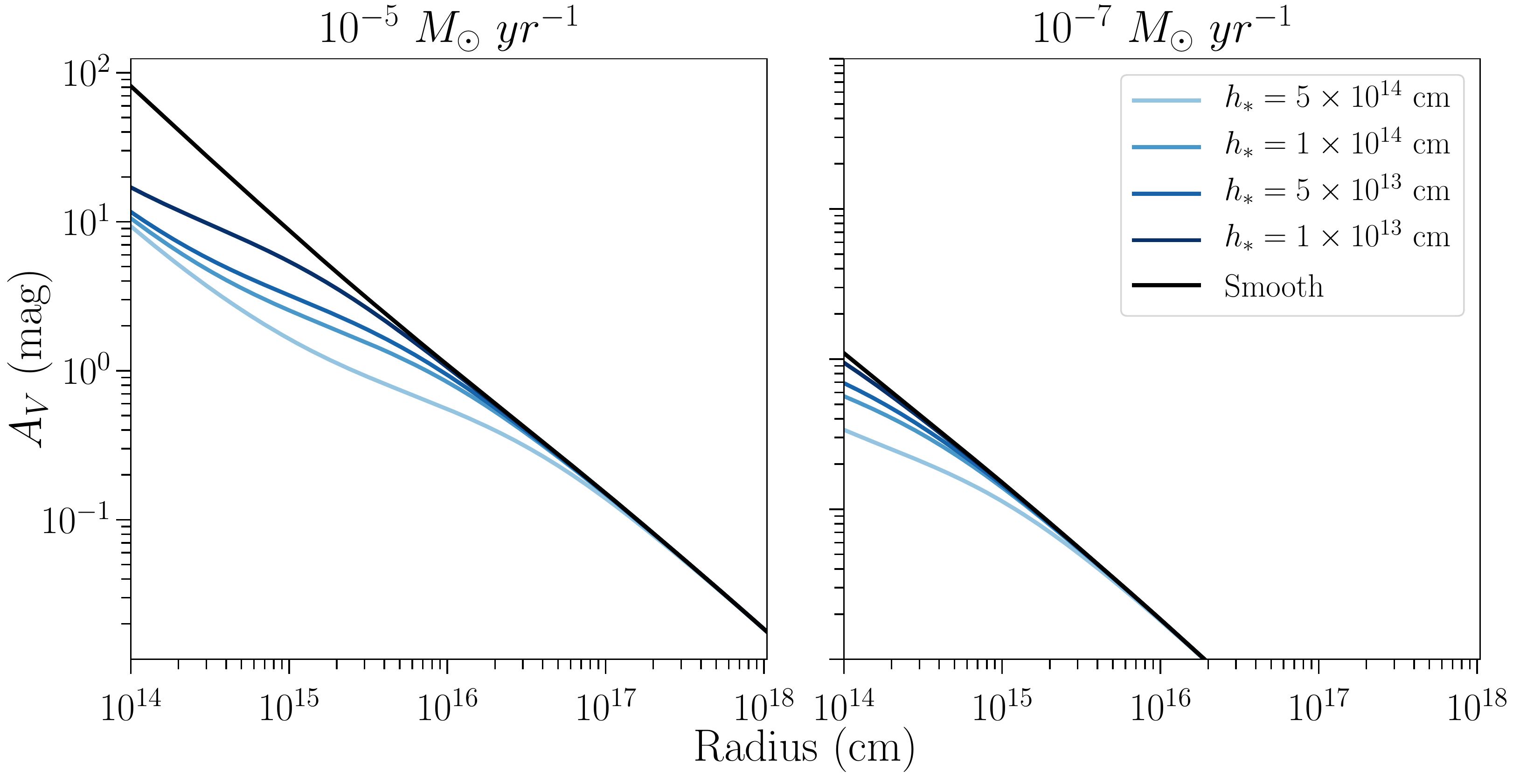}
\caption{As Fig. \ref{fig:Porosity-AV-1C-varH}, but for a two-component outflow with $f_\mathrm{ic} = 0.1$.
For reference, $1\ R_* = 5 \times 10^{13}$ cm.
}
\label{fig:Porosity-AV-2C-varH}
\end{figure}

\begin{figure}
\centering
\includegraphics[width=1.\columnwidth]{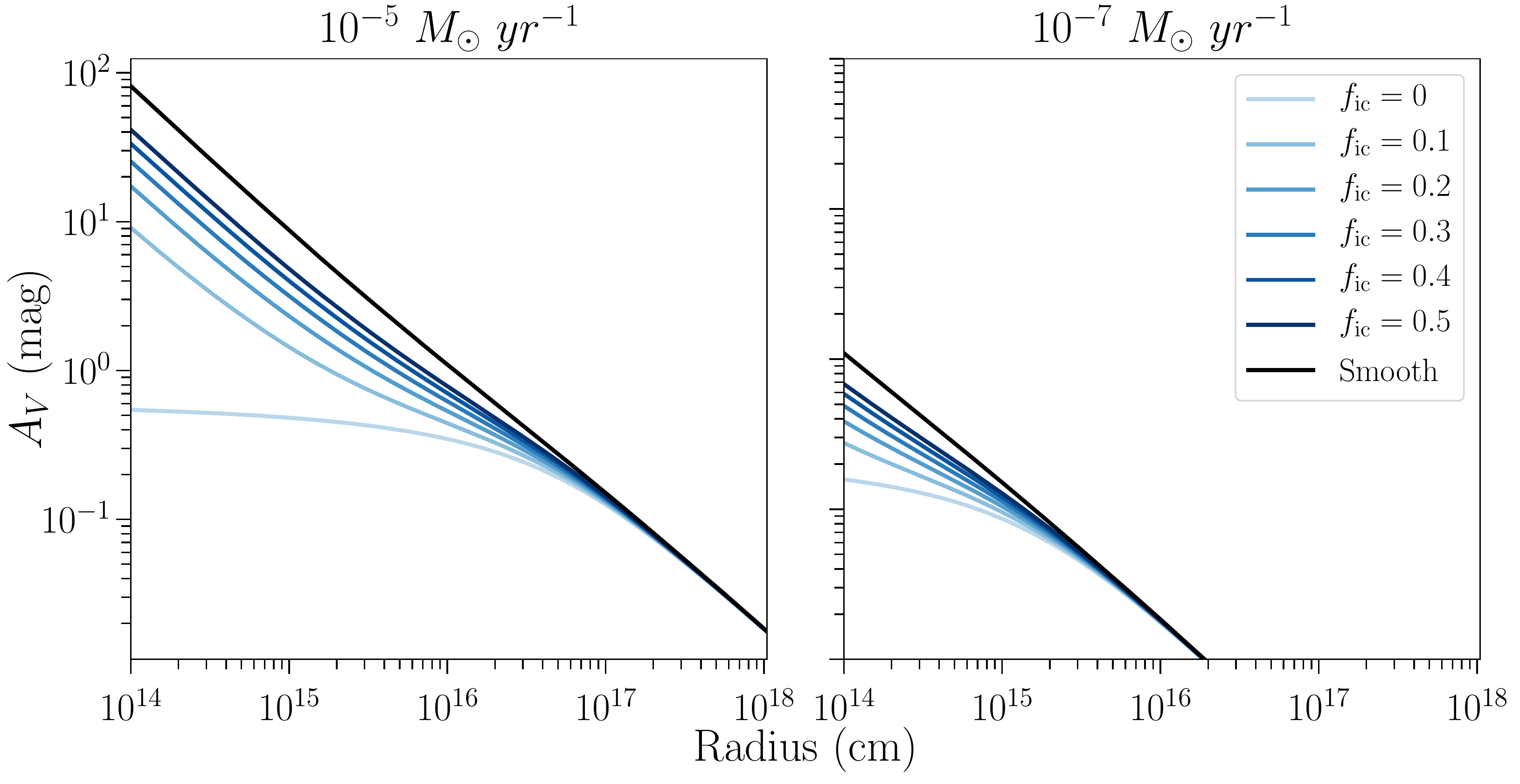}
\caption{Visual extinction $A_V$ throughout a two-component outflow with $h_* = 10^{15}$ cm and $\dot{M} = 10^{-5}\ \mathrm{M}_\odot\ \mathrm{yr}^{-1}$ (left panel) and $\dot{M} = 10^{-7}\ \mathrm{M}_\odot\ \mathrm{yr}^{-1}$ (right panel).
Results for a smooth, uniform outflow are shown in black. Results for different values of $f_\mathrm{ic}$ are shown in blue. 
For reference, $1\ R_* = 5 \times 10^{13}$ cm.
}
\label{fig:Porosity-AV-2C-varFic}
\end{figure}

In a two-component outflow, material is located both inside the clumps and the inter-clump component. Both components will now contribute to the total opacity of the clumpy outflow. The density of a two-component outflow is given by Eq. \ref{Eq:Density-2C}. {Using Eqs. \ref{Eq:Fic}, \ref{Eq:H-lfvol}, and \ref{Eq:H-Rdependence}}, the clump optical depth (Eq. \ref{Eq:TauClump-General}) then is given by 
\begin{equation}			\label{Eq:TauClump-2C}
\begin{split}
	\tau_\mathrm{cl} &= \kappa \  \langle\rho\rangle\  h\ \left( 1 - \left( 1 - f_\mathrm{vol} \right) f_\mathrm{ic} \right) \\
					&= \tau_*\ R_*^{1/3}\ h_*\ \left( 1 - \left( 1 - f_\mathrm{vol} \right) f_\mathrm{ic} \right)\ r^{-4/3}.
\end{split}
\end{equation}
The bridging law between the effective opacity of the clumpy medium and that of a smooth outflow now needs to be slightly modified. Following \citet{Sundqvist2014}, this is given by
\begin{equation}            \label{Eq:BridgingLaw-2C}
    \frac{\kappa_\mathrm{eff}}{\kappa} = \frac{1 +\tau_\mathrm{cl}\ f_\mathrm{ic}}{1+\tau_\mathrm{cl}}.
\end{equation}
Note that as $\tau_\mathrm{cl}$ goes to zero, the smooth opacity and optical depths are recovered.
This is again recovered when $f_\mathrm{ic} = 1$, corresponding to two equal components.
When the inter-clump component is effectively void ($f_\mathrm{ic} = 0$), both the clump optical depth and the bridging law reduce to their one-component expressions.
In the limit where $\tau_\mathrm{cl} \gg 1$, we find that $\kappa_\mathrm{eff}/\kappa = f_\mathrm{ic}$. 
Unlike a one-component outflow, where the effective extinction $\kappa \langle\rho\rangle$ saturates to $1/h$, we now find that the ratio of the effective opacity of the clumpy outflow to that of the smooth outflow $\kappa_\mathrm{eff}/\kappa$ saturates to $f_\mathrm{ic}$. 
The opacity of a two-component outflow hence always depends on $\kappa$, implying a larger effective extinction compared to a one-component outflow (Figs. \ref{fig:Porosity-AV-1C-varH}, \ref{fig:Porosity-AV-2C-varH}, and \ref{fig:Porosity-AV-2C-varFic}).

The effective optical depth throughout a two-component outflow is then given by 
\begin{equation}           	\label{Eq:IntegralTauEff-2C}   
\begin{split}
    \tau_\mathrm{eff} &= \kappa \int_r^\infty  \langle\rho\rangle\ \frac{1 +\tau_\mathrm{cl}\ f_\mathrm{ic}}{1+\tau_\mathrm{cl}} \mathrm{d}r\\
                      &= \tau_*\ R_* \int_r^\infty \frac{1+\tau_*\ R_*^{1/3}\ h_*\ \left( 1 - \left( 1 - f_\mathrm{vol} \right) f_\mathrm{ic} \right)\ r^{-4/3}}{r^2 +\tau_*\ R_*^{1/3}\ h_*\ \left( 1 - \left( 1 - f_\mathrm{vol} \right) f_\mathrm{ic} \right)\ r^{2/3}}\ \mathrm{d}r,
\end{split}
\end{equation}
{where Eq. \ref{Eq:TauClump-2C} has been used for $\tau_\mathrm{cl}$.
}
If $h_* = 0$, this again reduces to the optical depth of the smooth wind $\tau = \tau_*\ R_*/r$.
The analytic solution of this integral can be found in Appendix \ref{subsect:App:TauEff:2C}.
In contrast to the one-component model above, the scaling of the effective optical depth of a two-component outflow is now influenced by two parameters: the porosity length $h_*$ and the inter-clump density contrast $f_\mathrm{ic}$. 

Similar to a one-component outflow, larger values of $h_*$ cause a larger decrease in visual extinction in the inner and intermediate regions of the outflow and the onset of the decrease in visual extinction (compared to the visual extinction of a smooth outflow) is located further out (Fig. \ref{fig:Porosity-AV-2C-varH}).
Smaller values of $f_\mathrm{ic}$ further lead to an increase in the clump optical depth and a decrease in the ratio $\kappa_\mathrm{eff}/\kappa$ (Fig. \ref{fig:Porosity-AV-2C-varFic}). 
A larger fraction of the interstellar radiation field is able to reach the inner and intermediate regions of the outflow, leading to a smaller visual extinction in these regions compared to that of a smooth outflow.
{In analogy with the one-component case, the effect of porosity is less pronounced for lower mass-loss rates.
}

\subsection{Implementation into the UdfA CSE model}                                        \label{subsect:Porosity:Implementation}

Three extra parameters are thus necessary to fully include the effect of clumping and porosity in the chemical model: $f_\mathrm{ic}$, $f_\mathrm{vol}$, and $l_*$. These are provided as input parameters to the model. 
A model is calculated for each component separately: one for the clump component, and, if present, one for the inter-clump component.

In all models, the mean density of the corresponding smooth outflow $\langle\rho\rangle$ is calculated using the mass-loss rate and outflow velocity (Eq. \ref{Eq:Density-2C}).
The mean density $\langle\rho\rangle$ is then used to calculate the optical depth of the corresponding smooth outflow. This value goes into the calculation of the effective optical depth of the clumpy outflow $\tau_\mathrm{eff}$ using Eq. \ref{Eq:TauEff-2C}. 
In practice, this equation is also used for the one-component outflow, as it corresponds to the special case $f_\mathrm{ic} = 0$ (see Eq. \ref{Eq:TauEff-1C}).
Note that for a two-component model, our formulation assumes that clumps and inter-clump medium are exposed to the same UV radiation field, so that the same effective optical depth can be used in both components.

The density of the clump component is derived from Eq. \ref{Eq:Density-2C}:
\begin{equation}        \label{Eq:Rho-CL}
    \rho_\mathrm{cl} =  \frac{1 - \left( 1-f_\mathrm{vol}\right) f_\mathrm{ic} }{f_\mathrm{vol}}\ \langle \rho \rangle\ \geq\ \langle\rho\rangle.
\end{equation}
For a one-component model, this reduces to $\rho_\mathrm{cl} = {\langle \rho \rangle}/{f_\mathrm{vol}}$. 
The density of the inter-clump component is derived from Eq. \ref{Eq:Fic}:
\begin{equation}        \label{Eq:Rho-IC}
    \rho_\mathrm{ic} =  f_\mathrm{ic}\ \langle \rho \rangle\ \leq\ \langle\rho\rangle.
\end{equation}

Note that the use of $\tau_\mathrm{eff}$ (instead of $\tau$) in the calculations affects only the rate of photodissociation reactions induced by UV photons. However, the specific density of the component ($\rho_\mathrm{cl}$ or $\rho_\mathrm{ic}$) is used when calculating the rates of two-body reactions. Since cosmic-ray photons and particles are assumed to have a radially independent incidence, the overall ionisation of the outflow depends on its density distribution. 

{Finally then}, the resulting mean number density of species X $\langle n_\mathrm{X} \rangle(r)$ in the clumpy outflow is calculated using $f_\mathrm{vol}\ n_\mathrm{cl,X}(r) + \left( 1 - f_\mathrm{vol} \right) n_\mathrm{ic,X}(r)$, as in Eq. \ref{Eq:Density-2C}. 
Its  fractional abundance relative to H$_2$ $\langle y_\mathrm{X} \rangle(r)$ is therefore given by
\begin{equation}		\label{Eq:FracAbun}
	\langle y_\mathrm{X} \rangle(r) = y_\mathrm{cl,X} + f_\mathrm{ic}\ \left( 1-f_\mathrm{vol} \right)\ \left( y_\mathrm{ic,X} - y_\mathrm{cl,X} \right),
\end{equation}
{with $y_\mathrm{cl,X}$ and $y_\mathrm{ic,X}$ its abundance relative to the H$_2$ abundance of the clump and inter-clump component, respectively.
}

\subsection{Relevant parameter ranges}                                        \label{subsect:Porosity:ParameterRange}

The parameter $l_*$ denotes the characteristic length scale of the clumps at the stellar radius. Since it is physically unrealistic to have clumps with a larger volume than that of the star appear at the stellar surface, we have that $l_* \leq R_*$. 
The parameters $f_\mathrm{vol}$ and $f_\mathrm{ic}$ are both allowed to range between 0 and 1. 

In addition, we here assume that the clumps become optically thin as the outflow merges with the interstellar medium, meaning that porosity effects become negligible. 
{The merger occurs where the density of the outflow becomes comparable to that of the interstellar medium, between $10^{17} - 10^{18}$ cm depending on the mass-loss rate.
}
{We find that this assumption imposes a mass-loss rate dependent upper limit on $h_*$.}
{By requiring that the visual extinction of a clumpy outflow is equal to that of its corresponding smooth outflow near the merger, we find that $h_* \lesssim 1 \times 10^{15}$ cm for $\dot{M} = 1 \times 10^{-5}\ \mathrm{M}_\odot\ \mathrm{yr}^{-1}$, $h_* \lesssim 5 \times 10^{15}$ cm for $\dot{M} = 10^{-6}\ \mathrm{M}_\odot\ \mathrm{yr}^{-1}$, and $h_* \lesssim 1 \times 10^{16}$ cm for $\dot{M} = 10^{-7}\ \mathrm{M}_\odot\ \mathrm{yr}^{-1}$.
}
{The assumption of optically thin clumps as the outflow merges with the interstellar medium hence introduces an extra constraint on $l_*$ and $f_\mathrm{vol}$.}
By using this maximal value for $h_*$, a lower limit can be found for $f_\mathrm{vol}$ assuming a certain value of $l_*$, and vice versa.

\begin{table}
	\caption{{Porosity lengths corresponding to the outflows modelled in Sect. \ref{sect:Results}. For the one-component outflows, all values of $h_*$ are modelled. 
	For the two-component outflows, only the porosity lengths corresponding to $l_* = 1 \times 10^{13}$ cm are modelled. 
	}} 
	\centering
    \begin{tabular}{r r r}
    \hline \hline 
    \noalign{\smallskip}
	$l_*	$ $\left[ \mathrm{cm} \right]$ & $f_\mathrm{vol}$	& $h_*$ $\left[ \mathrm{cm} \right]$ \\
	\hline
	\noalign{\smallskip}
	$5 \times 10^{12}$	& 0.05	& $5 \times 10^{14}$ \\
						& 0.20	& $2.5 \times 10^{13}$ \\
						& 0.40	& $1.25 \times 10^{13}$ \\
	\noalign{\smallskip}
	$1 \times 10^{13}$	& 0.05	& $2 \times 10^{14}$ \\
						& 0.20	& $5 \times 10^{13}$ \\
						& 0.40	& $2.5 \times 10^{13}$ \\
	\noalign{\smallskip}
 	$5 \times 10^{13}$	& 0.05	& $1 \times 10^{15}$ \\
						& 0.20	& $2.5 \times 10^{14}$ \\
						& 0.40	& $1.25 \times 10^{14}$ \\
   \hline
    \end{tabular}
    \label{table:Results-param}    
\end{table}

\section{Effect on chemistry}                                        \label{sect:Results}

\begin{figure*}
\centering
\includegraphics[width=1.\textwidth]{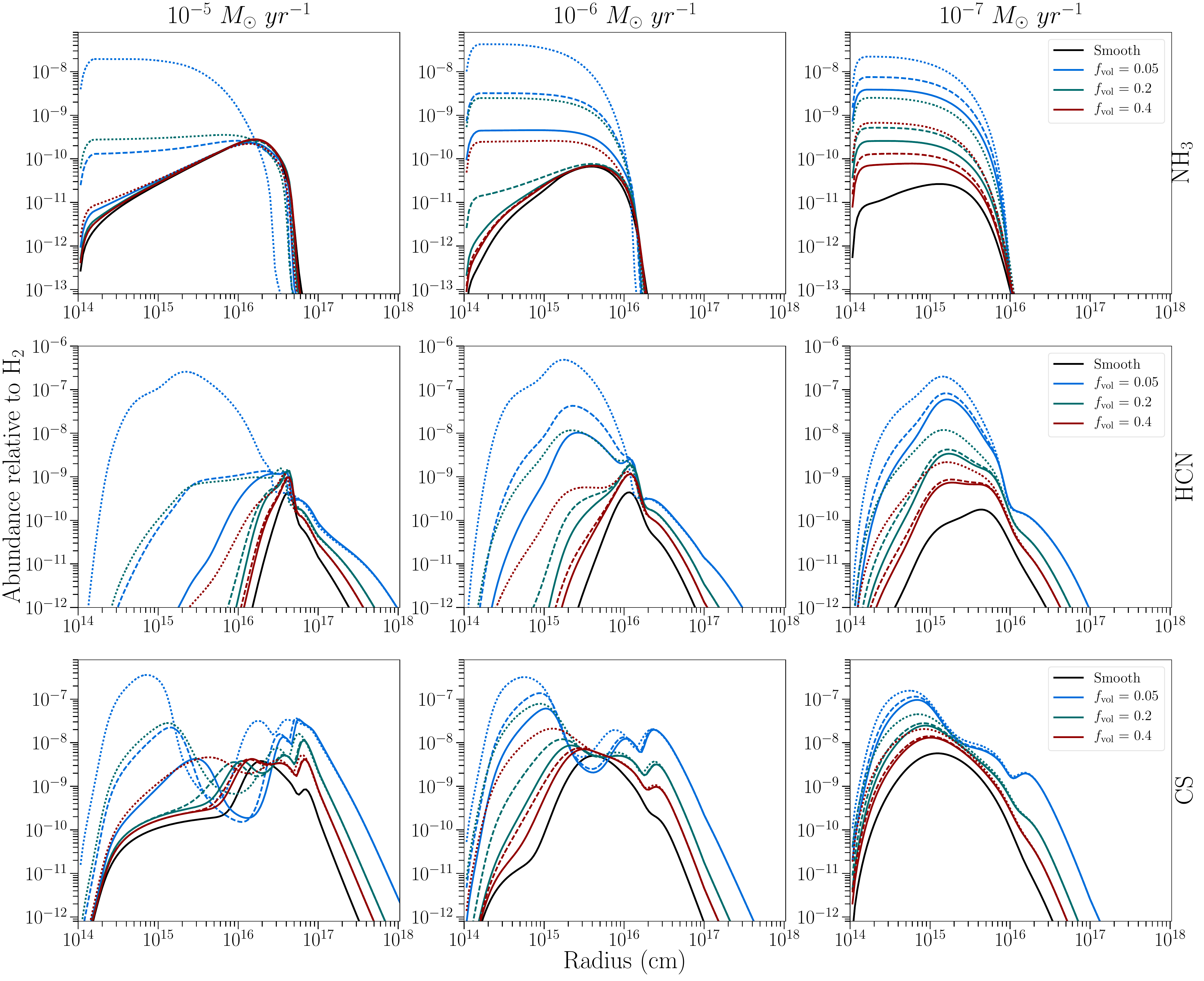}
\caption{Abundance of NH$_3$ (upper panels), HCN (middle panels), and CS (lower panels) relative to H$_2$ throughout a one-component O-rich outflow with different mass-loss rates $\dot{M}$ and clump volume filling factors $f_\mathrm{vol}$.
Solid black line: calculated abundance for a smooth, uniform outflow. 
Solid coloured line: characteristic clump scale $l_* = 5 \times 10^{12}$ cm, {porosity length $h_* = 1 \times 10^{14}, 2.5 \times 10^{13}, 1.25 \times 10^{13}$ cm for $f_\mathrm{vol} = 0.05, 0.2, 0.4$, respectively.} 
Dashed coloured line: $l_* = 10^{13}$ cm, {$h_* = 2 \times 10^{14}, 5 \times 10^{13}, 2.5 \times 10^{13}$ cm for $f_\mathrm{vol} = 0.05, 0.2, 0.4$, respectively.}
Dotted coloured line: $l_* = 5 \times 10^{13}$ cm, {$h_* = 1 \times 10^{15}, 2.5 \times 10^{14}, 1.25 \times 10^{14}$ cm for $f_\mathrm{vol} = 0.05, 0.2, 0.4$, respectively.}
{
Note that models with $f_\mathrm{vol} = 0.2,\ l_* = 5 \times 10^{12}$ cm (green, solid) and $f_\mathrm{vol} = 0.4,\ l_* =  1 \times 10^{13}$ cm (red, dashed) have the same porosity length $h_* = 2.5 \times 10^{13}$ cm.
}
For reference, $1\ R_* = 5 \times 10^{13}$ cm.
}
\label{fig:Results-OneComp-Orich}
\end{figure*}

We have studied the effect of clumping and porosity on both an O-rich and C-rich outflow. In both cases, a one-component and a two-component outflow are considered. 
The only physical variation between the models are the mass-loss rate and the description of the specific clumpiness of the outflow{, i.e. the values of the parameters $f_\mathrm{vol}$, $f_\mathrm{ic}$, and $l_*$}.
Despite being widely detected, there are few observational constraints on the specific density structures observed. 
Imaging of 22-GHz water masers suggests the presence of clouds that are 10 - 100 times more dense than the outflow, with filling factors of less than one percent \citep{Richards2012}.
From images of the molecular shell of IRC+10216, molecular clumps with sizes of $3.9 \times 10^{15}$ cm up to $15.6 \times 10^{15}$ cm have been detected at a {radial} distance of about {24} arcsec or $\sim {4.6} \times 10^{16}$ cm, for a distance of 130 pc \citep{Keller2017}. 
The {upper} limit of this size range is set by the beam size of these observations, hence smaller clumps may exist as well.  Applying our clump expansion assumption (Eq. 13) on the above size range, we then have 
${4 \times 10^{13} < l_* < 1.5 \times 10^{14}}$ cm.
Because the observational constraints on the specific clumpiness and porosity of observed outflows are very limited, we cannot estimate direct values for $f_\mathrm{vol}$, $f_\mathrm{ic}$, and $l_*$.
We therefore explore their parameter spaces.

\begin{table*}
    \caption{Column density [cm$^{-2}]$ of NH$_3$, HCN, and CS in a smooth O-rich outflow with different mass-loss rates, together with column density ratios relative to the smooth outflow for specific one-component outflows. The corresponding abundance profiles are shown in Fig. \ref{fig:Results-OneComp-Orich}. 
{Note that the models with $f_\mathrm{vol} = 0.2,\ l_* = 5 \times 10^{12}$ cm and $f_\mathrm{vol} = 0.4,\ l_* = 1 \times 10^{13}$ cm have the same porosity length $h_* = 2.5 \times 10^{13}$ cm.}
    }
    \centering
    \begin{tabular}{c c c c c c c c c c c}
    \hline \hline 
    \noalign{\smallskip}
    	$\dot{M}$ & Species & & {NH$_3$} & &  & {HCN} & &  & {CS} & \\
    \cmidrule(lr){1-1} \cmidrule(lr){2-2} \cmidrule(lr){3-5} \cmidrule(lr){6-8}  \cmidrule(lr){9-11} 
    	\noalign{\smallskip}
{\multirow{5}{*}{\rotatebox[origin=r]{90}{$10^{-5}\ \mathrm{M}_\odot\ \mathrm{yr}^{-1}$}}} &    Smooth & \multicolumn{3}{c}{ 6.3e+11  cm$^{-2}$} &  \multicolumn{3}{c}{ 4.0e+10  cm$^{-2}$} &  \multicolumn{3}{c}{ 2.9e+12  cm$^{-2}$} \\
    \cmidrule(lr){2-2} \cmidrule(lr){3-5} \cmidrule(lr){6-8}  \cmidrule(lr){9-11} 
&	$f_\mathrm{vol}$ & 0.05 & 0.2 & 0.4 & 0.05 & 0.2 & 0.4 & 0.05 & 0.2 & 0.4 \\	
    \cmidrule(lr){2-2} \cmidrule(lr){3-5} \cmidrule(lr){6-8}  \cmidrule(lr){9-11} 
& $l_* = 5 \times 10^{12}$ cm & 1.3e+00 & 1.1e+00 & 1.1e+00 & 1.4e+01 & 4.1e+00 & 2.3e+00 & 6.6e+00 & 2.0e+00 & 1.6e+00 \\
& $l_* = 1 \times 10^{13}$ cm & 9.7e+00 & 1.1e+00 & 1.1e+00 & 9.3e+01 & 4.8e+00 & 2.5e+00 & 4.9e+01 & 2.3e+00 & 1.7e+00 \\
& $l_* = 5 \times 10^{13}$ cm & 1.3e+03 & 2.0e+01 & 1.4e+00 & 5.0e+04 & 9.6e+01 & 5.1e+00 & 1.3e+03 & 7.4e+01 & 7.5e+00 \\
	\noalign{\smallskip}
	\hline
    	\noalign{\smallskip}
{\multirow{5}{*}{\rotatebox[origin=r]{90}{$10^{-6}\ \mathrm{M}_\odot\ \mathrm{yr}^{-1}$}}} &    Smooth & \multicolumn{3}{c}{ 3.0e+10  cm$^{-2}$} &  \multicolumn{3}{c}{ 2.0e+10  cm$^{-2}$} &  \multicolumn{3}{c}{ 1.3e+12  cm$^{-2}$} \\
    \cmidrule(lr){2-2} \cmidrule(lr){3-5} \cmidrule(lr){6-8}  \cmidrule(lr){9-11} 
&	$f_\mathrm{vol}$ & 0.05 & 0.2 & 0.4 & 0.05 & 0.2 & 0.4 & 0.05 & 0.2 & 0.4 \\	
    \cmidrule(lr){2-2} \cmidrule(lr){3-5} \cmidrule(lr){6-8}  \cmidrule(lr){9-11} 
& $l_* = 5 \times 10^{12}$ cm & 6.2e+01 & 1.4e+00 & 1.2e+00 & 1.7e+02 & 5.2e+00 & 2.6e+00 & 4.5e+01 & 3.6e+00 & 2.0e+00 \\
& $l_* = 1 \times 10^{13}$ cm & 4.4e+02 & 3.1e+00 & 1.3e+00 & 9.1e+02 & 7.5e+00 & 2.9e+00 & 1.3e+02 & 7.1e+00 & 2.6e+00 \\
& $l_* = 5 \times 10^{13}$ cm & 5.7e+03 & 3.4e+02 & 3.5e+01 & 1.7e+04 & 3.0e+02 & 1.4e+01 & 4.5e+02 & 8.3e+01 & 2.0e+01 \\
	\noalign{\smallskip}
	\hline
    	\noalign{\smallskip}
{\multirow{5}{*}{\rotatebox[origin=r]{90}{$10^{-7}\ \mathrm{M}_\odot\ \mathrm{yr}^{-1}$}}} &    Smooth & \multicolumn{3}{c}{ 1.1e+10  cm$^{-2}$} &  \multicolumn{3}{c}{ 7.9e+09  cm$^{-2}$} &  \multicolumn{3}{c}{ 6.4e+11  cm$^{-2}$} \\
    \cmidrule(lr){2-2} \cmidrule(lr){3-5} \cmidrule(lr){6-8}  \cmidrule(lr){9-11} 
&	$f_\mathrm{vol}$ & 0.05 & 0.2 & 0.4 & 0.05 & 0.2 & 0.4 & 0.05 & 0.2 & 0.4 \\	
    \cmidrule(lr){2-2} \cmidrule(lr){3-5} \cmidrule(lr){6-8}  \cmidrule(lr){9-11} 
& $l_* = 5 \times 10^{12}$ cm & 2.8e+02 & 1.9e+01 & 5.4e+00 & 4.6e+02 & 2.7e+01 & 6.5e+00 & 2.1e+01 & 5.1e+00 & 2.6e+00 \\
& $l_* = 1 \times 10^{13}$ cm & 5.4e+02 & 3.7e+01 & 9.3e+00 & 7.4e+02 & 3.6e+01 & 7.9e+00 & 2.7e+01 & 6.2e+00 & 2.9e+00 \\
& $l_* = 5 \times 10^{13}$ cm & 1.6e+03 & 1.8e+02 & 4.7e+01 & 2.4e+03 & 1.4e+02 & 2.6e+01 & 4.4e+01 & 1.2e+01 & 5.3e+00 \\
\noalign{\smallskip}
    	\hline 	\hline
    \end{tabular}%
    \label{table:Results-CD-1C-O}
\end{table*}

For the one-component outflow, we show the resulting abundance profiles and column densities for a selection of molecules for $f_\mathrm{vol} = $ 0.05, 0.2, and 0.4, corresponding to an outflow where the clumps take up a small fraction, almost a quarter, and almost half of the total volume of the outflow.
The parameter $l_*$ has values of $5 \times 10^{12}$ cm, $10^{13}$ cm, and $5 \times 10^{13}$ cm, corresponding to clumps with a characteristic length scale of 0.1 and 0.2 times the stellar radius, and to the extreme case where $l_* = R_*$.
{The corresponding porosity lengths $h_*$ for the different models are listed in Table \ref{table:Results-param}. 
Note that the outflows with $f_\mathrm{vol} = 0.2$, $l_* = 5 \times 10^{12}$ cm and $f_\mathrm{vol} = 0.4$, $l_* = 1 \times 10^{13}$ cm have the same porosity length, but different clump overdensities.
}

Figs. \ref{fig:Results-OneComp-Orich} and \ref{fig:Results-OneComp-Crich}, and Tables \ref{table:Results-CD-1C-O} and \ref{table:Results-CD-1C-C} show the results for respectively O- and C-rich one-component outflows.

\begin{figure*}
\centering
\includegraphics[width=1.\textwidth]{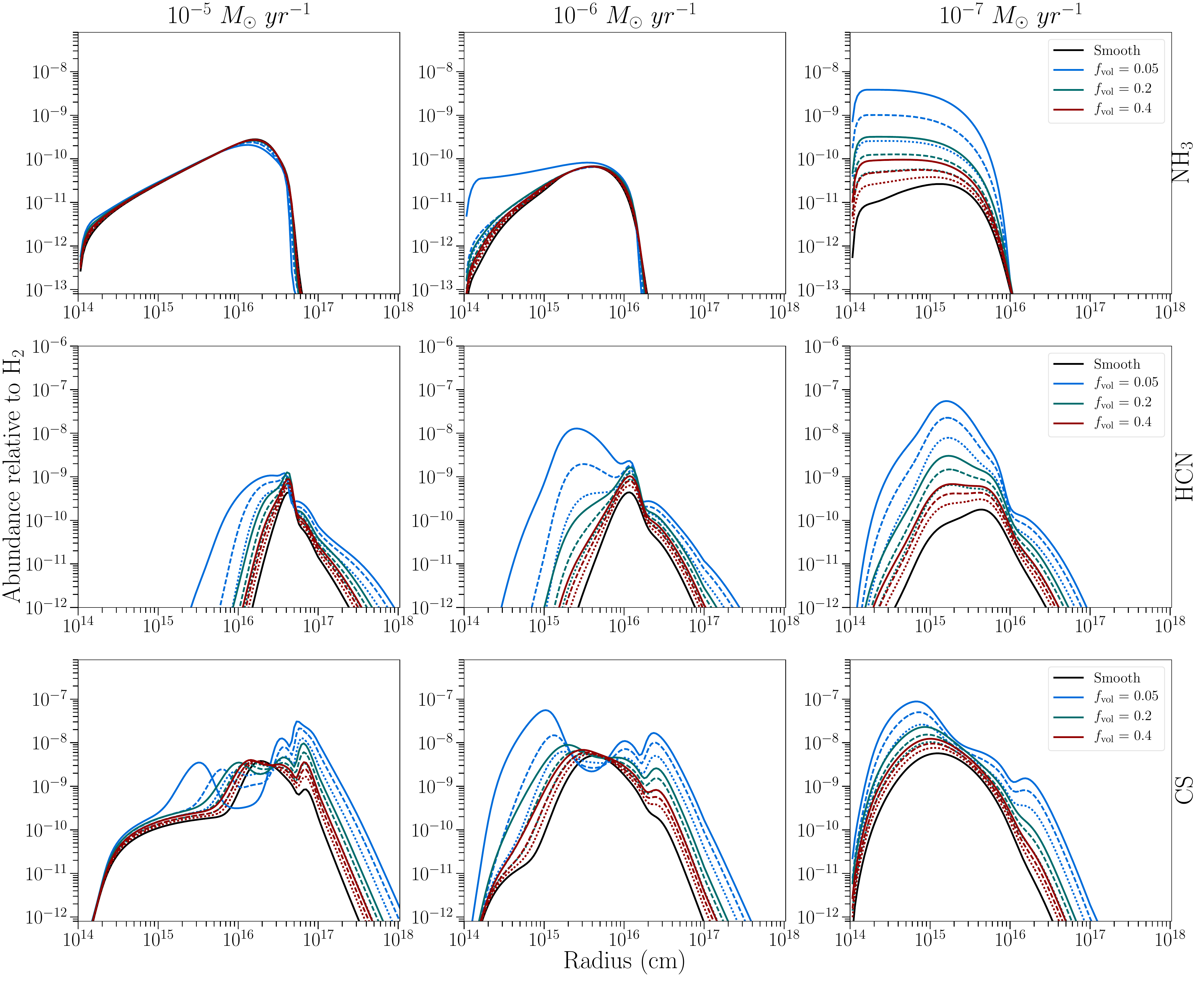}
\caption{Abundance of NH$_3$ (upper panels), HCN (middle panels), and CS (lower panels) relative to H$_2$ throughout a two-component O-rich outflow with different mass-loss rates $\dot{M}$ and clump volume filling factors $f_\mathrm{vol}$. The characteristic size of the clumps at the stellar radius is $l_* = 10^{13}$ cm.
{
Blue lines: porosity length $h_* = 2 \times 10^{14}$ cm.
Green lines: $h_* = 5 \times 10^{13}$ cm.
Red lines: $h_* = 2.5 \times 10^{13}$ cm.
}
Solid black line: calculated abundance for a smooth, uniform outflow. 
Solid coloured line: density contrast between the inter-clump and smooth outflow $f_\mathrm{ic} = 0.1$. Dashed coloured line: $f_\mathrm{ic} = 0.3$. Dotted coloured line: $f_\mathrm{ic} = 0.5$.
{Note that the models with $f_\mathrm{vol} = 0.4$ (red) have the same porosity length as the one-component outflows with $f_\mathrm{vol} = 0.2,\ l_* 5 \times 10^{12}$ cm and $f_\mathrm{vol} = 0.4,\ l_* = 1 \times 10^{13}$ cm.}
For reference, $1\ R_* = 5 \times 10^{13}$ cm.
}
\label{fig:Results-TwoComp-Orich}
\end{figure*}

For the two-component outflow, we show abundance profiles and column densities for $f_\mathrm{vol}$ = 0.05, 0.2, and 0.4 and for $f_\mathrm{ic}$ = 0.1, 0.3 and 0.5 (Figs. \ref{fig:Results-TwoComp-Orich} and \ref{fig:Results-TwoComp-Crich}, and Tables \ref{table:Results-CD-2C-O} and \ref{table:Results-CD-2C-C}), corresponding to a very rarefied, a moderately rarefied, and an inter-clump component with a density equal to half the density of the corresponding smooth outflow.
The parameter $l_*$ is now fixed at $10^{13}$ cm {to show the effect of the parameter $f_\mathrm{ic}$}. 
{The corresponding porosity lengths $h_*$ are listed in Table \ref{table:Results-param}.
}
Figs. \ref{fig:Results-TwoComp-Orich} and \ref{fig:Results-TwoComp-Crich}, and Tables \ref{table:Results-CD-2C-O} and \ref{table:Results-CD-2C-C} show the results for respectively O- and C-rich two-component outflows.
The visual extinction and decrease in interstellar UV radiation field throughout these one-component and two-component outflows are shown in Figs. \ref{fig:App:AVresults:1C} and Figs. \ref{fig:App:AVresults:2C}, respectively.

There are several general trends with $f_\mathrm{vol}$, $f_\mathrm{ic}$, and $l_*$ visible in the abundance profiles and column densities.
Larger values of $l_*$, or similarly $h_* = l_*/f_\mathrm{vol}$, increase the production of unexpected species in the inner region, as they lead to a larger UV radiation field in the inner wind, inducing photodissociation of parent species. This results in a larger column density and a larger abundance in the inner region of the unexpected species.
A larger value of $l_*$ or $h_*$ also shifts the photodissociation radius of all species in the outer region more inward.
{Smaller values of $f_\mathrm{ic}$ and $f_\mathrm{vol}$ generally result in an increase in column density of the investigated molecules relative to the corresponding smooth outflow}. 
{This is due to both clumping and porosity. The increased overdensity of the clumps for smaller values of $f_\mathrm{ic}$ and $f_\mathrm{vol}$ (Eq. \ref{Eq:Rho-CL}) causes the chemistry to occur at a faster rate and the abundances to increase, since the reaction rates are proportional to the density of the reactants. Smaller values of of $f_\mathrm{ic}$ and $f_\mathrm{vol}$ also correspond to more UV radiation reaching the inner wind (Eq. \ref{Eq:BridgingLaw-2C}, Figs. \ref{fig:Porosity-AV-1C-varH} and \ref{fig:Porosity-AV-2C-varFic})
}
``Highly porous'' outflows, with both a large clump overdensity and porosity length, therefore have the largest effect on the circumstellar chemistry.

The effect of clumping and porosity extends of course beyond these general trends, affecting the chemical formation and destruction pathways of all species. This is reflected in the shape of their abundance profiles.
In Sect{s.} \ref{subsect:Results-Orich} and \ref{subsect:Results-Crich}, we discuss the effect on the column densities and abundance profiles of some key unexpected molecules in detail for both the O-rich and the C-rich outflow.
These species have been chosen as they have been detected with an abundance larger than expected by TE in the inner region (see {Sect. \ref{sect:Introduction}} for an overview), and as the changes to their chemical formation and destruction pathways induced by {clumping} and porosity are representative for the other species present in the outflow.

{The species discussed in Sects. \ref{subsect:Results-Orich} and \ref{subsect:Results-Crich} have also been modelled by \citet{Agundez2010}. In order to facilitate a comparison, the column densities and abundance profiles of the other species modelled by \citet{Agundez2010} can be found in Appendices \ref{subsect:App:AddMolecules:Orich} and \ref{subsect:App:AddMolecules:Crich}, respectively, together with two other species that showcase the effects of clumping and porosity on the chemistry.
The column densities and abundance profiles of so far undetected molecules that experience a large increase in predicted abundance due to a clumpy outflow, making them potentially detectable, are given in Appendix \ref{sect:App:Predictions}.
}

\subsection{Oxygen-rich outflows}                            \label{subsect:Results-Orich}

{Under the assumption of TE chemistry,} C- and N-bearing molecules are not expected to be {abundantly} present in the inner region of O-rich outflows \citep{Cherchneff2006,Agundez2010,Gobrecht2016}. In the following, we investigate the effect of clumping and porosity on the chemical pathways involved in the formation and destruction of NH$_3$, HCN, and CS in the inner, intermediate, and outer regions of the stellar wind.
{The location of these regions depends on the mass-loss rate. The inner region ends roughly around $\sim 10^{15}$ cm. The intermediate region ends where photodissociation becomes the dominant destruction process, around $\sim 3 \times 10^{16}$ cm. These boundaries shift inward with increasing mass-loss rate.
}

{We compare our results to observed abundances in the O-rich AGB stars IK Tau, TX Cam, and R Dor.
IK Tau and TX Cam have high mass-loss rates of $\sim 5 \times 10^{-6}$ M$_\odot$ yr$^{-1}$ \citep{Decin2010} and $\sim 3 \times 10^{-6}$ M$_\odot$ yr$^{-1}$ \citep{Bujarrabal1994}, respectively.
R Dor has a low mass-loss rate of $\sim 1 \times 10^{-7}$ M$_\odot$ yr$^{-1}$ \citep{Olofsson2002}.
Only a rough comparison to these observations is possible, since the physical parameters and input species of our chemical model are not tailored to these stars.
Note that not all molecules have been observed in all three stars.
}

\subsubsection{NH$_3$}										\label{subsubsect:Results-Orich-NH3}

The NH$_3$ abundance in the inner region can reach up to a few times $10^{-8}$ relative to H$_2$ {in highly porous one-component outflows}, an increase of three orders of magnitude relative to a smooth outflow. {In such outflows, the NH$_3$} column density can increase up to {three} orders of magnitude relative to a smooth outflow.
The abundance profiles are shown in the upper panel of Figs. \ref{fig:Results-OneComp-Orich} and \ref{fig:Results-TwoComp-Orich}. The column densities are listed in Tables \ref{table:Results-CD-1C-O} and \ref{table:Results-CD-2C-O}.

For highly porous outflows, the abundance of NH$_3$ reaches its maximum value close to the {starting radius ($10^{14}$ cm) of the model} and is constant throughout the inner and intermediate regions. This differs from a smooth outflow, or an outflow where the increase in the UV field in the inner region is not significant enough, where the NH$_3$ abundance increases throughout the inner and intermediate regions.

{The observed inner wind abundance of NH$_3$ in IK Tau is $6 \times 10^{-7}$ relative to H$_2$ \citep{Wong2017}. In the outflows of other oxygen-rich evolved stars, NH$_3$ has a similar abundance of $\sim 10^{-6}$ \citep{Wong2017}. 
Our models do not produce such large abundances for any mass-loss rate considered, but including a highly porous clumpy outflow does lead to inner wind abundances about two orders of magnitude larger.
For high mass-loss rate outflows, a one-component outflow is needed to achieve such an increase.
}

\paragraph{Inner region}

NH$_3$ is predominantly produced via the hydrogenation of N through reactions with H$_2$.
In outflows with a well-shielded inner region, N is produced only through
\begin{equation*}
\mathrm{N_2 + He^+ \rightarrow N^+ + N + e^-}.
\end{equation*}
As He$^+$ is only produced via cosmic-ray ionisation, the formation of N is independent of the UV field.
However, for outflows with a larger UV field present in the inner region, either through a lower mass-loss rate or a larger porosity, N is also produced via photodissociation of the parent species N$_2$. This yields a higher abundance of N, and subsequently of NH$_3$.
The effect of $f_\mathrm{vol}$, $f_\mathrm{ic}$, and $l_*$ on the UV radiation field in the inner region is therefore strongly noticeable in Figs. \ref{fig:Results-OneComp-Orich} and \ref{fig:Results-TwoComp-Orich}.

\paragraph{Intermediate region}


The sharp decline of NH$_3$ {at the end of the intermediate region} is caused by photodissociation.
{In one-component outflows with larger mass-loss rates, the sharp decline shifts inward with increasing $h_*$, equivalent to increasing $l_*$ and decreasing $f_\mathrm{vol}$. This is due to the strong decrease in extinction in this region (Fig. \ref{fig:Porosity-AV-1C-varH}). 
}
Low mass-loss rate outflows {and two-component outflows} do not experience as large a decrease in visual extinction in this region. The onset of photodissociation for these outflows even moves outward with decreasing $f_\mathrm{vol}$, as the clumps are self-shielded.

\begin{table*}
    \caption{Column density [cm$^{-2}]$ of NH$_3$, HCN, and CS in a smooth O-rich outflow with different mass-loss rates, together with column density ratios relative to the smooth outflow for specific two-component outflows. The corresponding abundance profiles are shown in Fig. \ref{fig:Results-TwoComp-Orich}. 
{Note that models with $f_\mathrm{vol} = 0.4$ have the same porosity length $h_* = 2.5 \times 10^{13}$ cm as the one-component models with $f_\mathrm{vol} = 0.2,\ l_* = 5 \times 10^{12}$ cm and $f_\mathrm{vol} = 0.4,\ l_* = 1 \times 10^{13}$ cm.}
    }
    \centering
    \begin{tabular}{c c c c c c c c c c c}
    \hline \hline 
    \noalign{\smallskip}
    	$\dot{M}$ & Species & & {NH$_3$} & &  & {HCN} & &  & {CS} & \\
    \cmidrule(lr){1-1} \cmidrule(lr){2-2} \cmidrule(lr){3-5} \cmidrule(lr){6-8}  \cmidrule(lr){9-11} 
    	\noalign{\smallskip}
{\multirow{5}{*}{\rotatebox[origin=r]{90}{$10^{-5}\ \mathrm{M}_\odot\ \mathrm{yr}^{-1}$}}} &    Smooth & \multicolumn{3}{c}{ 6.3e+11  cm$^{-2}$} &  \multicolumn{3}{c}{ 4.0e+10  cm$^{-2}$} &  \multicolumn{3}{c}{ 2.9e+12  cm$^{-2}$} \\
    \cmidrule(lr){2-2} \cmidrule(lr){3-5} \cmidrule(lr){6-8}  \cmidrule(lr){9-11} 
&	$f_\mathrm{vol}$ & 0.05 & 0.2 & 0.4 & 0.05 & 0.2 & 0.4 & 0.05 & 0.2 & 0.4 \\	
    \cmidrule(lr){2-2} \cmidrule(lr){3-5} \cmidrule(lr){6-8}  \cmidrule(lr){9-11} 
& $f_\mathrm{ic} = 0.1$ & 1.1e+00 & 1.1e+00 & 1.1e+00 & 1.5e+01 & 3.9e+00 & 2.2e+00 & 3.8e+00 & 1.9e+00 & 1.5e+00 \\
& $f_\mathrm{ic} = 0.3$ & 1.1e+00 & 1.1e+00 & 1.1e+00 & 7.3e+00 & 2.7e+00 & 1.7e+00 & 2.1e+00 & 1.6e+00 & 1.3e+00 \\
& $f_\mathrm{ic} = 0.5$ & 1.1e+00 & 1.1e+00 & 1.2e+00 & 4.0e+00 & 1.9e+00 & 1.5e+00 & 1.6e+00 & 1.3e+00 & 1.3e+00 \\
	\noalign{\smallskip}
	\hline
	\noalign{\smallskip}
{\multirow{5}{*}{\rotatebox[origin=r]{90}{$10^{-6}\ \mathrm{M}_\odot\ \mathrm{yr}^{-1}$}}} &    Smooth & \multicolumn{3}{c}{ 3.0e+10  cm$^{-2}$} &  \multicolumn{3}{c}{ 2.0e+10  cm$^{-2}$} &  \multicolumn{3}{c}{ 1.3e+12  cm$^{-2}$} \\
    \cmidrule(lr){2-2} \cmidrule(lr){3-5} \cmidrule(lr){6-8}  \cmidrule(lr){9-11} 
&	$f_\mathrm{vol}$ & 0.05 & 0.2 & 0.4 & 0.05 & 0.2 & 0.4 & 0.05 & 0.2 & 0.4 \\	
    \cmidrule(lr){2-2} \cmidrule(lr){3-5} \cmidrule(lr){6-8}  \cmidrule(lr){9-11} 
& $f_\mathrm{ic} = 0.1$ & 5.8e+00 & 1.3e+00 & 1.2e+00 & 2.0e+02 & 5.2e+00 & 2.5e+00 & 3.6e+01 & 4.2e+00 & 2.2e+00 \\
& $f_\mathrm{ic} = 0.3$ & 1.4e+00 & 1.2e+00 & 1.2e+00 & 2.7e+01 & 3.1e+00 & 1.9e+00 & 7.0e+00 & 2.3e+00 & 1.7e+00 \\
& $f_\mathrm{ic} = 0.5$ & 1.2e+00 & 1.2e+00 & 1.2e+00 & 6.5e+00 & 2.1e+00 & 1.5e+00 & 2.8e+00 & 1.7e+00 & 1.4e+00 \\
	\noalign{\smallskip}
	\hline
	\noalign{\smallskip}
{\multirow{5}{*}{\rotatebox[origin=r]{90}{$10^{-7}\ \mathrm{M}_\odot\ \mathrm{yr}^{-1}$}}} &    Smooth & \multicolumn{3}{c}{ 1.1e+10  cm$^{-2}$} &  \multicolumn{3}{c}{ 7.9e+09  cm$^{-2}$} &  \multicolumn{3}{c}{ 6.4e+11  cm$^{-2}$} \\
    \cmidrule(lr){2-2} \cmidrule(lr){3-5} \cmidrule(lr){6-8}  \cmidrule(lr){9-11} 
&	$f_\mathrm{vol}$ & 0.05 & 0.2 & 0.4 & 0.05 & 0.2 & 0.4 & 0.05 & 0.2 & 0.4 \\	
    \cmidrule(lr){2-2} \cmidrule(lr){3-5} \cmidrule(lr){6-8}  \cmidrule(lr){9-11} 
& $f_\mathrm{ic} = 0.1$ & 2.8e+02 & 2.3e+01 & 6.7e+00 & 4.5e+02 & 2.5e+01 & 6.2e+00 & 2.0e+01 & 4.9e+00 & 2.5e+00 \\
& $f_\mathrm{ic} = 0.3$ & 7.3e+01 & 9.1e+00 & 3.7e+00 & 1.7e+02 & 1.2e+01 & 3.8e+00 & 1.1e+01 & 3.2e+00 & 1.9e+00 \\
& $f_\mathrm{ic} = 0.5$ & 1.9e+01 & 3.9e+00 & 2.3e+00 & 5.6e+01 & 5.6e+00 & 2.4e+00 & 5.3e+00 & 2.1e+00 & 1.6e+00 \\
     \noalign{\smallskip}
    	\hline \hline
    \end{tabular}%
    \label{table:Results-CD-2C-O}
\end{table*}

\subsubsection{HCN}											\label{subsubsect:Results-Orich-HCN}

{In highly porous one-component outflows, the HCN abundance can reach up to a few times 10$^{-7}$ relative to H$_2$ in the inner region, an increase of 10 orders of magnitude relative to a smooth outflow. The HCN column density can increase in such outflows up to four orders of magnitude relative to a smooth outflow. 
}
The abundance profiles are shown in the middle panel of Figs. \ref{fig:Results-OneComp-Orich} and \ref{fig:Results-TwoComp-Orich}. The column densities are listed in Tables \ref{table:Results-CD-1C-O} and \ref{table:Results-CD-2C-O}.

The overdensity of the clumps increases the overall HCN abundance, but the strength of the UV radiation field in the inner region plays a crucial role in its formation. 
{The importance of the UV radiation field can already be deduced by comparing the HCN abundance in smooth outflows for different mass-loss rates. Its abundance in the inner region is larger for lower mass-loss rates, as these outflows have an overall lower visual extinction.
}

{HCN has been observed in IK Tau with an inner wind abundance of $4.4 \times 10^{-7}$ by \citet{Decin2010} and $9.8 \times 10^{-7}$ by \citet{Bujarrabal1994}. In TX Cam, it has been observed with a peak abundance of $2.2 \times 10^{-6}$ \citep{Bujarrabal1994}.
Its inner wind abundance in R Dor is $5.0 \times 10^{-7}$ \citep{VandeSandeRDor}. 
Our models with $\dot{M} = 1 \times 10^{-6}$ M$_\odot$ yr$^{-1}$ produce an abundance similar to the inner wind abundance of IK Tau and TX Cam only for highly porous one-component outflows. 
The observed inner wind abundance of R Dor is reproduced within a factor of five with $\dot{M} = 1 \times 10^{-7}$ M$_\odot$ yr$^{-1}$ by both highly porous one- and two-component outflows.
}

\paragraph{Inner region}

In the inner region, HCN is mainly produced via
\begin{equation*}
\mathrm{H_2 + CN \rightarrow HCN + H}.
\end{equation*}
The abundance of CN depends on the UV radiation field in the inner region, as it is mainly produced via
\begin{equation*}
\mathrm{C + NO \rightarrow CN + O}
\end{equation*}
and
\begin{equation*}
\mathrm{N + CS \rightarrow CN + S}.
\end{equation*}
The abundances of all four reactants increase with increased UV radiation field. C and N are produced in greater abundance due to the contribution of the photodissociation of the parent species CO and N$_2$. NO and CS are produced via reactions between the parent species SO and N and C respectively, and are therefore also more abundant.

Moreover, an increased UV radiation field opens up additional formation pathways that contribute significantly to the formation of HCN. The abundances of all reactants increase with the UV radiation field.
The main additional reactions are 
\begin{equation*}
\mathrm{N + CH_2 \rightarrow HCN + H},
\end{equation*}
\begin{equation*}
\mathrm{N + CH_3 \rightarrow HCN + H_2},
\end{equation*}
and
\begin{equation*}
\mathrm{H + H_2CN \rightarrow HCN + H_2},
\end{equation*}
where H$_2$CN is produced via N + CH$_3$.
CH$_2$ and CH$_3$ are produced via hydrogenation of C through reactions with H$_2$ and are therefore more abundant as well.

\paragraph{Intermediate region}

The HCN abundance increases and forms a bump on the abundance profile near the end of the intermediate region. The feature is more prominent for higher mass-loss rates.
In this region, HCN is mainly produced via
\begin{equation*}
\mathrm{CH + NO  \rightarrow HCN + O}.
\end{equation*}
The bump is visible in the abundance profile of CH, and is transferred to the profile of HCN.
CH is produced through CH$_3^+$ + e$^-$ {in this region}, where CH$_3^+$ is the product of hydrogenation of C$^+$ through reactions with H$_2$. 
In the inner region, C$^+$ is predominantly produced via He$^+$ + CO, and is therefore independent on the UV radiation field.
{The C$^+$ production reaction} shifts to the photoionisation of C {near the end of the intermediate region}. 
The shift to photoionisation of C {occurs closer to the central star for larger UV radiation fields and} increases {the} C$^+$ abundance. { This increase creates the bump feature, which is transferred to the abundance profile of HCN through CH$_3^+$ and CH.}

The end of the feature corresponds to the decrease in NO. In the intermediate reaction, NO is predominantly produced via 
\begin{equation*}
\mathrm{OH + N \rightarrow NO + H},
\end{equation*}
where OH is produced through photodissociation of the parent species H$_2$O. The decline in NO abundance corresponds to the photodissociation radius of H$_2$O, decreasing the abundance of OH and therefore NO.
The earlier onset of photodissociation of H$_2$O in highly porous outflows with a high mass-loss rate and low mass-loss rate outflows shifts the peak in NO abundance inward, and hence makes the bump in the HCN profile less pronounced.

\begin{figure*}
\centering
\includegraphics[width=1.\textwidth]{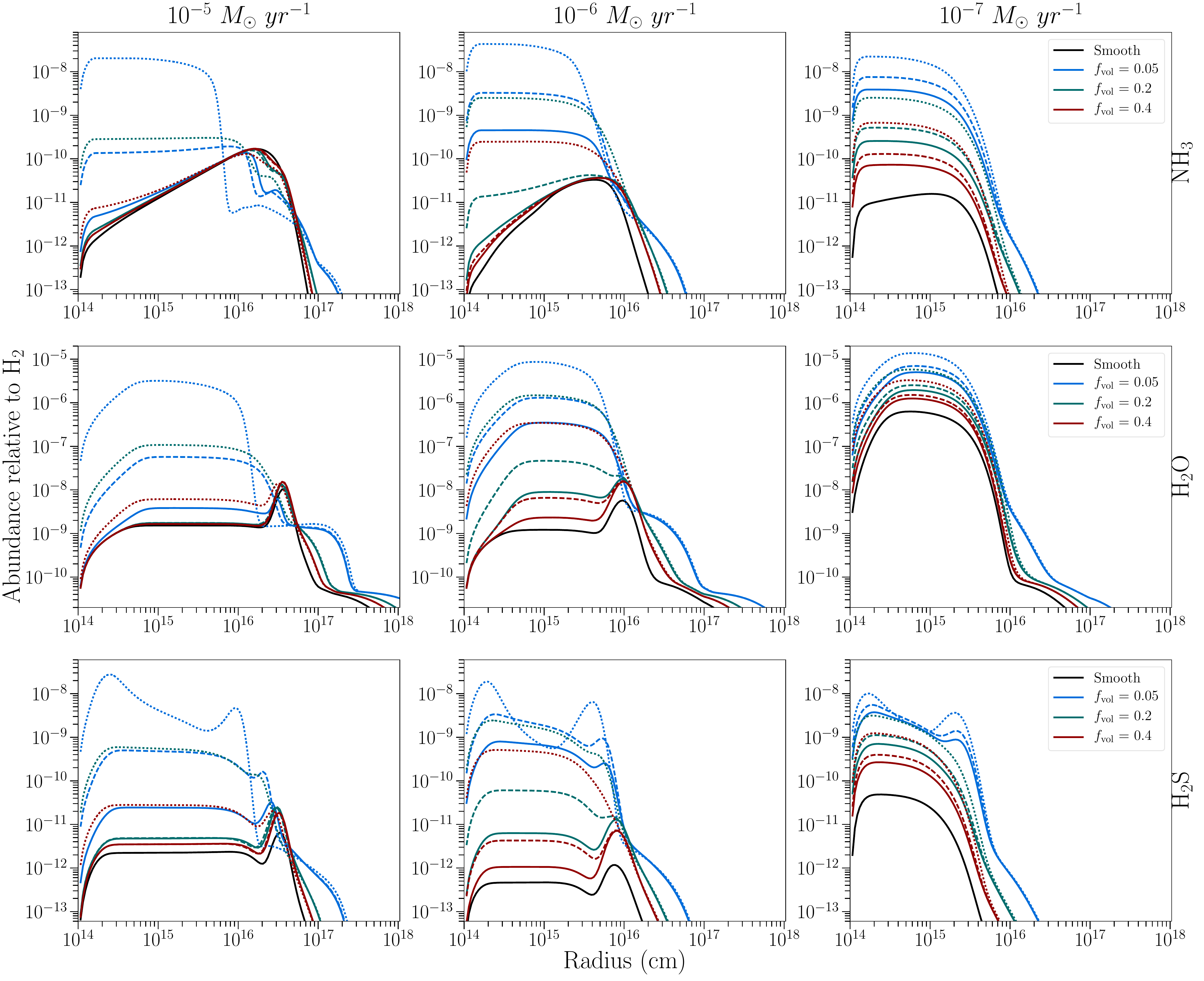}
\caption{Abundance of NH$_3$ (upper panels), H$_2$O (middle panels), and H$_2$S (lower panels) relative to H$_2$ throughout a one-component C-rich outflow with different mass-loss rates $\dot{M}$ and clump volume filling factors $f_\mathrm{vol}$.
Solid black line: calculated abundance for a smooth, uniform outflow. 
Solid coloured line: characteristic clump scale $l_* = 5 \times 10^{12}$ cm, {porosity length $h_* = 1 \times 10^{14}, 2.5 \times 10^{13}, 1.25 \times 10^{13}$ cm for $f_\mathrm{vol} = 0.05, 0.2, 0.4$, respectively.} 
Dashed coloured line: $l_* = 10^{13}$ cm, {$h_* = 2 \times 10^{14}, 5 \times 10^{13}, 2.5 \times 10^{13}$ cm for $f_\mathrm{vol} = 0.05, 0.2, 0.4$, respectively.}
Dotted coloured line: $l_* = 5 \times 10^{13}$ cm, {$h_* = 1 \times 10^{15}, 2.5 \times 10^{14}, 1.25 \times 10^{14}$ cm for $f_\mathrm{vol} = 0.05, 0.2, 0.4$, respectively.}
{
Note that models with $f_\mathrm{vol} = 0.2,\ l_* = 5 \times 10^{12}$ cm (green, solid) and $f_\mathrm{vol} = 0.4,\ l_* =  1 \times 10^{13}$ cm (red, dashed) have the same porosity length $h_* = 2.5 \times 10^{13}$ cm.
}
For reference, $1\ R_* = 5 \times 10^{13}$ cm.
}
\label{fig:Results-OneComp-Crich}
\end{figure*}

\paragraph{Outer region}

The slope of the decline in HCN abundance differs for different values of $f_\mathrm{vol}$.
At the start of the outer region, reactions with C$^+$ contribute significantly to the destruction of HCN, along with its photodissociation. The C$^+$ abundance in this region is smaller for smaller values of $f_\mathrm{vol}$, as the increased self-shielding of the clumps makes the photoionisation of C less efficient.
The smaller C$^+$ abundance leads to a slower decline in HCN abundance.

\subsubsection{CS}											\label{subsubsect:Results-Orich-CS}

{A CS abundance of up to a few times 10$^{-7}$ relative to H$_2$ is reached in the inner regions of highly porous one-component outflows, an increase of three orders of magnitude relative to a smooth outflow. For such outflows, the CS column density increases up to three orders of magnitude relative to a smooth outflow.
}
The abundance profiles are shown in the lower panel of Figs. \ref{fig:Results-OneComp-Orich} and \ref{fig:Results-TwoComp-Orich}. The column densities are listed in Tables \ref{table:Results-CD-1C-O} and \ref{table:Results-CD-2C-O}.

Both the overdensity of the clumps and the UV radiation field throughout the outflow have a large influence on the abundance of CS throughout the outflow. 
The overdensity of the clumps increases the overall abundance. 
For very overdense clumps, the abundance drops significantly in the intermediate region. 

{The CS abundance in the inner wind of IK Tau lies between $8.1 \times 10^{-8}$ and $3.0 \times 10^{-7}$ \citep{Bujarrabal1994,Decin2010,Kim2010}. 
In TX Cam, its peak abundance lies between $1 \times 10^{-6}$ {and} $5.0 \times 10^{-7}$ \citep{Lindqvist1988,Olofsson1991,Bujarrabal1994}.
Our models with $\dot{M} = 1 \times 10^{-6}$ M$_\odot$ yr$^{-1}$ have peak abundances similar to these observed values in both one-component and two-component highly porous outflows. In models with $\dot{M} = 1 \times 10^{-5}$ M$_\odot$ yr$^{-1}$, such abundances are only produced in highly porous one-component outflows.
}

\paragraph{Inner region}

The formation of CS in the inner region of the wind occurs predominantly through
\begin{equation*}
\mathrm{C + SO \rightarrow CS + O}.
\end{equation*}
The abundance of C increases for larger UV fields in the inner region (see Sect. \ref{subsubsect:Results-Orich-HCN}), increasing the CS abundance.

 \begin{table*}
    \caption{Column density [cm$^{-2}]$ of NH$_3$, H$_2$O, and H$_2$S in a smooth C-rich outflow with different mass-loss rates, together with column density ratios relative to the smooth outflow for specific one-component outflows. The corresponding abundance profiles are shown in Fig. \ref{fig:Results-OneComp-Crich}. 
{Note that the models with $f_\mathrm{vol} = 0.2,\ l_* = 5 \times 10^{12}$ cm and $f_\mathrm{vol} = 0.4,\ l_* = 1 \times 10^{13}$ cm have the same porosity length $h_* = 2.5 \times 10^{13}$ cm.}
    }
    \centering
    \begin{tabular}{c c c c c c c c c c c}
    \hline \hline 
    \noalign{\smallskip}
    	 $\dot{M}$ & Species & & {NH$_3$} & &  & {H$_2$O} & &  & {H$_2$S} & \\
    \cmidrule(lr){1-1} \cmidrule(lr){2-2} \cmidrule(lr){3-5} \cmidrule(lr){6-8}  \cmidrule(lr){9-11} 
    	\noalign{\smallskip}
{\multirow{5}{*}{\rotatebox[origin=r]{90}{$10^{-5}\ \mathrm{M}_\odot\ \mathrm{yr}^{-1}$}}} &    Smooth & \multicolumn{3}{c}{ 3.3e+11  cm$^{-2}$} &  \multicolumn{3}{c}{ 4.1e+13  cm$^{-2}$} &  \multicolumn{3}{c}{ 6.8e+10  cm$^{-2}$} \\
    \cmidrule(lr){2-2} \cmidrule(lr){3-5} \cmidrule(lr){6-8}  \cmidrule(lr){9-11} 
&	$f_\mathrm{vol}$ & 0.05 & 0.2 & 0.4 & 0.05 & 0.2 & 0.4 & 0.05 & 0.2 & 0.4 \\	
    \cmidrule(lr){2-2} \cmidrule(lr){3-5} \cmidrule(lr){6-8}  \cmidrule(lr){9-11} 
& $l_* = 5 \times 10^{12}$ cm & 1.4e+00 & 1.1e+00 & 1.1e+00 & 1.8e+00 & 1.1e+00 & 1.0e+00 & 9.2e+00 & 2.0e+00 & 1.5e+00 \\
& $l_* = 1 \times 10^{13}$ cm & 1.8e+01 & 1.1e+00 & 1.1e+00 & 2.3e+01 & 1.1e+00 & 1.0e+00 & 1.8e+02 & 2.0e+00 & 1.5e+00 \\
& $l_* = 5 \times 10^{13}$ cm & 2.6e+03 & 3.8e+01 & 1.8e+00 & 1.5e+03 & 4.8e+01 & 3.0e+00 & 7.3e+03 & 2.4e+02 & 1.2e+01 \\
	\noalign{\smallskip}
	\hline
	\noalign{\smallskip}
{\multirow{5}{*}{\rotatebox[origin=r]{90}{$10^{-6}\ \mathrm{M}_\odot\ \mathrm{yr}^{-1}$}}} &    Smooth & \multicolumn{3}{c}{ 1.5e+10  cm$^{-2}$} &  \multicolumn{3}{c}{ 3.8e+12  cm$^{-2}$} &  \multicolumn{3}{c}{ 1.5e+09  cm$^{-2}$} \\
    \cmidrule(lr){2-2} \cmidrule(lr){3-5} \cmidrule(lr){6-8}  \cmidrule(lr){9-11} 
&	$f_\mathrm{vol}$ & 0.05 & 0.2 & 0.4 & 0.05 & 0.2 & 0.4 & 0.05 & 0.2 & 0.4 \\	
    \cmidrule(lr){2-2} \cmidrule(lr){3-5} \cmidrule(lr){6-8}  \cmidrule(lr){9-11} 
& $l_* = 5 \times 10^{12}$ cm & 1.3e+02 & 1.5e+00 & 1.3e+00 & 1.5e+02 & 3.9e+00 & 1.5e+00 & 1.4e+03 & 1.2e+01 & 2.2e+00 \\
& $l_* = 1 \times 10^{13}$ cm & 9.1e+02 & 5.0e+00 & 1.3e+00 & 6.4e+02 & 2.1e+01 & 3.1e+00 & 5.4e+03 & 1.2e+02 & 8.4e+00 \\
& $l_* = 5 \times 10^{13}$ cm & 1.2e+04 & 7.0e+02 & 7.0e+01 & 4.9e+03 & 8.3e+02 & 1.9e+02 & 2.2e+04 & 4.5e+03 & 1.0e+03 \\
	\noalign{\smallskip}
	\hline
	\noalign{\smallskip}
{\multirow{5}{*}{\rotatebox[origin=r]{90}{$10^{-7}\ \mathrm{M}_\odot\ \mathrm{yr}^{-1}$}}} &    Smooth & \multicolumn{3}{c}{ 9.9e+09  cm$^{-2}$} &  \multicolumn{3}{c}{ 2.5e+14  cm$^{-2}$} &  \multicolumn{3}{c}{ 2.7e+10  cm$^{-2}$} \\
    \cmidrule(lr){2-2} \cmidrule(lr){3-5} \cmidrule(lr){6-8}  \cmidrule(lr){9-11} 
&	$f_\mathrm{vol}$ & 0.05 & 0.2 & 0.4 & 0.05 & 0.2 & 0.4 & 0.05 & 0.2 & 0.4 \\	
    \cmidrule(lr){2-2} \cmidrule(lr){3-5} \cmidrule(lr){6-8}  \cmidrule(lr){9-11} 
& $l_* = 5 \times 10^{12}$ cm & 3.5e+02 & 2.3e+01 & 6.3e+00 & 7.5e+00 & 2.9e+00 & 1.9e+00 & 7.0e+01 & 1.4e+01 & 5.5e+00 \\
& $l_* = 1 \times 10^{13}$ cm & 6.7e+02 & 4.6e+01 & 1.1e+01 & 1.1e+01 & 4.0e+00 & 2.4e+00 & 1.0e+02 & 2.3e+01 & 8.2e+00 \\
& $l_* = 5 \times 10^{13}$ cm & 1.9e+03 & 2.2e+02 & 5.8e+01 & 2.4e+01 & 1.0e+01 & 6.0e+00 & 1.6e+02 & 6.3e+01 & 2.6e+01 \\
     \noalign{\smallskip}
    	\hline \hline
    \end{tabular}%
    \label{table:Results-CD-1C-C}
\end{table*}

\paragraph{Intermediate region}

The large dip in CS abundance is caused by its destruction through
\begin{equation*}
\mathrm{OH + CS \rightarrow OCS + H}.
\end{equation*}
The region where this reaction is the main CS destruction mechanism corresponds to the peak in OH abundance, created by photodissociation of the parent species H$_2$O. 
The CS abundance increases again as OH is photodissociated, which occurs closer to the central star for a larger porosity.
The decrease in abundance is larger for smaller values of $f_\mathrm{vol}$, as the reaction can occur more efficiently due to the overdensity of the clumps.

\paragraph{Outer region}

Before the photodissociation of CS, its abundance first increases. The associated feature is shifted inwards for highly porous outflows.

Near the end of the intermediate region and at the beginning of the outer region, CS is mainly produced via
\begin{equation*}
\mathrm{C + SO \rightarrow CS + O}.
\end{equation*}
At the location of the trough, the reaction
\begin{equation*}
\mathrm{C + HS \rightarrow CS + H}
\end{equation*}
contributes to the production of CS. This region corresponds to a peak in HS abundance, caused by the reaction
\begin{equation*}
\mathrm{H + SiS^+ \rightarrow HS + Si^+},
\end{equation*}
where SiS$^+$ is mainly produced in this region by
\begin{equation*}
\mathrm{C^+ + SiS  \rightarrow SiS^+ + C}.
\end{equation*}
The location of the peak in abundance can hence be traced back to the abundance of C$^+$ (see Sect. \ref{subsubsect:Results-Orich-HCN}).

\subsection{Carbon-rich outflows}                            \label{subsect:Results-Crich}

In the inner region of C-rich outflows, O- and N{-}bearing molecules are not expected to be {abundantly} present {under the assumption of} TE chemistry \citep{Cherchneff2006,Agundez2010}.
In the following, we investigate the effect of clumping and porosity on the chemical pathways involved in the formation and destruction of NH$_3$, H$_2$O, and H$_2$S in the the inner, intermediate, and outer regions of the stellar wind.
{As with the O-rich outflows, the location of these regions depends on the mass-loss rate. The inner region ends roughly around $\sim 10^{15}$ cm. The end of the intermediate region, around $\sim 3 \times 10^{16}$ cm, corresponds to where photodissociation becomes the dominant destruction process. These boundaries shift inward with increasing mass-loss rate.
}

{
We compare our results to the observed abundances in the well-studied C-rich AGB star IRC+10216 (CW Leo), which has a high mass-loss rate of $2-4 \times 10^{-5}$ M$_\odot$ yr$^{-1}$ \citep{DeBeck2012,Cernicharo2015}. 
Only a rough comparison to these observations is possible, since the physical parameters and input species of our chemical model are not tailored to IRC+10216.
}

\subsubsection{NH$_3$}										\label{subsubsect:Results-Crich-NH3}

\begin{figure*}
\centering
\includegraphics[width=1.\textwidth]{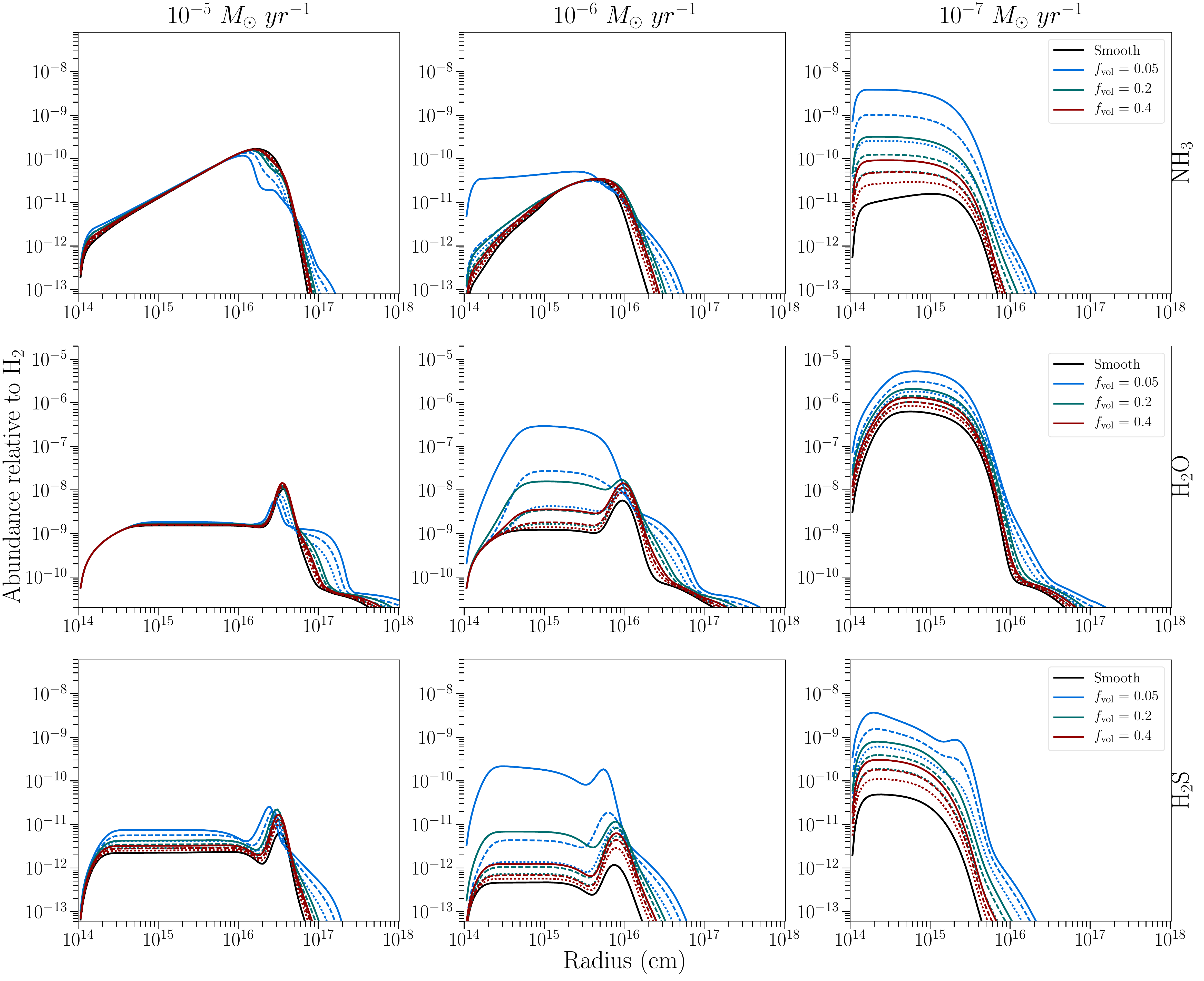}
\caption{Abundance of NH$_3$ (upper panels), H$_2$O (middle panels), and H$_2$S (lower panels) relative to H$_2$ throughout a two-component C-rich outflow with different mass-loss rates $\dot{M}$ and clump volume filling factors $f_\mathrm{vol}$. The characteristic size of the clumps at the stellar radius is $l_* = 10^{13}$ cm.
{
Blue lines: porosity length $h_* = 2 \times 10^{14}$ cm.
Green lines: $h_* = 5 \times 10^{13}$ cm.
Red lines: $h_* = 2.5 \times 10^{13}$ cm.
}
Solid black line: calculated abundance for a smooth, uniform outflow. 
Solid coloured line: density contrast between the inter-clump and smooth outflow $f_\mathrm{ic} = 0.1$. Dashed coloured line: $f_\mathrm{ic} = 0.3$. Dotted coloured line: $f_\mathrm{ic} = 0.5$.
{Note that the models with $f_\mathrm{vol} = 0.4$ (red) have the same porosity length as the one-component outflows with $f_\mathrm{vol} = 0.2,\ l_* = 5 \times 10^{12}$ cm and $f_\mathrm{vol} = 0.4,\ l_* = 1 \times 10^{13}$ cm.}
For reference, $1\ R_* = 5 \times 10^{13}$ cm.
}
\label{fig:Results-TwoComp-Crich}
\end{figure*}

As in O-rich chemistry, {the inner region NH$_3$ abundance in C-rich outflows can reach up to a few times $10^{-8}$ relative to H$_2$ in highly porous outflows. This corresponds to an increase of three orders of magnitude relative to a smooth outflow. The NH$_3$ column density of such outflows can increase up to {four} orders of magnitude relative to a smooth outflow.
}
The abundance profiles are shown in the upper panel of Figs. \ref{fig:Results-OneComp-Crich} and \ref{fig:Results-TwoComp-Crich}. The column densities are listed in Tables \ref{table:Results-CD-1C-C} and \ref{table:Results-CD-2C-C}.

For highly porous outflows, the NH$_3$ abundance reaches its maximum value close to the start of the model and is constant throughout the inner and intermediate regions, similar to the O-rich outflow.
The destruction of NH$_3$ at the end of the intermediate region is influenced by the overdensity of the clumps and the UV radiation field present.

{In IRC+10216, NH$_3$ has an observed inner wind abundance of $1.7 \times 10^{-7}$ \citep{Keady1993}. Similar to the O-rich outflows, our models do not produce such a large abundance, but including a highly porous clumpy outflow does increase the inner wind abundance significantly. For $\dot{M} = 10^{-5}$ M$_\odot$ yr $^{-1}$, a one-component outflow is again needed to achieve such an increase.
}

\paragraph{Inner region}

The formation of NH$_3$ in the inner region is similar to that in an O-rich outflow.
The extra N-bearing parent species HCN does not influence the formation of NH$_3$ in the inner region, as its photodissociation products are CN and H.

\begin{table*}
    \caption{Column density [cm$^{-2}]$ of NH$_3$, H$_2$O, and H$_2$S in a smooth C-rich outflow with different mass-loss rates, together with column density ratios relative to the smooth outflow for specific two-component outflows. The corresponding abundance profiles are shown in Fig. \ref{fig:Results-TwoComp-Crich}. 
{Note that models with $f_\mathrm{vol} = 0.4$ have the same porosity length $h_* = 2.5 \times 10^{13}$ cm as the one-component models with $f_\mathrm{vol} = 0.2,\ l_* = 5 \times 10^{12}$ cm and $f_\mathrm{vol} = 0.4,\ l_* = 1 \times 10^{13}$ cm.
}
    }
    \centering
    \begin{tabular}{c c c c c c c c c c c}
    \hline \hline 
    \noalign{\smallskip}
    	 $\dot{M}$ & Species & & {NH$_3$} & &  & {H$_2$O} & &  & {H$_2$S} & \\
    \cmidrule(lr){1-1} \cmidrule(lr){2-2} \cmidrule(lr){3-5} \cmidrule(lr){6-8}  \cmidrule(lr){9-11} 
    	\noalign{\smallskip}
{\multirow{5}{*}{\rotatebox[origin=r]{90}{$10^{-5}\ \mathrm{M}_\odot\ \mathrm{yr}^{-1}$}}} &    Smooth & \multicolumn{3}{c}{ 3.3e+11  cm$^{-2}$} &  \multicolumn{3}{c}{ 4.1e+13  cm$^{-2}$} &  \multicolumn{3}{c}{ 6.8e+10  cm$^{-2}$} \\
    \cmidrule(lr){2-2} \cmidrule(lr){3-5} \cmidrule(lr){6-8}  \cmidrule(lr){9-11} 
&	$f_\mathrm{vol}$ & 0.05 & 0.2 & 0.4 & 0.05 & 0.2 & 0.4 & 0.05 & 0.2 & 0.4 \\	
    \cmidrule(lr){2-2} \cmidrule(lr){3-5} \cmidrule(lr){6-8}  \cmidrule(lr){9-11} 
& $f_\mathrm{ic} = 0.1$ & 1.1e+00 & 1.1e+00 & 1.1e+00 & 1.1e+00 & 1.1e+00 & 1.1e+00 & 2.9e+00 & 1.8e+00 & 1.4e+00 \\
& $f_\mathrm{ic} = 0.3$ & 1.1e+00 & 1.1e+00 & 1.1e+00 & 1.1e+00 & 1.1e+00 & 1.1e+00 & 2.2e+00 & 1.5e+00 & 1.3e+00 \\
& $f_\mathrm{ic} = 0.5$ & 1.1e+00 & 1.1e+00 & 1.2e+00 & 1.0e+00 & 1.1e+00 & 1.2e+00 & 1.7e+00 & 1.3e+00 & 1.3e+00 \\
	\noalign{\smallskip}
	\hline
	\noalign{\smallskip}
{\multirow{5}{*}{\rotatebox[origin=r]{90}{$10^{-6}\ \mathrm{M}_\odot\ \mathrm{yr}^{-1}$}}} &    Smooth & \multicolumn{3}{c}{ 1.5e+10  cm$^{-2}$} &  \multicolumn{3}{c}{ 3.8e+12  cm$^{-2}$} &  \multicolumn{3}{c}{ 1.5e+09  cm$^{-2}$} \\
    \cmidrule(lr){2-2} \cmidrule(lr){3-5} \cmidrule(lr){6-8}  \cmidrule(lr){9-11} 
&	$f_\mathrm{vol}$ & 0.05 & 0.2 & 0.4 & 0.05 & 0.2 & 0.4 & 0.05 & 0.2 & 0.4 \\	
    \cmidrule(lr){2-2} \cmidrule(lr){3-5} \cmidrule(lr){6-8}  \cmidrule(lr){9-11} 
& $f_\mathrm{ic} = 0.1$ & 1.0e+01 & 1.4e+00 & 1.2e+00 & 1.1e+02 & 6.3e+00 & 1.9e+00 & 3.4e+02 & 1.2e+01 & 2.5e+00 \\
& $f_\mathrm{ic} = 0.3$ & 1.4e+00 & 1.3e+00 & 1.2e+00 & 8.7e+00 & 1.8e+00 & 1.3e+00 & 7.4e+00 & 2.2e+00 & 1.5e+00 \\
& $f_\mathrm{ic} = 0.5$ & 1.2e+00 & 1.2e+00 & 1.2e+00 & 1.9e+00 & 1.3e+00 & 1.3e+00 & 2.7e+00 & 1.6e+00 & 1.4e+00 \\
	\noalign{\smallskip}
	\hline
	\noalign{\smallskip}
{\multirow{5}{*}{\rotatebox[origin=r]{90}{$10^{-7}\ \mathrm{M}_\odot\ \mathrm{yr}^{-1}$}}} &    Smooth & \multicolumn{3}{c}{ 9.9e+09  cm$^{-2}$} &  \multicolumn{3}{c}{ 2.5e+14  cm$^{-2}$} &  \multicolumn{3}{c}{ 2.7e+10  cm$^{-2}$} \\
    \cmidrule(lr){2-2} \cmidrule(lr){3-5} \cmidrule(lr){6-8}  \cmidrule(lr){9-11} 
&	$f_\mathrm{vol}$ & 0.05 & 0.2 & 0.4 & 0.05 & 0.2 & 0.4 & 0.05 & 0.2 & 0.4 \\	
    \cmidrule(lr){2-2} \cmidrule(lr){3-5} \cmidrule(lr){6-8}  \cmidrule(lr){9-11} 
& $f_\mathrm{ic} = 0.1$ & 3.4e+02 & 2.8e+01 & 8.0e+00 & 8.1e+00 & 3.3e+00 & 2.1e+00 & 6.9e+01 & 1.6e+01 & 6.3e+00 \\
& $f_\mathrm{ic} = 0.3$ & 9.0e+01 & 1.1e+01 & 4.3e+00 & 4.6e+00 & 2.3e+00 & 1.8e+00 & 3.0e+01 & 8.0e+00 & 3.8e+00 \\
& $f_\mathrm{ic} = 0.5$ & 2.3e+01 & 4.5e+00 & 2.5e+00 & 2.7e+00 & 1.7e+00 & 1.5e+00 & 1.2e+01 & 4.0e+00 & 2.4e+00 \\
     \noalign{\smallskip}
    	\hline \hline
    \end{tabular}%
    \label{table:Results-CD-2C-C}
\end{table*}

\paragraph{Intermediate region}

For higher mass-loss rate outflows with a large porosity, the NH$_3$ abundance declines rapidly near the end of the intermediate region. The decline is due to the onset of the reaction
\begin{equation*}
\mathrm{NH_3 + CN \rightarrow HCN + NH_2},
\end{equation*} 
which contributes significantly to its destruction as the CN abundance peaks in this region. 
The peak in CN abundance is caused by the photodissociation of HC$_3$N, which is formed mainly through
\begin{equation*}
\mathrm{CN + C_2H_2 \rightarrow HC_3N + H}.
\end{equation*} 
Since CN is produced through photodissociation of the parent species HCN, the peak in abundance of its daughter species HC$_3$N shifts closer to the central star for larger porosity lengths, as does the photodissociation of HC$_3$N.
The decline in NH$_3$ abundance ends when the CN abundance is reduced through the reaction
\begin{equation*}
\mathrm{O + CN \rightarrow N + CO},
\end{equation*}  
as the O abundance increases due to the photodissociation of the parent species SiO.

Highly porous outflows with a higher mass-loss rate also show an increase in NH$_3$ abundance after its sharp decline near a few times $10^{16}$ cm.
In this region, NH$_3$ is predominantly produced through reactions with NH$_4^+$, which originates from the hydrogenation of N$^+$ through reactions with H$_2$. N$^+$ is solely formed through reactions with He$^+$ with N$_2$ and is therefore unaffected by porosity. However, as the UV radiation field increases in this region, the abundances of the intermediary products NH$^+$, NH$_2^+$, and NH$_3^+$ increase through photoionisation of their neutral counterparts. This leads to an increased abundance of NH$_4^+$ and subsequently of NH$_3$.

\paragraph{Outer region}

At the start of the outer region, reactions with C$^+$ contribute to the destruction of NH$_3$.
The different slope of the decline abundance with different values of $f_\mathrm{vol}$ at the beginning of the outer region has a similar origin as that of HCN in the O-rich outflow (Sect. \ref{subsubsect:Results-Orich-HCN}).

\subsubsection{H$_2$O}										\label{subsubsect:Results-Crich-H2O}

The H$_2$O abundance can reach up to $10^{-5}$ relative to H$_2$ in the inner region {in highly porous one-component outflows and in highly porous two-component outflows with a small mass-loss rate}, an increase of three orders of magnitude relative to a smooth outflow. {In such outflows, the H$_2$O} column density can increase up to three orders of magnitude relative to a smooth outflow.
The abundance profiles are shown in the middle panel of Figs. \ref{fig:Results-OneComp-Crich} and \ref{fig:Results-TwoComp-Crich}. The column densities are listed in Tables \ref{table:Results-CD-1C-C} and \ref{table:Results-CD-2C-C}.

The peak in abundance around $\sim 1-3 \times 10^{16}$ cm, visible in the smooth abundance profile, disappears as $f_\mathrm{vol}$ decreases.
Outflows with very overdense clumps show a steep decline in this region, followed by a plateau-feature. 

{The inner wind H$_2$O abundance of IRC+10216 lies around $1 \times 10^{-7}$ \citep{Decin2010b}. 
\citet{Lombaert2016} find that the inner wind abundance lies between $\sim 10^{-6} - 10^{-4}$ for a sample of C-rich stars with mass-loss rates between $\sim 10^{-7} - 10^{-5}$ M$_\odot$ yr$^{-1}$. 
Only highly porous one-component outflows with $\dot{M} = 10^{-5}$ M$_\odot$ yr$^{-1}$ produce an inner wind abundance close to these observations. Models with $\dot{M} = 10^{-6}$ and $10^{-7}$ M$_\odot$ yr$^{-1}$ produce inner wind abundances within the range probed by \citet{Lombaert2016} for both one-component and two-component outflows, where the models have a maximum abundance of $\sim 1 \times 10^{-5}$.
} 

\paragraph{Inner region}

H$_2$O is formed in the inner region through hydrogenation of O through reactions with H$_2$, similar to the formation of NH$_3$.
O is produced in greater abundance in outflows with a large UV radiation field in the inner region, either through a low mass-loss rate or a large porosity, as the parent species CO and SiO are photodissociated in this region.

\paragraph{Intermediate region}

In the intermediate region, the formation of water shifts from a chemistry initiated by the proton transfer of H$_3^+$ with O atoms to one in which the dissociative recombination of CH$_2$CO$^+$ dominates
\begin{equation*}
\mathrm{CH_2CO^+ + e^- \rightarrow H_2O + C_2}.
\end{equation*} 
However, there are no laboratory measurements of the dissociative recombination of CH$_2$CO$^+$. The channel forming H$_2$O is one of three equal channels suggested by \citet{Prasad1980}. Given the re-arrangement needed to get H$_2$O out of the ion, which is structured as H$_2$CCO$^+$, it may have zero probability of occuring. 

Nevertheless, the CH$_2$CO$^+$ abundance peaks near the end of the intermediate region. It is formed by the photoionisation of CH$_2$CO, which also peaks in the same region. The peak in abundance is very sensitive to the temperature of the outflow, as several reactions with a negative temperature dependency play a large role in the chain of reactions forming CH$_2$CO:
\begin{equation*}
\mathrm{C_2H_2 + {h\nu} \rightarrow C_2H_2^+ + e^-},
\end{equation*} 
\begin{equation*}
\mathrm{H_2 + C_2H_2^+ \rightarrow C_2H_4^+ + {h\nu}}\ \ \ (\beta = -2.01),
\end{equation*} 
\begin{equation*}
\mathrm{C_2H_4^+ + O \rightarrow CH_3^+ + H},
\end{equation*} 
\begin{equation*}
\mathrm{CH_3^+ + CO \rightarrow CH_3CO^+ + {h\nu}	}\ \ \ (\beta = -1.2),
\end{equation*} 
\begin{equation*}
\mathrm{CH_3CO^+ + e^- \rightarrow CH_2CO + H},
\end{equation*} 
with $\beta$ the exponent denoting the temperature dependency of the reaction rate in the {modified} Arrhenius formula \citep{McElroy2013}.
For larger values of the exponent $\epsilon$ of the temperature power-law, corresponding to a lower temperatures throughout the outflow, the peak disappears.

For large UV radiation fields, the CH$_2$CO$^+$ peak shifts closer to the central star and becomes comparable in abundance to H$_3$O$^+$, which removes the peak in H$_2$O abundance.

\paragraph{Outer region}

At the start of the outer region, reactions with C$^+$ contribute to the destruction of H$_2$O.
The effect of $f_\mathrm{vol}$ on the slope of the decline in H$_2$O abundance at the beginning of the outer region has a similar origin as that of NH$_3$ and HCN in the O-rich outflow {(Sect. \ref{subsect:Results-Orich})}.

\subsubsection{H$_2$S}										\label{subsubsect:Results-Crich-H2S}

{In highly porous one-component outflows, the inner wind abundance of H$_2$S can reach up to $10^{-8}$ relative to H$_2$, an increase of three orders of magnitude relative to a smooth outflow. The H$_2$S column density in such outflows can increase up to four orders of magnitude relative to a smooth outflow.
}
The abundance profiles are shown in the lower panel of Figs. \ref{fig:Results-OneComp-Crich} and \ref{fig:Results-TwoComp-Crich}. The column densities are listed in Tables \ref{table:Results-CD-1C-C} and \ref{table:Results-CD-2C-C}.

For one-component outflows with $f_\mathrm{vol} = 0.05$, the abundance shows a large dip in the inner and intermediate regions.
The H$_2$S abundance profile also shows a plateau-feature after a sharp decline around $10^{16}$ cm.

{The inner wind abundance of H$_2$S in IRC+10216 is $4 \times 10^{-9}$ \citep{Agundez2012}. In models with $\dot{M} = 10^{-5}$ M$_\odot$ yr$^{-1}$, such an inner wind abundance is produced in highly porous one-component outflows.
}

\paragraph{Inner region}

In the inner region, H$_2$S is formed through the sequential hydrogenation of S through reactions with H$_2$. 
In outflows with a well-shielded inner region, S is formed via cosmic-ray induced photodissociation of the parent species CS and SiS.
As the UV radiation field increases, the S abundance increases as these parent species are also photodissociated. The increased N abundance through photodissociation of the parent species N$_2$ moreover leads to an increased contribution to S through the reaction
\begin{equation*}
\mathrm{N + CS \rightarrow S + CN}.
\end{equation*} 

For outflows with a large porosity, the reaction
\begin{equation*}
\mathrm{N + HS \rightarrow NS + H}
\end{equation*} 
destroys a fraction of the HS after $\sim 2 \times 10^{14}$ cm. The onset of the reaction corresponds to the start of the decrease of HS abundance, and coincides with the increase in N abundance.

\paragraph{Intermediate region}

The decrease in H$_2$S abundance before the peak at the end of the intermediate region is also caused by the photodissociation of the parent species CS. It therefore has the same origin as the extended dip in H$_2$S abundance for outflows with a larger porosity, but is more spatially confined due to the later onset of the CS photodissociation.

The peak in H$_2$S abundance corresponds to the contribution of the reaction
\begin{equation*}
\mathrm{H + SiS^+ \rightarrow HS + Si^+}
\end{equation*} 
to the H$_2$S production, as the abundance of SiS$^+$ reaches its maximal value in this region and forms a peak at the same location as that in the HS {abundance}, and consequently also the H$_2$S abundance.
The SiS$^+$ peak is caused by reactions between C$^+$ and the parent species SiS, and therefore shifts inward for smaller values of $f_\mathrm{vol}$ (see Sect. \ref{subsubsect:Results-Orich-HCN}).
The SiS$^+$ abundance then decreases through dissociative recombination with electrons, marking the end of the peak in abundance.

The origin of the peak in H$_2$S abundance is similar to that in the H$_2$O abundance (Sect. \ref{subsubsect:Results-Crich-H2S}). The rates of the reactions between SiS and C$^+$ and S$^+$ have a negative temperature dependency ($\beta = -0.5$ for both). Again, the peak disappears for larger values of the exponent $\epsilon$ of the temperature power-law, corresponding to lower temperatures.

As the mass-loss rate decreases and/or the porosity of the outflow decreases, the peak in the H$_2$S abundance disappears. This is due to the earlier onset of its photodissociation in these outflows.

\paragraph{Outer region}

The different slope of the decline in H$_2$S abundance with different values of $f_\mathrm{vol}$ at the beginning of the outer region is again due to the C$^+$ abundance (see e.g. Sect. \ref{subsubsect:Results-Orich-NH3}).
In this region, H$_2$S is formed via reactions between H$_2$S$^+$ and various anions of C-chains. The different slopes are also present in the decline of the H$_2$S$^+$ abundance, as it is mainly produced by the reaction
\begin{equation*}
\mathrm{H_2CO+ S^+ \rightarrow H_2S^+ + CO},
\end{equation*} 
where S$^+$ is produced via the photoionisation of S. Before H$_2$CO is photodissociated, it is destroyed through
 \begin{equation*}
\mathrm{H_2CO+ C^+ \rightarrow CO + CH_2^+},
\end{equation*} 
causing different slopes for different values of $f_\mathrm{vol}$, which are transferred to the H$_2$S abundance profile.

\section{Discussion}                                     \label{sect:Discussion}

In Sect. \ref{subsect:Disc-EffectParam}, we discuss the general influence of the parameters $f_\mathrm{vol}$, $f_\mathrm{ic}$, and $l_*$ on the chemistry throughout the outflow.
We compare our results to those of \citet{Agundez2010}, who have also modelled the effect of a clumpy outflow in Sect. \ref{subsect:Disc-CompChem}, and to those of shock-induced non-equilibrium chemistry models in Sect. \ref{subsect:Disc-CompNEChem}.
{We compare our modelled inner wind abundances to observations in Sect. \ref{subsect:Disc-CompObs}. 
Finally, molecules that experience a large increase in abundance or change in their abundance profiles due to a clumpy outflow are discussed in Sect. \ref{subsect:Disc-Pred}. 
}

\subsection{Influence of the parameters $f_\mathrm{vol}$, $f_\mathrm{ic}$, and $l_*$}         \label{subsect:Disc-EffectParam}

The {porosity of the outflow} is determined by the porosity length $h_* = l_*/f_\mathrm{vol}$ and $f_\mathrm{ic}$.
The porosity length has the largest influence on {the decrease in} visual extinction, for a certain value of $f_\mathrm{ic}$.
{For a certain value of $h_*$, the visual extinction increases with $f_\mathrm{ic}$. 
A larger value of $f_\mathrm{ic}$ corresponds to more material in the inter-clump component, hindering radiation travelling through the porous channels between the clumps. 
}
An increase in the UV radiation field in the inner region leads to the photodissociation of parent species in this region, which is well-shielded in a smooth outflow. The elements which are normally locked up in stable parent species in smooth outflows (O in C-rich outflows, C in O-rich outflows, and N in both) are therefore released closer to the central star, allowing for an increased formation of the unexpected species in the inner region. These species are, under the assumption of TE chemistry, expected either to be absent in the inner wind or to have a smaller abundance than observed in the outer regions.
More generally, an increased UV radiation field influences the chemistry throughout the outflow, as photodissociation of both parent and daughter species occurs closer to the central star, opening up new chemical pathways.

{Clumping also influences the overall abundance of the unexpected species.
For the same value of $h_*$, smaller values of $f_\mathrm{vol}$ and $f_\mathrm{ic}$ will generally lead to larger inner region abundances and column densities. 
The effect of porosity on the increase in abundance is however much larger. 
Highly porous outflows, characterised by both a large clump overdensity and a large porosity length, therefore have the largest influence on the chemistry throughout the outflow.}

{Since the inner wind abundances of the unexpected species are mainly determined by the porosity of the outflow, one-component outflows yield a larger increase in abundance than their corresponding two-component outflows. 
For higher mass-loss rate outflows, the inter-clump component needs to be effectively void in order for porosity to increase the inner wind abundance of the unexpected species. 
}

\subsection{Comparison to chemical model of \citet{Agundez2010}}                            \label{subsect:Disc-CompChem}

The effect of a clumpy outflow on its chemistry has previously been studied by \citet{Agundez2010}.
In their model, the outflow is also split into two components: a smooth well-shielded component and a clumpy component, for which a fraction of solid angle of arrival of interstellar photons $f_\Omega$ is free of matter. 
The smooth component experiences the same UV radiation field as a uniform outflow, whereas a certain fraction of photons is allowed to reach the inner regions of the clumpy component uninhibitedly.
During the modelling, both components have the same density, namely that of the corresponding smooth outflow. The final abundances are calculated by taking a weighted average of the models of the components, using the mass fraction of the total circumstellar mass that is located inside the clump component $f_m$. 

The model of \citet{Agundez2010} hence differs from our model in both its implementation of the density distribution and the alteration of the UV radiation field. Whereas we take the density distribution into account while calculating the chemical network, \citet{Agundez2010} include it via post-processing. Both components experience the same effective UV radiation field in our model, while the components of \citet{Agundez2010} experience two distinct UV radiation fields.
Finally, the chemical reaction network used by \citet{Agundez2010} is different to the UDfA \textsc{Rate12} reaction network.

\citet{Agundez2010} assumed that $f_m = 0.1$ and $f_\Omega = 0.25$, chosen in order to roughly model the clumpy outflow of the C-rich star CW Leo. 
The smooth component is hence a major contributor to their final abundances, while at the same time a significant fraction of the interstellar radiation is still able to reach the inner wind. This combination is not possible in our model, as a larger density of the inter-clump component leads to a smaller decrease in visual extinction. 
Translated into our parameters, the $f_m$ and $f_\Omega$ parameters would correspond to $f_\mathrm{vol} \approx 1 - f_\Omega = {0.75}$ and $f_\mathrm{ic} \approx 1 - f_m = 0.9$. However, in order to have a similar effect on the UV radiation field, we would need that $f_\mathrm{ic} = 0$.
It is therefore not possible to make a direct comparison to one of the models shown in Sect. \ref{sect:Results}.

Nevertheless, we can perform a qualitative comparison and point out several clear differences.
Firstly, the shapes of the abundance profiles of \citet{Agundez2010} are only roughly similar to those shown in this paper, with a large discrepancy for H$_2$CO, C$_2$H$_2$, HCN, and CS in the O-rich outflow, and for OH and H$_2$S in the C-rich outflow. 
{Secondly, the modelled inner wind abundances of \citet{Agundez2010} can only be obtained with our model by assuming a one-component outflow with a large porosity length. In that case, a large fraction of UV photons is able to reach the inner region of these outflows, which matches the assumption by \citet{Agundez2010}. 
Moreover, the one-component outflow that reproduces the inner wind abundances of NH$_3$ in the O-rich outflow, yields abundances up to an order of magnitude higher for all other species shown in \citet{Agundez2010}. Similar for the C-rich outflow, where the one-component model that reproduces the inner wind abundances of NH$_3$ and H$_2$S yields abundances an order of magnitude larger for the other molecules shown.
}


\subsection{Comparison to non-equilibrium chemistry models}                            \label{subsect:Disc-CompNEChem}

In order to explain the presence and abundance of the unexpected species in the inner wind, shock-induced non-equilibrium chemistry in the inner region has been proposed as a possible solution. 
These models are able to reproduce the observed abundances in the inner regions, except for NH$_3$, which is predicted to have an abundance of $\sim 10^{-19}$ relative to H$_2$ \citep{Willacy1998,Duari1999,Gobrecht2016}. 
Although a direct comparison is not possible as we have not specifically modelled IK Tau, we do produce NH$_3$  {with a larger abundance} than TE.
{In non-equilibrium models, CH$_4$ is not expected to be present with an abundance larger than predicted by TE chemistry in O-rich outflows \citep{Duari1999}. Our models produce an inner wind abundance up to seven order of magnitude larger than predicted for this species (Figs. \ref{fig:App-OneComp-Orich} and \ref{fig:App-TwoComp-Orich}).}

The inner regions of C-rich outflows have not been studied using non-equilibrium chemical models to the same extent as O-rich outflows. 
However, the recently discovered distribution of CH$_3$CN in the C-rich AGB star CW Leo does not match the predictions of both {thermodynamic} equilibrium and non-equilibrium chemistry \citep{Agundez2015,Millar2016}. \citet{Agundez2015} suggest a clumpy outflow as a possible origin to this distribution. Similar to their best fitting abundance profile, we find a peak in the CH$_3$CN abundance around $10^{16}$ cm (Figs. \ref{fig:App-OneComp-Crich} and \ref{fig:App-TwoComp-Crich}).
Moreover, the presence of H$_2$O in the inner region of CW Leo was not predicted to be present by non-equilibrium chemistry \citep{Cherchneff2006}, and could only be modelled by manually adapting the rate of the reaction between SiO and H {\citep{Cherchneff2012}}. Although a direct comparison to the observations is again not possible, our models do produce H$_2$O in larger abundances than TE in the inner region.

\subsection{Comparison to observations}                         \label{subsect:Disc-CompObs}

{ 
For both O-rich and C-rich outflows, we find that the observations are best reproduced by highly porous, one-component outflows. 
A large UV field is therefore required to be present in the inner region for the deficient elements to be released in order to obtain abundances for the unexpected species similar to those observed. The overdensity of the clumps leads to a further increase in abundance, so that both porosity and clumping are critical in reproducing the observed abundances.
For lower mass-loss rates, the requirement of a highly porous, one-component outflow is less strict. The UV field in the inner region is still sufficiently large in these outflows if a fraction of the material resides in the inter-clump component.

The observed NH$_3$ abundances in both O-rich and C-rich AGB stars are not reproduced by our models. 
Our models also do not produce the upper limit of the observed H$_2$O abundance range in C-rich AGB stars. 
Despite this, we do find an increase of up to four orders of magnitude compared to a smooth outflow in the inner wind abundance of NH$_3$, and up to three orders of magnitude in the inner wind abundance of H$_2$O for highly porous, one-component outflows. 
However, an effectively void inter-clump medium is likely not realistic for the outflows of AGB stars. The density contrast between the clumped and inter-clump component of IRC+10216 is found to lie between 1.5 and 7.6 \citep{Mauron2000,Dinh-V-Trung2008}, its inter-clump component can therefore not be considered effectively void.

The requirement of an effectively void inter-clump component might be due to assumption of optically thin clumps at the edge of the outflow, which places an upper limit on the porosity length (Sect. \ref{subsect:Porosity:ParameterRange}). If the outer boundary condition were less strict, larger porosity lengths are possible, potentially allowing for an increased inter-clump density.


}

\subsection{Predicatbility of models}                         \label{subsect:Disc-Pred}

{

The clumpy outflow model can also be used to predict the effect of clumping and porosity on the abundances of unobserved molecules, making them potentially observable, and on their effect on the abundance profiles of already detected molecules.
We have identified a set of molecules whose abundance is strongly influenced by a clumpy outflow by requiring a peak fractional abundance larger than $1 \times 10^{-9}$ relative to H$_2$ and an increase in column density relative to the smooth outflow of at least two orders of magnitude. 
Their abundance profiles and column densities are shown in Appendix \ref{sect:App:Predictions}.
For the O-rich outflow, we find considerably changed abundance profiles and column densities for N$_2$O, C$_2$N, C$_3$H, C$_3$H$_2$ and OCS (Figs. \ref{fig:App-OneComp-Orich-Pred} and \ref{fig:App-TwoComp-Orich-Pred}, Tables \ref{table:App-CD-1C-O-Pred} and Tables \ref{table:App-CD-2C-O-Pred}).
For the C-rich outflow, considerable changes are found for CO$_2$, SO$_2$, HC$_9$N, NO and OCS (Figs. \ref{fig:App-OneComp-Crich-Pred} and \ref{fig:App-TwoComp-Crich-Pred}, Tables \ref{table:App-CD-1C-C-Pred} and Tables \ref{table:App-CD-2C-C-Pred}).

%

}

\section{Conclusion}                                     \label{sect:Conclusion}

We have determined the effect of a non-uniform, clumpy AGB outflow on its chemistry by implementing a porosity formalism in our one-dimensional chemical model. 
This theoretical formalism takes into account both the effect of porosity on the UV radiation field throughout the outflow (by adjusting its optical depth), and the effect of clumping in general on the reaction rates for two-body processes (by splitting the outflow into an overdense clumped and a rarefied inter-clump component). 
The specific clumpiness is determined by three parameters: the clump volume filling factor $f_\mathrm{vol}$, the density contrast $f_\mathrm{ic}$, and the characteristic size of the clumps at the stellar surface $l_*$. The first two parameters determine the relative densities of the clumped and inter-clump component. The porosity length $h_* = l_*/f_\mathrm{vol}$ and $f_\mathrm{ic}$ determine the decrease in visual extinction throughout the outflow, with $h_*$ having the largest influence.
In this paper, the clumps are assumed to have a spatially constant $f_\mathrm{vol}$, which implies that they expand as they move outward.
Note that, being a statistical description of a stochastic ensemble of clumps, the porosity formalism does not require any assumptions on the specific location of the clumps.

We have shown results for a one-component outflow, where all material is located inside clumps and the inter-clump component is effectively void, and a two-component outflow, where material is located in both the clumped and inter-clump component.
Both the overdensity of the clumps and the increase in interstellar UV radiation in the inner regions of the outflow alter the chemistry throughout the outflow, as they alter the abundances and the chemical formation and destruction pathways.
The decrease in visual extinction has the largest influence on the formation of the unexpected species in both O- and C-rich outflows. 
{Porosity, not clumping, is therefore the main mechanism determining the overall increase in abundance.}
The larger UV radiation field in the inner region liberates the deficient elements through photodissociation of parent species, enabling the formation of the unexpected species.
A highly porous outflow, characterised by a large clump overdensity and a large porosity length, therefore has the largest effect on the chemistry throughout the outflow. 

Our chemical model is able to produce all unexpected species in clumpy O- and C-rich outflows {with inner region abundances larger than expected by TE chemistry}. 
Their column densities and molecular abundances in the inner region increase with clump overdensity and porosity length, and decrease with inter-clump density contrast. 
Unlike the shock-induced non-equilibrium chemistry models, our models are able to produce NH$_3$ and H$_2$O in the inner region with a larger abundance than predicted by TE.
The differences between our results and those of \citet{Agundez2010}, who have also implemented a clumpy outflow in their model, are most likely due to a different modelling approach. However, both approaches are able to produce all unexpected species in the inner region. 

{
Highly porous, one-component outflows generally reproduce the observationally retrieved values of the unexpected species in higher mass-loss rate outflows. For lower mass-loss rate outflows, the inter-clump component does not need to be effectively void. 
Nonetheless, such models do not produce the observed inner wind abundance of NH$_3$ in O-rich and C-rich outflows  and the upper limit of the range of H$_2$O in C-rich outflows. 
Combining both shock-induced non-equilibrium chemistry and a clumpy outflow can potentially solve the discrepancy.
}

Chemical models that include a non-uniform density structure can lead to an increase the abundances of the unexpected species in the inner regions of both O- and C-rich outflows, as demonstrated in this paper. 
This emphasises the importance of including the effect of a clumpy outflow in the chemical modelling of AGB outflows.
{Even though it cannot account for the observed abundance of all unexpected species investigated in this paper,} the non-uniform density structure should be taken into account in order to accurately model the chemistry and to correctly interpret the observed abundance profiles.

\begin{acknowledgements}
MVdS, DK{, WH,} and LD acknowledge support from the ERC consolidator grant 646758 AEROSOL. 
TJM thanks STFC for support under grant reference ST/P000312/1.
{FDC acknowledges support from the EPSRC iCASE studentship programme, Intel Corporation and Cray Inc.}
{We thank the anonymous referee for a detailed reading and helping to improve this paper.}
\end{acknowledgements}

\bibliographystyle{aa}
\bibliography{chemistry}

\newpage

\begin{appendix}

\section{Opacity of the smooth outflow}        \label{Sect:App:Opacity}

The opacity of the smooth outflow is given by
\begin{equation}
	\kappa = {1.086}\ \frac{{2}\ \lbrack A_{UV}/A_V\rbrack}{1.87 \times 10^{21}}\  \frac{1}{\mu\ m_H} \approx {1200}\  \mathrm{cm}^2 \ \mathrm{g}^{-1}
\end{equation}
with $\mu $ = {2.68} the mean molecular mass {of the outflow relative to H$_2$, including the fractional abundance of He (Table \ref{table:Model-Parents}),} and $m_H$ the atomic mass unit.
We hence assume that the extinction throughout the outflow is equal to that throughout the ISM of $1.87 \times 10^{21}$ atoms cm$^{-2}$ mag$^{-1}$ \citep{Cardelli1989}, which is then scaled according to the ratio of UV and visual extinction, with $A_{UV}/A_V = 4.65$ the adopted ratio of UV and visual extinction \citep{Nejad1984}. 
{The factor of 1.086 originates from the conversion of extinction to optical depth.}

\section{Figures of the decrease in interstellar UV radiation field}        \label{sect:App:Figetau}

\begin{figure}[h!]
\centering
\includegraphics[width=1.\columnwidth]{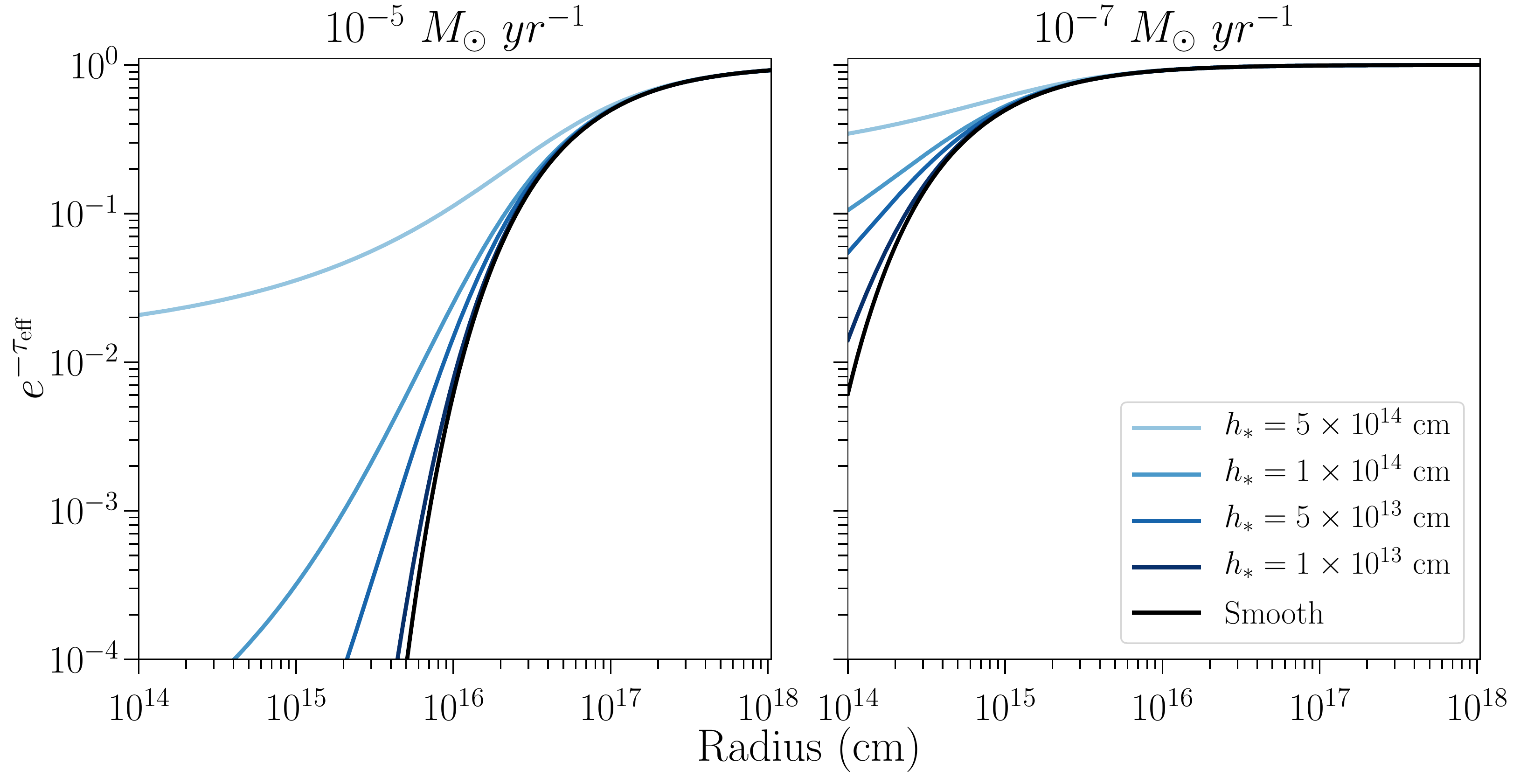}
\caption{Decrease in interstellar UV radiation field throughout a one-component outflow with $\dot{M} = 10^{-5}\ \mathrm{M}_\odot\ \mathrm{yr}^{-1}$ (left panel) and $\dot{M} = 10^{-7}\ \mathrm{M}_\odot\ \mathrm{yr}^{-1}$ (right panel).
Results for a smooth, uniform outflow are shown in black. Results for different values of $h_*$ are shown in blue. 
For reference, $1\ R_* = 5 \times 10^{13}$ cm.
}
\label{fig:Porosity-RF-1C-varH}
\end{figure}

\begin{figure}[h!]
\centering
\includegraphics[width=1.\columnwidth]{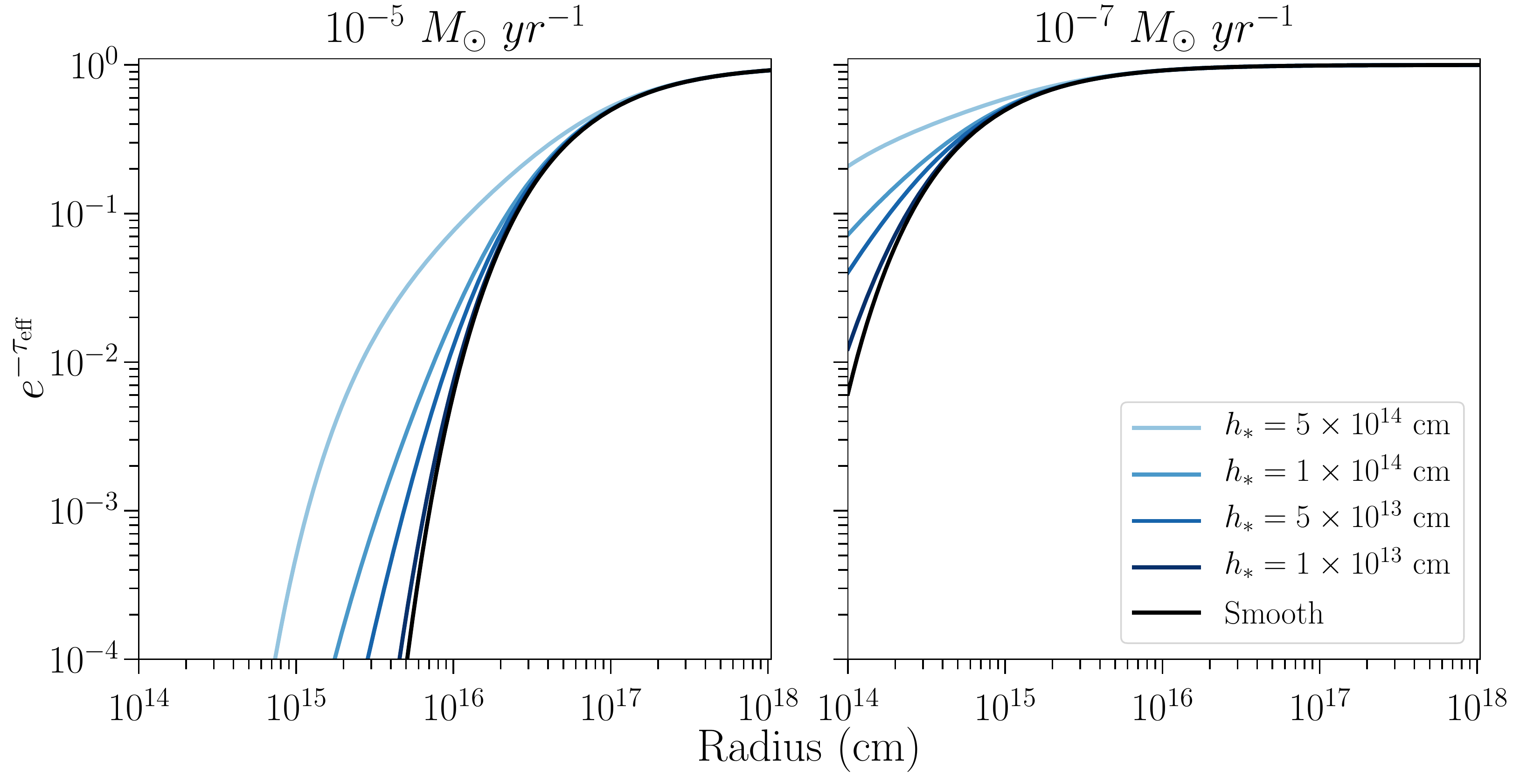}
\caption{As Fig. \ref{fig:Porosity-RF-1C-varH}, but for a two-component outflow with $f_\mathrm{ic} = 0.1$.
For reference, $1\ R_* = 5 \times 10^{13}$ cm.
}
\label{fig:Porosity-RF-2C-varH}
\end{figure}

\begin{figure}[h!]
\centering
\includegraphics[width=1.\columnwidth]{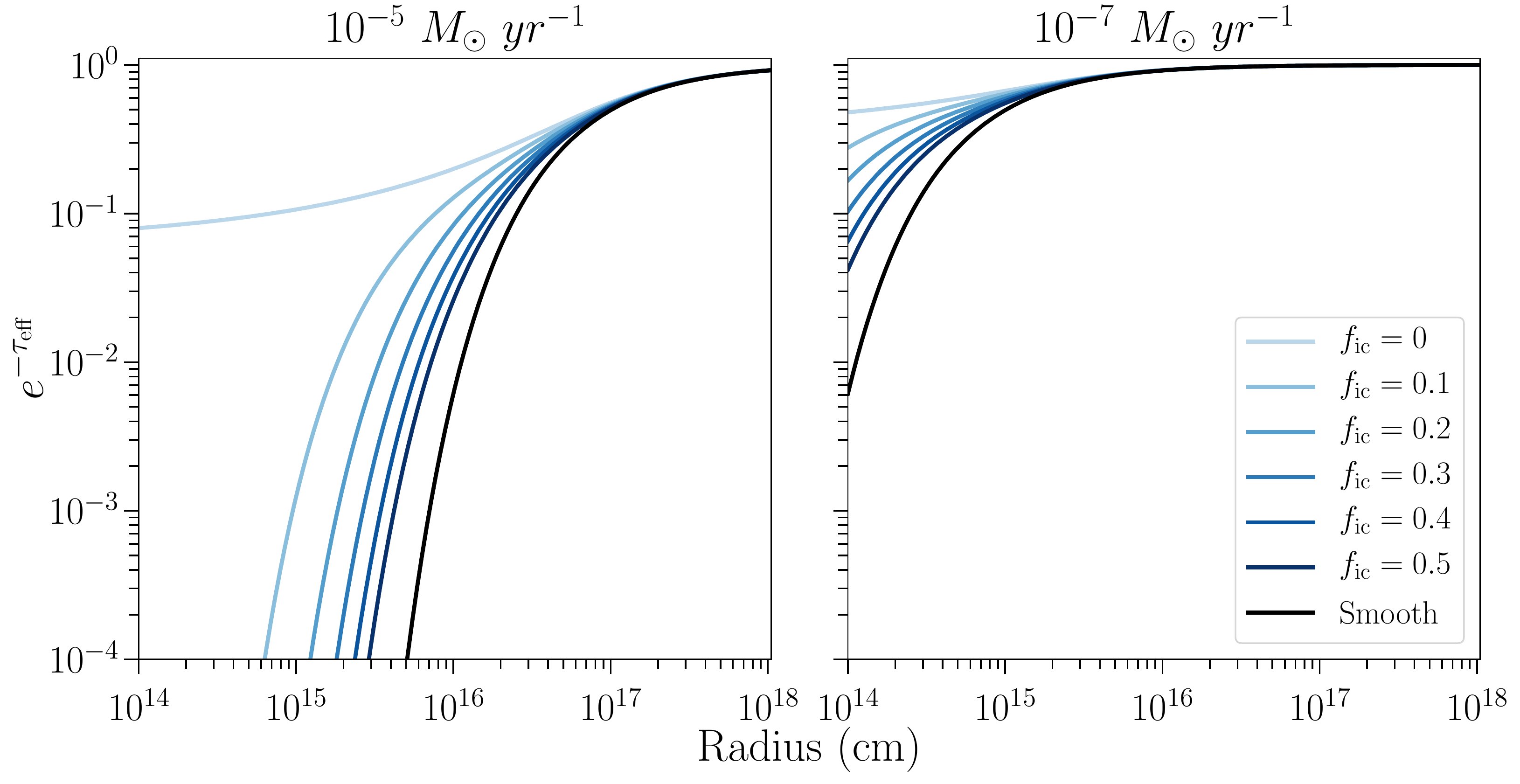}
\caption{Decrease in interstellar UV radiation field throughout a two-component outflow with $h = 10^{15}$ cm and and $\dot{M} = 10^{-5}\ \mathrm{M}_\odot\ \mathrm{yr}^{-1}$ (left panel) and $\dot{M} = 10^{-7}\ \mathrm{M}_\odot\ \mathrm{yr}^{-1}$ (right panel).
Results for a smooth, uniform outflow are shown in black. Results for different values of $f_\mathrm{ic}$ are shown in blue. 
For reference, $1\ R_* = 5 \times 10^{13}$ cm.
}

\label{fig:Porosity-RF-2C-varFic}
\end{figure}

Figs. \ref{fig:Porosity-RF-1C-varH} - \ref{fig:Porosity-RF-2C-varFic} show the decrease in interstellar UV radiation field $I(r)/I_0 = e^{-\tau_\mathrm{eff}(r)}$ throughout the outflow, with $I$ the UV radiation field throughout the outflow and $I_0$ the interstellar UV radiation field.

\section{Expression of the effective optical depth}        \label{sect:App:TauEff}

\subsection{One-component outflow}        \label{subsect:App:TauEff:1C}

The integral describing the effective optical depth $\tau_\mathrm{eff}$ for the one-component medium (Eq. \ref{Eq:IntegralTauEff-1C}) can be solved analytically. We have that
\begin{equation}			\label{Eq:TauEff-1C-notau}
\begin{split}
	\tau_\mathrm{eff} &= \tau_*\ R_* \int_r^\infty \frac{1}{r^2 +\tau_*\ R_*^{1/3}\ h_*\ r^{2/3}}\ \mathrm{d}r \\
					 &= \frac{3\pi \ \tau_*^{1/4} R_*^{3/4}}{2\sqrt{2}\ h_*^{3/4}} + \frac{3\ \tau_*^{1/4} R_*^{3/4}}{4\sqrt{2}\ h_*^{3/4}} \times \ \Bigg(2\ \mathrm{arctan} \left( 1 - \frac{\sqrt{2}\ r^{1/3}}{\left(\tau_*\ R_*^{1/3}h_*\right)^{1/4}} \right) \\
					& - 2\ \mathrm{arctan} \left( 1 + \frac{\sqrt{2}\ r^{1/3}}{\left(\tau_*\ R_*^{1/3}h _*\right)^{1/4}} \right) \\
					&+ \mathrm{log}\left( \sqrt{\tau_*\ R_*^{1/3}h_*} - \sqrt{2}\ \left(\tau_*\ R_*^{1/3}h_*\right)^{1/4}\ r^{1/3} + r^{2/3} \right) \\
					& - \mathrm{log}\left( \sqrt{\tau_*\ R_*^{1/3}h_*} + \sqrt{2}\ \left(\tau_*\ R_*^{1/3}h_*\right)^{1/4}\ r^{1/3} + r^{2/3} \right) \Bigg).
\end{split}
\end{equation}
Since $\tau = \tau_*\ R_*/r$, this can be rewritten as
\begin{equation}			\label{Eq:TauEff-1C}
\begin{split}
	\tau_\mathrm{eff} &= \frac{3\pi\ (\tau\ r)^{1/4} R_*^{1/2}}{2\sqrt{2}\ h_*^{3/4}} + \frac{3\ (\tau\ r)^{1/4} R_*^{1/2}}{4\sqrt{2}\ h_*^{3/4}}\ \times \\
	& \Bigg( 2\ \mathrm{arctan} \left( 1 - \frac{\sqrt{2}\ r^{1/12}}{\left(\tau\ h_*\ R_*^{-2/3}\right)^{1/4}}  \right) \\
					& - \mathrm{arctan} \left( 1 + \frac{\sqrt{2}\ r^{1/12}}{\left(\tau\ h_*\ R_*^{-2/3}\right)^{1/4}} \right) \\
					& + \mathrm{log}\left( \sqrt{\tau\ h_*\ R_*^{-2/3}\ r} - \sqrt{2}\ \left(\tau\  h_*\ R_*^{-2/3}\right)^{1/4}\ r^{7/12} + r^{2/3}\right) \\
					& - \mathrm{log}\left( \sqrt{\tau\ h_*\ R_*^{-2/3}\ r} + \sqrt{2}\ \left(\tau\  h_*\ R_*^{-2/3}\right)^{1/4}\ r^{7/12} + r^{2/3}\right) \Bigg).
\end{split}
\end{equation}

\subsection{Two-component outflow}        \label{subsect:App:TauEff:2C}

The integral describing the effective optical depth $\tau_\mathrm{eff}$ for the two-component medium (Eq. \ref{Eq:IntegralTauEff-2C}) can be solved analytically. We have that 
\begin{equation}				\label{Eq:TauEff-2C-notau}
\begin{split}
\tau_\mathrm{eff} &= \tau_*\ R_* \int_r^\infty \frac{1+\tau_*\ R_*^{1/3}\ h_*\ B\ r^{-4/3}}{r^2 +\tau_*\ R_*^{1/3}\ h_*\ B\ r^{2/3}}\ \mathrm{d}r \\
				 &= \frac{\tau_*\ R_*\ f_\mathrm{ic}}{r} + \frac{3\pi \ \tau_*^{1/4}\ R_*^{3/4} \left(1-f_\mathrm{ic}\right)}{2\sqrt{2}\ (h_*\ B)^{3/4}}\\
	& + \frac{3\ \tau_*^{1/4}\ R_*^{3/4} \left(1-f_\mathrm{ic}\right)}{4\sqrt{2}\ (h_*\ B)^{3/4}} \times \ \Bigg(2\ \mathrm{arctan} \left( 1 - \frac{\sqrt{2}\ r^{1/3}}{\left(\tau_*\ R_*^{1/3}h_*\ B\right)^{1/4}} \right) \\
					& - 2\ \mathrm{arctan} \left( 1 + \frac{\sqrt{2}\ r^{1/3}}{\left(\tau_*\ R_*^{1/3}h_*\ B\right)^{1/4}} \right) \\
					&+ \mathrm{log}\left( \sqrt{\tau_*\ R_*^{1/3}h_*\ B} - \sqrt{2}\ \left(\tau_*\ R_*^{1/3}h_*\ B\right)^{1/4}\ r^{1/3} + r^{2/3} \right) \\
					& - \mathrm{log}\left( \sqrt{\tau_*\ R_*^{1/3}h_*\ B} + \sqrt{2}\ \left(\tau_*\ R_*^{1/3}h_*\ B\right)^{1/4}\ r^{1/3} + r^{2/3} \right) \Bigg),
\end{split}
\end{equation}
with $B = 1 - \left( 1-f_\mathrm{vol} \right) f_\mathrm{ic}$. 
Since $\tau = \tau_*\ R_*/r$, this can be rewritten as
\begin{equation}				\label{Eq:TauEff-2C}
\begin{split}
	\tau_\mathrm{eff} &= \tau\ f_\mathrm{ic} + \frac{3\pi \ (\tau\ r)^{1/4} \left(1-f_\mathrm{ic}\right)}{2\sqrt{2}\ (h_*\ R_*^{-2/3} B)^{3/4}}\\
	& + \frac{3\ (\tau\ r)^{1/4} \left(1-f_\mathrm{ic}\right)}{4\sqrt{2}\ (h_*\ R_*^{-2/3} B)^{3/4}} \times \ \Bigg( 2\ \mathrm{arctan} \left( 1 - \frac{\sqrt{2}\ r^{1/12}}{\left(\tau\ h_*\ R_*^{-2/3} B \right)^{1/4}}  \right) \\
					& - 2\ \mathrm{arctan} \left( 1 + \frac{\sqrt{2}\ r^{1/12}}{\left(\tau\ h_*\ R_*^{-2/3} B\right)^{1/4}} \right) \\
					& + \mathrm{log}\left( \sqrt{\tau\ h_*\ R_*^{-2/3} B\ r} - \sqrt{2}\ \left(\tau\ h_*\ R_*^{-2/3} B\right)^{1/4}\ r^{7/12} + r^{2/3}\right) \\
					& - \mathrm{log}\left( \sqrt{\tau\ h_*\ R_*^{-2/3} B\ r} + \sqrt{2}\ \left(\tau\ h_*\ R_*^{-2/3} B\right)^{1/4}\ r^{7/12} + r^{2/3}\right)\Bigg).
\end{split}
\end{equation}
Note that Eqs. \ref{Eq:TauEff-2C-notau} and \ref{Eq:TauEff-2C} indeed reduce to the one-component equations for $f_\mathrm{ic} = 0$.

\section{Visual extinction and decrease in interstellar UV radiation field throughout the outflows considered in Sect. \ref{sect:Results}} \label{sect:App:AVresults}

\begin{figure*}[h!]
\centering
\includegraphics[width=.9\textwidth]{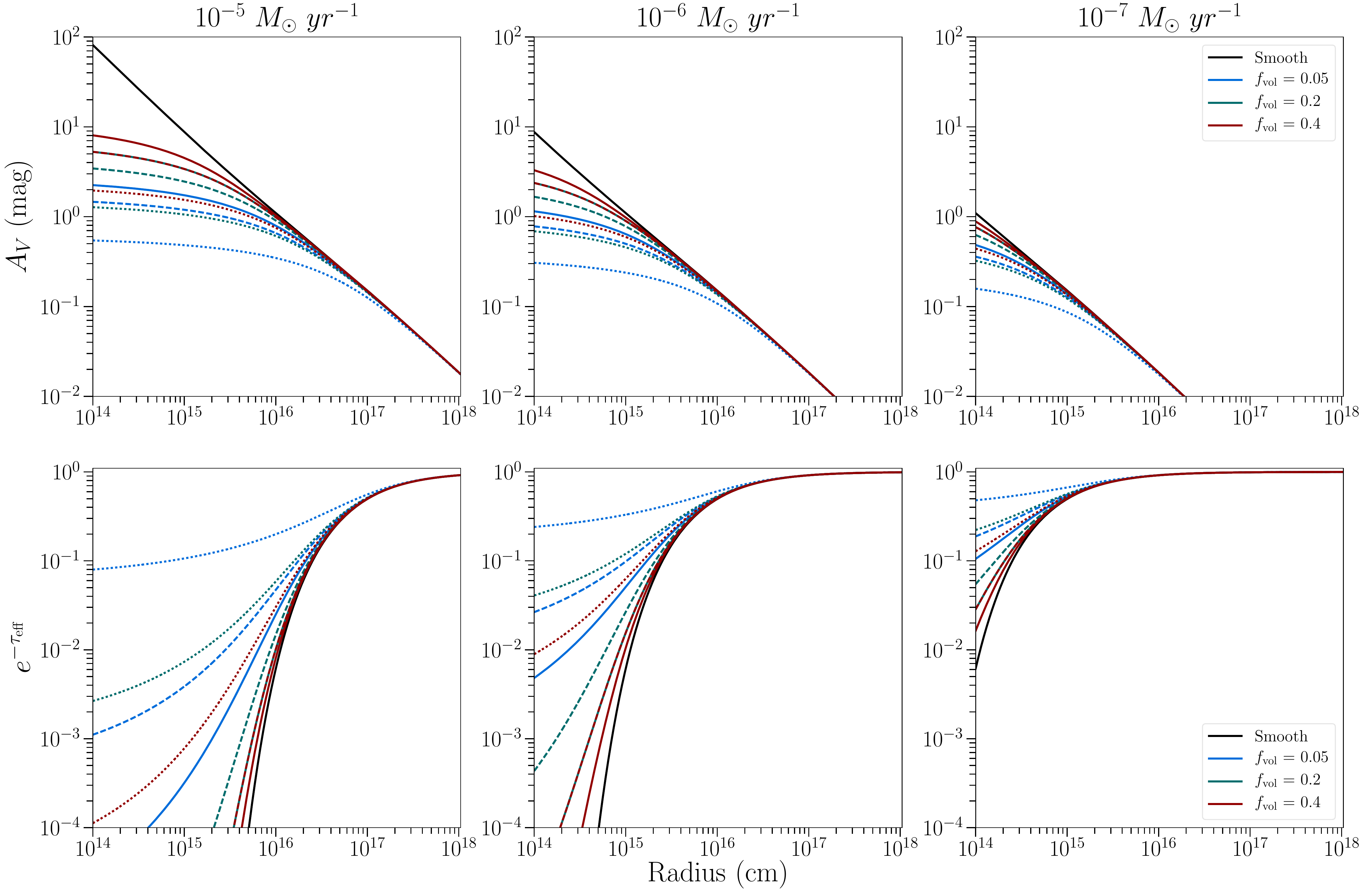}
\caption{Visual extinction and decrease in interstellar UV radiation field throughout the one-component outflows shown in Figs. \ref{fig:Results-OneComp-Orich} and \ref{fig:Results-OneComp-Crich}.
{Note that the models with $f_\mathrm{vol} = 0.2,\ l_* = 5 \times 10^{12}$ cm and $f_\mathrm{vol} = 0.4,\ l_* = 1 \times 10^{13}$ cm have the same porosity length $h_* = 2.5 \times 10^{13}$ cm.}
For reference, $1\ R_* = 5 \times 10^{13}$ cm.
}
\label{fig:App:AVresults:1C}
\end{figure*}

\begin{figure*}[h!]
\centering
\includegraphics[width=0.9\textwidth]{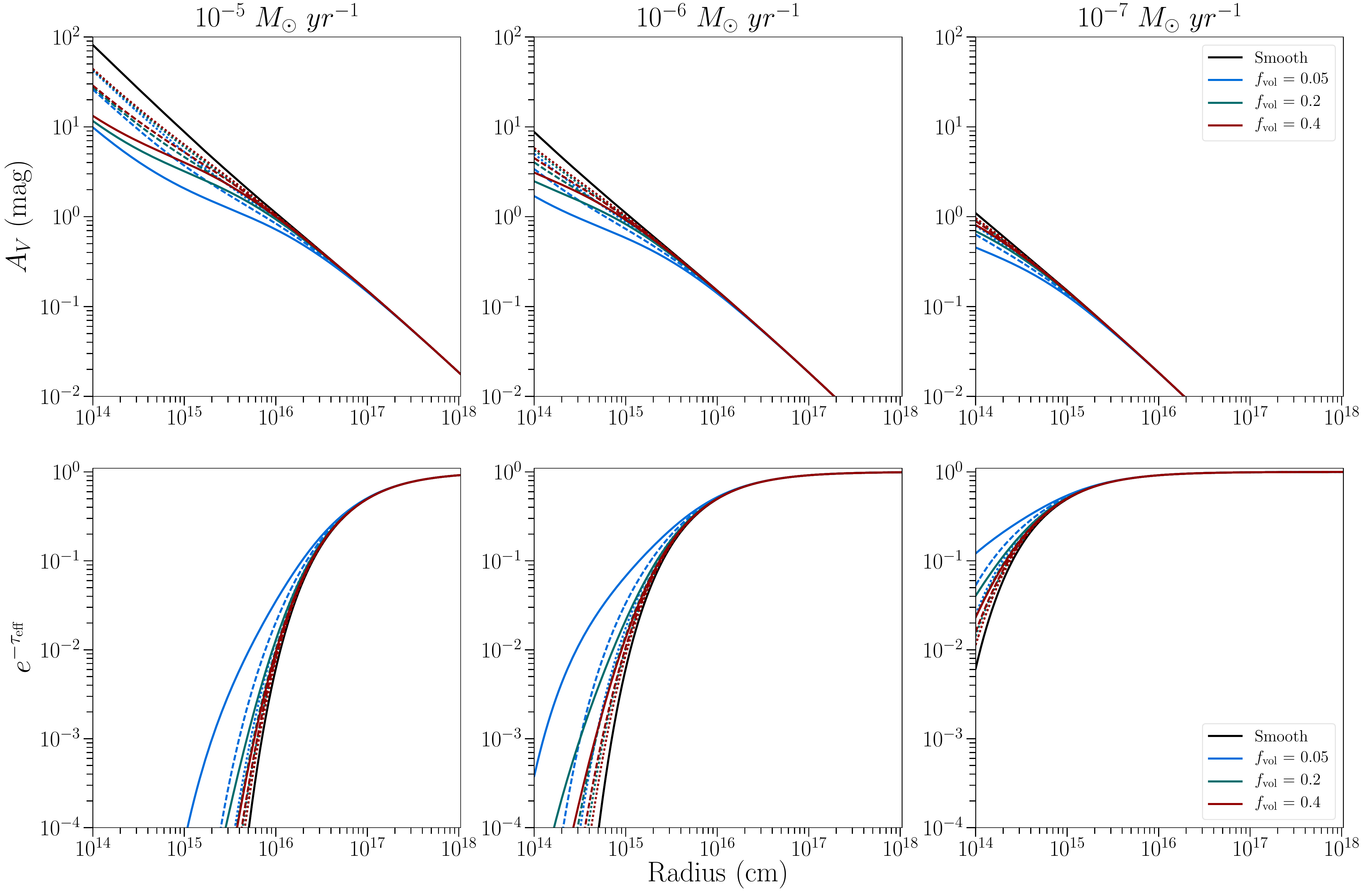}
\caption{Visual extinction and decrease in interstellar UV radiation field throughout the two-component outflows shown in Figs. \ref{fig:Results-TwoComp-Orich} and \ref{fig:Results-TwoComp-Crich}.
{Note that models with $f_\mathrm{vol} = 0.4$ have the same porosity length $h_* = 2.5 \times 10^{13}$ cm as the one-component models with $f_\mathrm{vol} = 0.2,\ l_* = 5 \times 10^{12}$ cm and $f_\mathrm{vol} = 0.4,\ l_* = 1 \times 10^{13}$ cm.}
For reference, $1\ R_* = 5 \times 10^{13}$ cm.
}
\label{fig:App:AVresults:2C}
\end{figure*}

Figs. \ref{fig:App:AVresults:1C} and \ref{fig:App:AVresults:2C} show the visual extinction and decrease in interstellar UV radiation field throughout respectively the one-component and two-component outflows considered in Sect. \ref{sect:Results}.

\section{Additional molecules}        \label{sect:App:AddMolecules}

{
In this Section, abundance profiles and number densities of additional molecules are shown. The effects of clumpiness and porosity on the chemical formation and destruction pathways of these molecules are similar to those presented in Sect. \ref{sect:Results}.
For both the O-rich and C-rich outflows, we show CH$_4$, H$_2$CO, and C$_2$H$_2$ in analogy with \citet{Agundez2010}. We additionally show CN and the parent species SO for the O-rich outflow (Sect. \ref{subsect:App:AddMolecules:Orich}) and the recently detected species CH$_3$CN and C$_4$H$_2$ for the C-rich outflow (Sect. \ref{subsect:App:AddMolecules:Crich}). 

}

\subsection{O-rich outflows}        \label{subsect:App:AddMolecules:Orich}

\begin{figure*}[ht]
\centering
\includegraphics[width=0.96\textwidth]{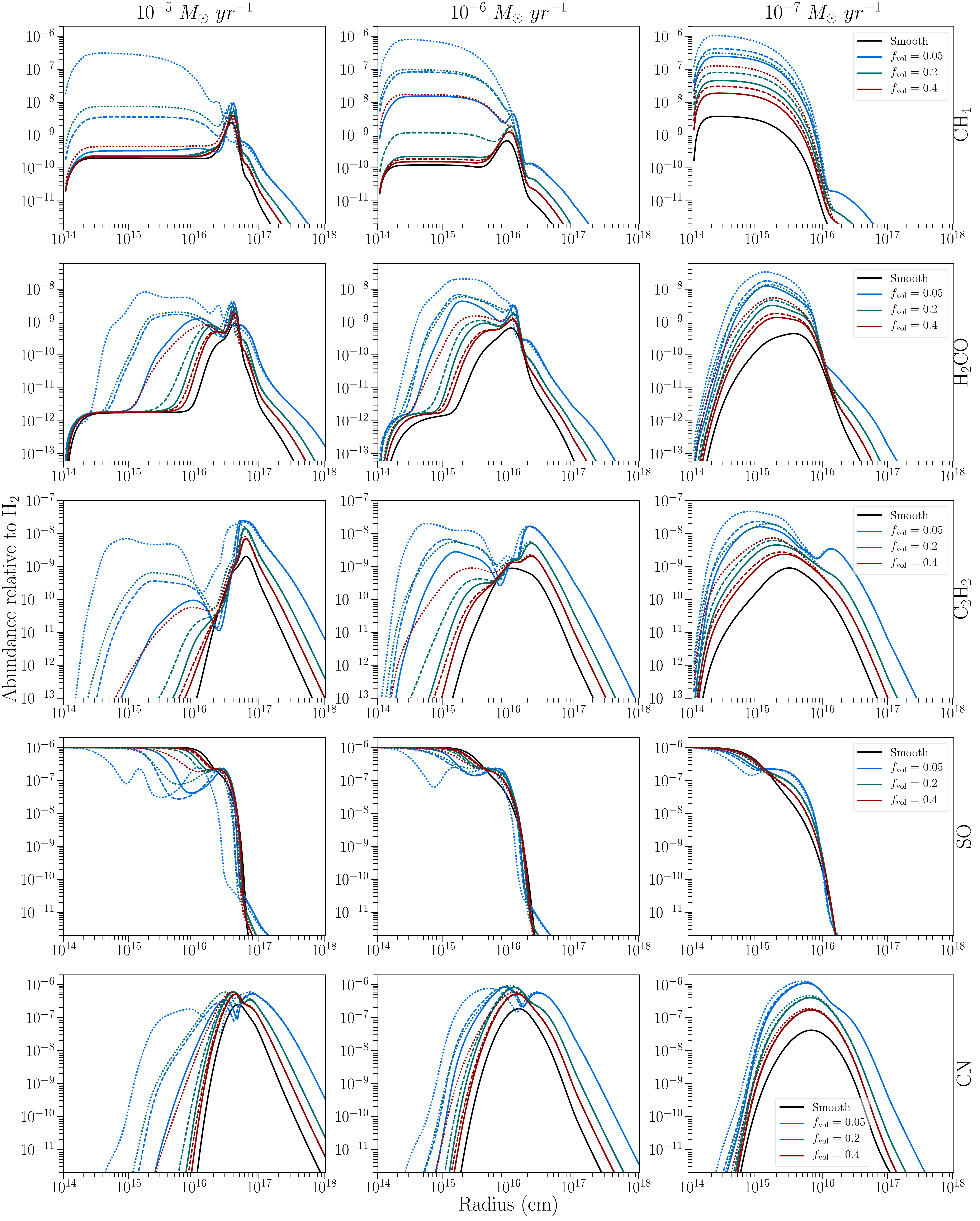}
\caption{Abundance of CH$_4$, H$_2$CO, C$_2$H$_2$, SO and CN relative to H$_2$ throughout one-component O-rich outflow with different mass-loss rates $\dot{M}$ and clump volume filling factors $f_\mathrm{vol}$.
Solid black line: calculated abundance for a smooth, uniform outflow. 
Solid coloured line: characteristic clump scale $l_* = 5 \times 10^{12}$ cm, {porosity length $h_* = 1 \times 10^{14}, 2.5 \times 10^{13}, 1.25 \times 10^{13}$ cm for $f_\mathrm{vol} = 0.05, 0.2, 0.4$, respectively.} 
Dashed coloured line: $l_* = 10^{13}$ cm, {$h_* = 2 \times 10^{14}, 5 \times 10^{13}, 2.5 \times 10^{13}$ cm for $f_\mathrm{vol} = 0.05, 0.2, 0.4$, respectively.}
Dotted coloured line: $l_* = 5 \times 10^{13}$ cm, {$h_* = 1 \times 10^{15}, 2.5 \times 10^{14}, 1.25 \times 10^{14}$ cm for $f_\mathrm{vol} = 0.05, 0.2, 0.4$, respectively.}
{
Note that models with $f_\mathrm{vol} = 0.2,\ l_* = 5 \times 10^{12}$ cm (green, solid) and $f_\mathrm{vol} = 0.4,\ l_* =  1 \times 10^{13}$ cm (red, dashed) have the same porosity length $h_* = 2.5 \times 10^{13}$ cm.
}
For reference, $1\ R_* = 5 \times 10^{13}$ cm.
}
\label{fig:App-OneComp-Orich}
\end{figure*}

\begin{figure*}[ht]
\centering
\includegraphics[width=0.96\textwidth]{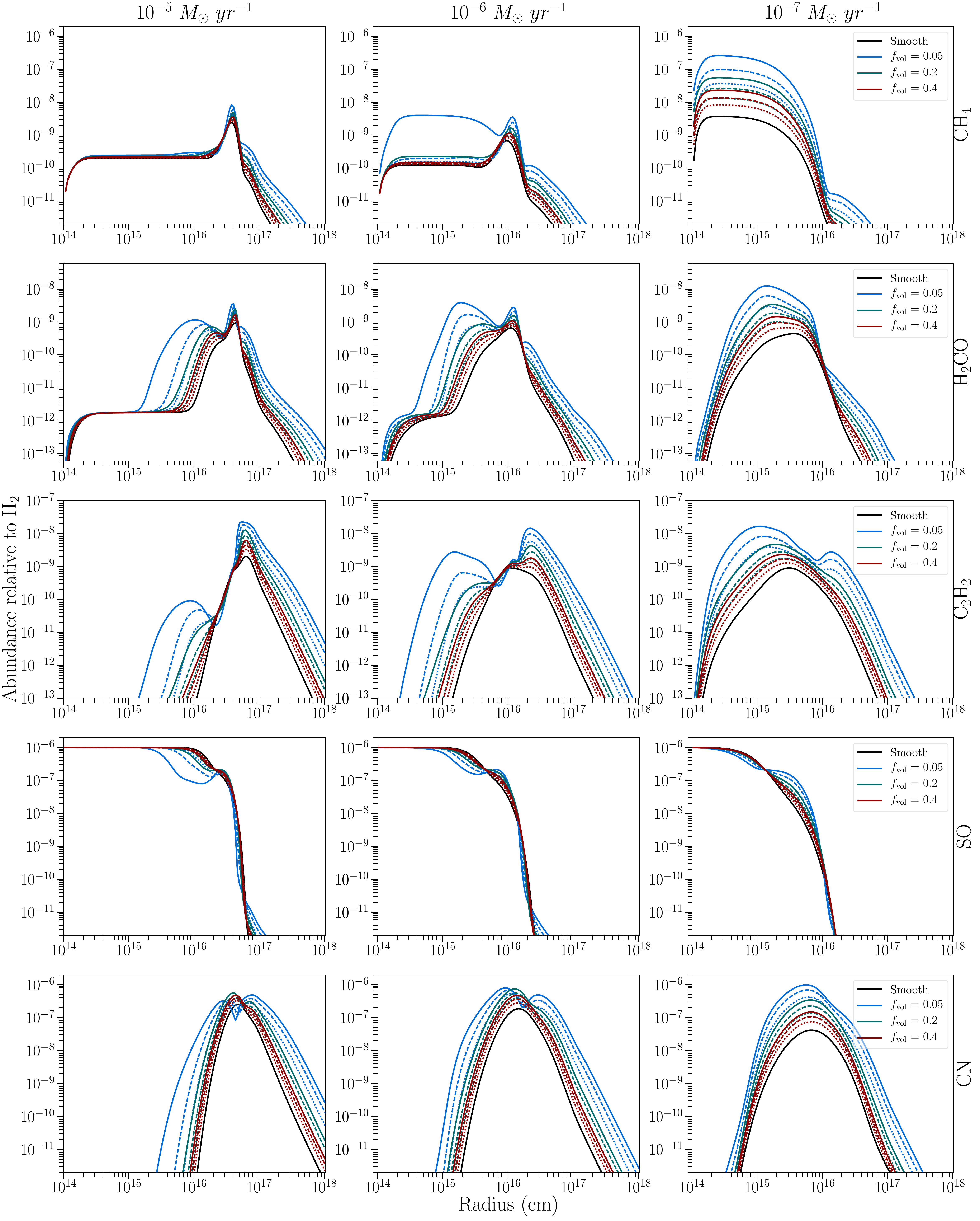}
\caption{Abundance of CH$_4$, H$_2$CO, C$_2$H$_2$, SO, and CN relative to H$_2$ throughout a two-component O-rich outflow with different mass-loss rates $\dot{M}$ and clump volume filling factors $f_\mathrm{vol}$. The characteristic size of the clumps at the stellar radius is $l_* = 10^{13}$ cm.
{
Blue lines: porosity length $h_* = 2 \times 10^{14}$ cm.
Green lines: $h_* = 5 \times 10^{13}$ cm.
Red lines: $h_* = 2.5 \times 10^{13}$ cm.
}
Solid black line: calculated abundance for a smooth, uniform outflow. 
Solid coloured line: density contrast between the inter-clump and smooth outflow $f_\mathrm{ic} = 0.1$. Dashed coloured line: $f_\mathrm{ic} = 0.3$. Dotted coloured line: $f_\mathrm{ic} = 0.5$.
{Note that the models with $f_\mathrm{vol} = 0.4$ (red) have the same porosity length as the one-component outflows with $f_\mathrm{vol} = 0.2,\ l_* = 5 \times 10^{12}$ cm and $f_\mathrm{vol} = 0.4,\ l_* = 1 \times 10^{13}$ cm.}
For reference, $1\ R_* = 5 \times 10^{13}$ cm.
}
\label{fig:App-TwoComp-Orich}
\end{figure*}

{
Figs. \ref{fig:App-OneComp-Orich} and \ref{fig:App-TwoComp-Orich} show the abundance profiles of CH$_4$, H$_2$CO, C$_2$H$_2$, SO, and CN for respectively a one-component and two-component O-rich outflow.
The corresponding column densities are listed in Tables \ref{table:App-CD-1C-O} and \ref{table:App-CD-2C-O}.

CH$_4$ and C$_2$H$_2$ are expected to have very low abundances assuming TE \citep{Agundez2010}.
Similar to the molecules presented in Sect. \ref{sect:Results}, we find that a clumpy outflow can increase their inner wind abundances significantly. The largest effects are again seen for one-component, highly porous outflows.

H$_2$CO has been detected in IK Tau with an abundance between $10^{-7} - 10^{-8}$ \citep{VelillaPrieto2017}.
Our models with $\dot{M} = 10^{-6}$ M$_\odot$ yr$^{-1}$ give a peak abundance similar to the observed range for highly porous one-component outflows.

SO is a parent species in the O-rich outflow, and plays an important role in the formation of CS in the inner wind (Sect. \ref{subsubsect:Results-Orich-CS}).
The decrease in SO abundance is linked to the C abundance (Sect. \ref{subsubsect:Results-Orich-HCN}), and shows that parent species can also be affected by the clumpiness of the outflow.

CN plays a large role in the formation of HCN (Sect. \ref{subsubsect:Results-Orich-HCN}).
In IK Tau, it is observed in the intermediate wind (at $\sim 10^{16}$ cm) with an abundance between $2 \times 10^{-10} - 6 \times 10^{-8}$, with a peak abundance of  $3 \times 10^{-7}$ around $5 \times 10^{16}$ cm \citep{Decin2010}.
In TX Cam, CN has a peak abundance of $2 \times 10^{-7}$ \citep{Olofsson1991}, with an upper limit of $5 \times 10^{-6}$ \citep{Charnley1997}.
Both one-component and two-component models with $\dot{M} = 10^{-6}$ M$_\odot$ yr$^{-1}$ have peak abundances within an order of magnitude of those of IK Tau and TX Cam. Note that this does not only hold for highly porous outflows.
The location of the peak CN abundance in the outflow of IK Tau is reproduced by one-component and two-component highly porous models with $\dot{M} = 10^{-6}$ M$_\odot$ yr$^{-1}$.
}

\begin{sidewaystable}[ht]
    \caption{Column density [cm$^{-2}]$ of CH$_4$, H$_2$CO, C$_2$H$_2$, SO, and CN in a smooth O-rich outflow with different mass-loss rates, together with column density ratios relative to the smooth outflow for specific one-component outflows. The corresponding abundance profiles are shown in Fig. \ref{fig:App-OneComp-Orich}. 
{Note that the models with $f_\mathrm{vol} = 0.2,\ l_* = 5 \times 10^{12}$ cm and $f_\mathrm{vol} = 0.4,\ l_* = 1 \times 10^{13}$ cm have the same porosity length $h_* = 2.5 \times 10^{13}$ cm.}
    }
    \centering
    \resizebox{0.95\textwidth}{!}{%
    \begin{tabular}{c c c c c c c c c c c c c c c c c  }
    \hline \hline 
    \noalign{\smallskip}
    	$\dot{M}$ & Species & & {CH$_4$} & &  & {H$_2$CO} & &  & {C$_2$H$_2$} & &  & {SO} & &  & {CN} & \\
    \cmidrule(lr){1-1} \cmidrule(lr){2-2} \cmidrule(lr){3-5} \cmidrule(lr){6-8}  \cmidrule(lr){9-11}  \cmidrule(lr){12-14} \cmidrule(lr){15-17}  
    	\noalign{\smallskip}
{\multirow{5}{*}{\rotatebox[origin=r]{90}{$10^{-5}\ \mathrm{M}_\odot\ \mathrm{yr}^{-1}$}}} &    Smooth & \multicolumn{3}{c}{ 7.8e+12  cm$^{-2}$} &  \multicolumn{3}{c}{ 1.6e+11  cm$^{-2}$} &  \multicolumn{3}{c}{ 1.6e+11  cm$^{-2}$} &  \multicolumn{3}{c}{ 5.6e+16  cm$^{-2}$} &  \multicolumn{3}{c}{ 2.9e+13  cm$^{-2}$} \\
    \cmidrule(lr){2-2} \cmidrule(lr){3-5} \cmidrule(lr){6-8}  \cmidrule(lr){9-11}  \cmidrule(lr){12-14} \cmidrule(lr){15-17}  
	& $f_\mathrm{vol}$ & 0.05 & 0.2 & 0.4 & 0.05 & 0.2 & 0.4 & 0.05 & 0.2 & 0.4 & 0.05 & 0.2 & 0.4 & 0.05 & 0.2 & 0.4 \\	
    \cmidrule(lr){2-2} \cmidrule(lr){3-5} \cmidrule(lr){6-8}  \cmidrule(lr){9-11}  \cmidrule(lr){12-14} \cmidrule(lr){15-17}  
& $l_* = 5 \times 10^{12}$ cm & 1.5e+00 & 1.2e+00 & 1.1e+00 & 9.4e+00 & 2.7e+00 & 1.8e+00 & 1.2e+01 & 4.6e+00 & 2.4e+00 & 9.7e-01 & 1.0e+00 & 1.0e+00 & 3.9e+00 & 3.2e+00 & 2.3e+00 \\
& $l_* = 1 \times 10^{13}$ cm & 1.5e+01 & 1.2e+00 & 1.1e+00 & 3.4e+01 & 3.3e+00 & 1.9e+00 & 2.3e+01 & 4.9e+00 & 2.5e+00 & 9.5e-01 & 9.9e-01 & 1.0e+00 & 5.1e+00 & 3.5e+00 & 2.4e+00 \\
& $l_* = 5 \times 10^{13}$ cm & 1.3e+03 & 3.2e+01 & 2.1e+00 & 2.9e+02 & 4.4e+01 & 5.8e+00 & 5.8e+02 & 2.9e+01 & 3.7e+00 & 7.4e-01 & 9.5e-01 & 9.8e-01 & 1.5e+01 & 7.0e+00 & 3.7e+00 \\
	\noalign{\smallskip}
	\hline
    \noalign{\smallskip}
{\multirow{5}{*}{\rotatebox[origin=r]{90}{$10^{-6}\ \mathrm{M}_\odot\ \mathrm{yr}^{-1}$}}} &    Smooth & \multicolumn{3}{c}{ 5.0e+11  cm$^{-2}$} &  \multicolumn{3}{c}{ 5.0e+10  cm$^{-2}$} &  \multicolumn{3}{c}{ 6.8e+10  cm$^{-2}$} &  \multicolumn{3}{c}{ 5.4e+15  cm$^{-2}$} &  \multicolumn{3}{c}{ 8.8e+12  cm$^{-2}$} \\
     \cmidrule(lr){2-2} \cmidrule(lr){3-5} \cmidrule(lr){6-8}  \cmidrule(lr){9-11}  \cmidrule(lr){12-14} \cmidrule(lr){15-17}  
	& $f_\mathrm{vol}$ & 0.05 & 0.2 & 0.4 & 0.05 & 0.2 & 0.4 & 0.05 & 0.2 & 0.4 & 0.05 & 0.2 & 0.4 & 0.05 & 0.2 & 0.4 \\	
     \cmidrule(lr){2-2} \cmidrule(lr){3-5} \cmidrule(lr){6-8}  \cmidrule(lr){9-11}  \cmidrule(lr){12-14} \cmidrule(lr){15-17}  
& $l_* = 5 \times 10^{12}$ cm & 9.8e+01 & 1.7e+00 & 1.3e+00 & 3.4e+01 & 4.8e+00 & 2.4e+00 & 3.0e+01 & 3.2e+00 & 1.9e+00 & 9.4e-01 & 9.9e-01 & 9.9e-01 & 9.0e+00 & 5.3e+00 & 3.0e+00 \\
& $l_* = 1 \times 10^{13}$ cm & 5.4e+02 & 8.1e+00 & 1.5e+00 & 7.2e+01 & 6.9e+00 & 2.8e+00 & 9.8e+01 & 4.2e+00 & 2.0e+00 & 8.9e-01 & 9.8e-01 & 9.9e-01 & 1.1e+01 & 5.7e+00 & 3.2e+00 \\
& $l_* = 5 \times 10^{13}$ cm & 5.1e+03 & 6.8e+02 & 1.2e+02 & 3.0e+02 & 6.3e+01 & 1.2e+01 & 5.1e+02 & 7.5e+01 & 8.5e+00 & 6.9e-01 & 9.0e-01 & 9.6e-01 & 2.0e+01 & 1.0e+01 & 4.6e+00 \\
	\noalign{\smallskip}
	\hline
    \noalign{\smallskip}
{\multirow{5}{*}{\rotatebox[origin=r]{90}{$10^{-7}\ \mathrm{M}_\odot\ \mathrm{yr}^{-1}$}}} &    Smooth & \multicolumn{3}{c}{ 3.0e+12  cm$^{-2}$} &  \multicolumn{3}{c}{ 3.3e+10  cm$^{-2}$} &  \multicolumn{3}{c}{ 5.2e+10  cm$^{-2}$} &  \multicolumn{3}{c}{ 4.7e+14  cm$^{-2}$} &  \multicolumn{3}{c}{ 5.0e+11  cm$^{-2}$} \\
    \cmidrule(lr){2-2} \cmidrule(lr){3-5} \cmidrule(lr){6-8}  \cmidrule(lr){9-11}  \cmidrule(lr){12-14} \cmidrule(lr){15-17}  
	& $f_\mathrm{vol}$ & 0.05 & 0.2 & 0.4 & 0.05 & 0.2 & 0.4 & 0.05 & 0.2 & 0.4 & 0.05 & 0.2 & 0.4 & 0.05 & 0.2 & 0.4 \\	
    \cmidrule(lr){2-2} \cmidrule(lr){3-5} \cmidrule(lr){6-8}  \cmidrule(lr){9-11}  \cmidrule(lr){12-14} \cmidrule(lr){15-17}  
& $l_* = 5 \times 10^{12}$ cm & 6.4e+01 & 1.2e+01 & 5.1e+00 & 4.0e+01 & 9.2e+00 & 3.9e+00 & 4.9e+01 & 8.9e+00 & 3.7e+00 & 9.2e-01 & 9.8e-01 & 9.9e-01 & 3.4e+01 & 1.0e+01 & 4.2e+00 \\
& $l_* = 1 \times 10^{13}$ cm & 1.1e+02 & 2.2e+01 & 8.4e+00 & 6.2e+01 & 1.4e+01 & 5.3e+00 & 8.2e+01 & 1.4e+01 & 4.9e+00 & 8.8e-01 & 9.7e-01 & 9.8e-01 & 3.7e+01 & 1.1e+01 & 4.3e+00 \\
& $l_* = 5 \times 10^{13}$ cm & 2.7e+02 & 8.5e+01 & 3.5e+01 & 1.4e+02 & 4.7e+01 & 1.8e+01 & 2.2e+02 & 5.8e+01 & 1.8e+01 & 7.8e-01 & 9.0e-01 & 9.4e-01 & 4.9e+01 & 1.4e+01 & 5.1e+00 \\
	\noalign{\smallskip}  
    	\hline \hline
    \end{tabular}%
    }
    \label{table:App-CD-1C-O}
\end{sidewaystable}

\begin{sidewaystable}[ht]
    \caption{Column density [cm$^{-2}]$ of CH$_4$, H$_2$CO, C$_2$H$_2$, SO, and CN in a smooth O-rich outflow with different mass-loss rates, together with column density ratios relative to the smooth outflow for specific two-component outflows. The corresponding abundance profiles are shown in Fig. \ref{fig:App-TwoComp-Orich}. 
{Note that models with $f_\mathrm{vol} = 0.4$ have the same porosity length $h_* = 2.5 \times 10^{13}$ cm as the one-component models with $f_\mathrm{vol} = 0.2,\ l_* = 5 \times 10^{12}$ cm and $f_\mathrm{vol} = 0.4,\ l_* 1 = \times 10^{13}$ cm.}
    }
    \centering
    \resizebox{0.95\textwidth}{!}{%
    \begin{tabular}{c c c c c c c c c c c c c c c c c  }
    \hline \hline 
    \noalign{\smallskip}
    	$\dot{M}$ & Species & & {CH$_4$} & &  & {H$_2$CO} & &  & {C$_2$H$_2$} & &  & {SO} & &  & {CN} & \\
    \cmidrule(lr){1-1} \cmidrule(lr){2-2} \cmidrule(lr){3-5} \cmidrule(lr){6-8}  \cmidrule(lr){9-11}  \cmidrule(lr){12-14} \cmidrule(lr){15-17}  
    	\noalign{\smallskip}
{\multirow{5}{*}{\rotatebox[origin=r]{90}{$10^{-5}\ \mathrm{M}_\odot\ \mathrm{yr}^{-1}$}}} &    Smooth & \multicolumn{3}{c}{ 7.8e+12  cm$^{-2}$} &  \multicolumn{3}{c}{ 1.6e+11  cm$^{-2}$} &  \multicolumn{3}{c}{ 1.6e+11  cm$^{-2}$} &  \multicolumn{3}{c}{ 5.6e+16  cm$^{-2}$} &  \multicolumn{3}{c}{ 2.9e+13  cm$^{-2}$} \\
    \cmidrule(lr){2-2} \cmidrule(lr){3-5} \cmidrule(lr){6-8}  \cmidrule(lr){9-11}  \cmidrule(lr){12-14} \cmidrule(lr){15-17}  
	& $f_\mathrm{vol}$ & 0.05 & 0.2 & 0.4 & 0.05 & 0.2 & 0.4 & 0.05 & 0.2 & 0.4 & 0.05 & 0.2 & 0.4 & 0.05 & 0.2 & 0.4 \\	
    \cmidrule(lr){2-2} \cmidrule(lr){3-5} \cmidrule(lr){6-8}  \cmidrule(lr){9-11}  \cmidrule(lr){12-14} \cmidrule(lr){15-17}  
& $f_\mathrm{ic} = 0.1$ & 1.2e+00 & 1.1e+00 & 1.1e+00 & 8.7e+00 & 2.7e+00 & 1.7e+00 & 1.1e+01 & 4.1e+00 & 2.2e+00 & 9.8e-01 & 1.0e+00 & 1.0e+00 & 4.0e+00 & 3.1e+00 & 2.2e+00 \\
& $f_\mathrm{ic} = 0.3$ & 1.1e+00 & 1.1e+00 & 1.1e+00 & 4.0e+00 & 1.9e+00 & 1.5e+00 & 7.3e+00 & 2.9e+00 & 1.8e+00 & 1.0e+00 & 1.1e+00 & 1.1e+00 & 2.9e+00 & 2.3e+00 & 1.7e+00 \\
& $f_\mathrm{ic} = 0.5$ & 1.1e+00 & 1.1e+00 & 1.2e+00 & 2.4e+00 & 1.5e+00 & 1.3e+00 & 4.5e+00 & 2.0e+00 & 1.5e+00 & 1.0e+00 & 1.1e+00 & 1.2e+00 & 2.2e+00 & 1.8e+00 & 1.5e+00 \\
	\noalign{\smallskip}
	\hline
    \noalign{\smallskip}
{\multirow{5}{*}{\rotatebox[origin=r]{90}{$10^{-6}\ \mathrm{M}_\odot\ \mathrm{yr}^{-1}$}}} &    Smooth & \multicolumn{3}{c}{ 5.0e+11  cm$^{-2}$} &  \multicolumn{3}{c}{ 5.0e+10  cm$^{-2}$} &  \multicolumn{3}{c}{ 6.8e+10  cm$^{-2}$} &  \multicolumn{3}{c}{ 5.4e+15  cm$^{-2}$} &  \multicolumn{3}{c}{ 8.8e+12  cm$^{-2}$} \\
   \cmidrule(lr){2-2} \cmidrule(lr){3-5} \cmidrule(lr){6-8}  \cmidrule(lr){9-11}  \cmidrule(lr){12-14} \cmidrule(lr){15-17}  
	& $f_\mathrm{vol}$ & 0.05 & 0.2 & 0.4 & 0.05 & 0.2 & 0.4 & 0.05 & 0.2 & 0.4 & 0.05 & 0.2 & 0.4 & 0.05 & 0.2 & 0.4 \\	
    \cmidrule(lr){2-2} \cmidrule(lr){3-5} \cmidrule(lr){6-8}  \cmidrule(lr){9-11}  \cmidrule(lr){12-14} \cmidrule(lr){15-17}  
& $f_\mathrm{ic} = 0.1$ & 2.3e+01 & 1.7e+00 & 1.2e+00 & 3.0e+01 & 4.9e+00 & 2.4e+00 & 2.8e+01 & 3.1e+00 & 1.8e+00 & 9.5e-01 & 1.0e+00 & 1.0e+00 & 8.9e+00 & 4.9e+00 & 2.8e+00 \\
& $f_\mathrm{ic} = 0.3$ & 1.5e+00 & 1.2e+00 & 1.2e+00 & 1.1e+01 & 3.0e+00 & 1.8e+00 & 7.1e+00 & 2.1e+00 & 1.5e+00 & 9.9e-01 & 1.1e+00 & 1.1e+00 & 6.1e+00 & 3.4e+00 & 2.1e+00 \\
& $f_\mathrm{ic} = 0.5$ & 1.2e+00 & 1.2e+00 & 1.2e+00 & 4.8e+00 & 2.0e+00 & 1.5e+00 & 3.3e+00 & 1.6e+00 & 1.3e+00 & 1.0e+00 & 1.1e+00 & 1.2e+00 & 4.1e+00 & 2.3e+00 & 1.6e+00 \\
	\noalign{\smallskip}
	\hline
    \noalign{\smallskip}
{\multirow{5}{*}{\rotatebox[origin=r]{90}{$10^{-7}\ \mathrm{M}_\odot\ \mathrm{yr}^{-1}$}}} &    Smooth & \multicolumn{3}{c}{ 3.0e+12  cm$^{-2}$} &  \multicolumn{3}{c}{ 3.3e+10  cm$^{-2}$} &  \multicolumn{3}{c}{ 5.2e+10  cm$^{-2}$} &  \multicolumn{3}{c}{ 4.7e+14  cm$^{-2}$} &  \multicolumn{3}{c}{ 5.0e+11  cm$^{-2}$} \\
    \cmidrule(lr){2-2} \cmidrule(lr){3-5} \cmidrule(lr){6-8}  \cmidrule(lr){9-11}  \cmidrule(lr){12-14} \cmidrule(lr){15-17}  
	& $f_\mathrm{vol}$ & 0.05 & 0.2 & 0.4 & 0.05 & 0.2 & 0.4 & 0.05 & 0.2 & 0.4 & 0.05 & 0.2 & 0.4 & 0.05 & 0.2 & 0.4 \\	
    \cmidrule(lr){2-2} \cmidrule(lr){3-5} \cmidrule(lr){6-8}  \cmidrule(lr){9-11}  \cmidrule(lr){12-14} \cmidrule(lr){15-17}  
& $f_\mathrm{ic} = 0.1$ & 6.7e+01 & 1.5e+01 & 6.3e+00 & 4.2e+01 & 1.0e+01 & 4.3e+00 & 5.3e+01 & 9.7e+00 & 3.9e+00 & 9.2e-01 & 9.9e-01 & 1.0e+00 & 3.1e+01 & 8.9e+00 & 3.7e+00 \\
& $f_\mathrm{ic} = 0.3$ & 2.5e+01 & 7.2e+00 & 3.7e+00 & 2.0e+01 & 5.5e+00 & 2.8e+00 & 2.2e+01 & 5.2e+00 & 2.6e+00 & 9.7e-01 & 1.0e+00 & 1.1e+00 & 2.0e+01 & 5.7e+00 & 2.6e+00 \\
& $f_\mathrm{ic} = 0.5$ & 9.4e+00 & 3.6e+00 & 2.4e+00 & 8.7e+00 & 3.0e+00 & 2.0e+00 & 9.1e+00 & 2.9e+00 & 1.8e+00 & 1.0e+00 & 1.1e+00 & 1.2e+00 & 1.1e+01 & 3.4e+00 & 1.9e+00 \\
    	\hline \hline
    \end{tabular}%
    }
    \label{table:App-CD-2C-O}
\end{sidewaystable}

\subsection{C-rich outflows}        \label{subsect:App:AddMolecules:Crich}

\begin{figure*}[!ht]
\centering
\includegraphics[width=0.96\textwidth]{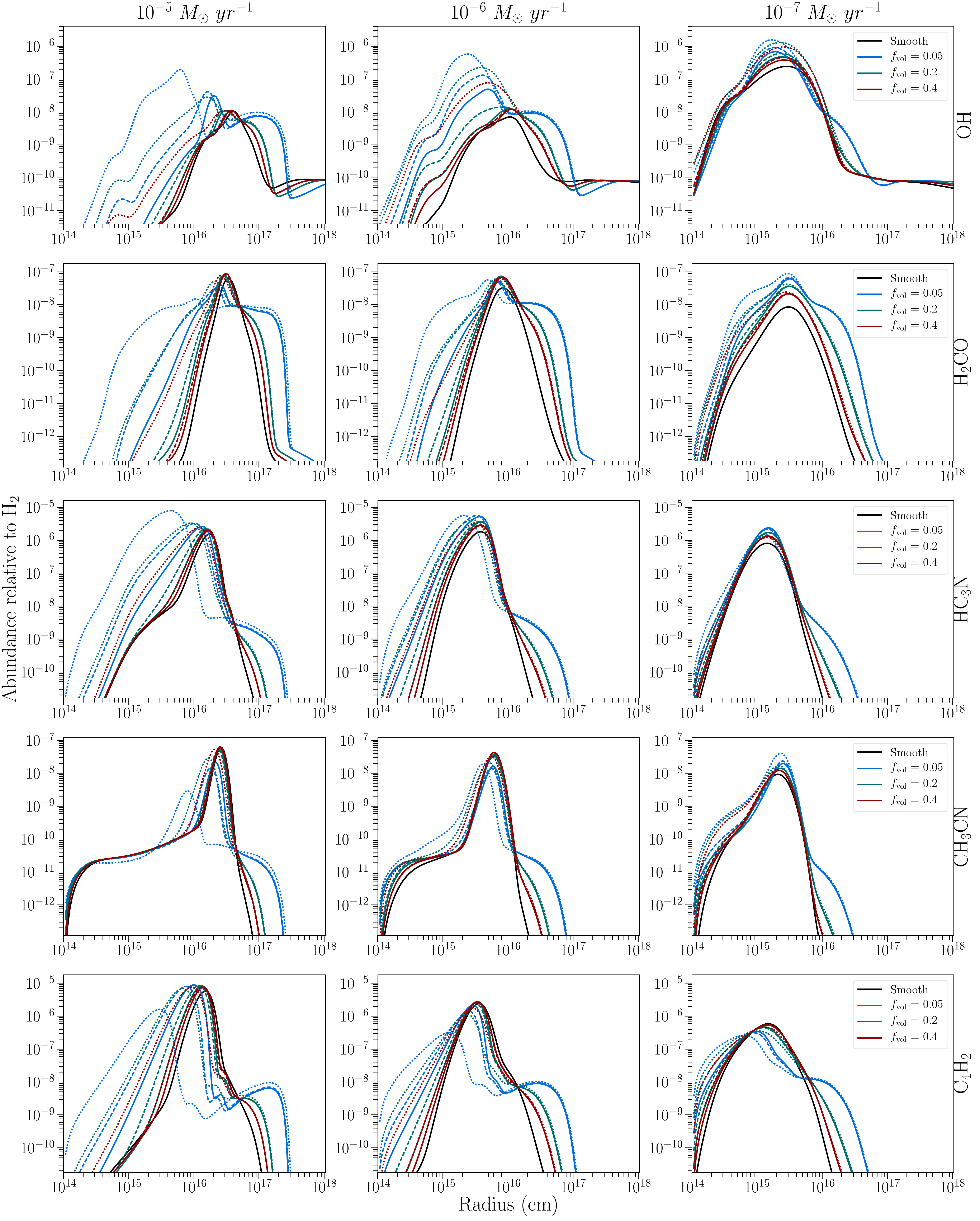}
\caption{Abundance of OH, H$_2$CO, HC$_3$N, CH$_3$CN, and C$_4$H$_2$ relative to H$_2$ throughout a one-component C-rich outflow with different mass-loss rates $\dot{M}$ and clump volume filling factors $f_\mathrm{vol}$.
Solid black line: calculated abundance for a smooth, uniform outflow. 
Solid coloured line: characteristic clump scale $l_* = 5 \times 10^{12}$ cm, {porosity length $h_* = 1 \times 10^{14}, 2.5 \times 10^{13}, 1.25 \times 10^{13}$ cm for $f_\mathrm{vol} = 0.05, 0.2, 0.4$, respectively.} 
Dashed coloured line: $l_* = 10^{13}$ cm, {$h_* = 2 \times 10^{14}, 5 \times 10^{13}, 2.5 \times 10^{13}$ cm for $f_\mathrm{vol} = 0.05, 0.2, 0.4$, respectively.}
Dotted coloured line: $l_* = 5 \times 10^{13}$ cm, {$h_* = 1 \times 10^{15}, 2.5 \times 10^{14}, 1.25 \times 10^{14}$ cm for $f_\mathrm{vol} = 0.05, 0.2, 0.4$, respectively.}
{
Note that models with $f_\mathrm{vol} = 0.2,\ l_* = 5 \times 10^{12}$ cm (green, solid) and $f_\mathrm{vol} = 0.4,\ l_* =  1 \times 10^{13}$ cm (red, dashed) have the same porosity length $h_* = 2.5 \times 10^{13}$ cm.
}
For reference, $1\ R_* = 5 \times 10^{13}$ cm.
}
\label{fig:App-OneComp-Crich}
\end{figure*}

\begin{figure*}[!hb]
\centering
\includegraphics[width=0.96\textwidth]{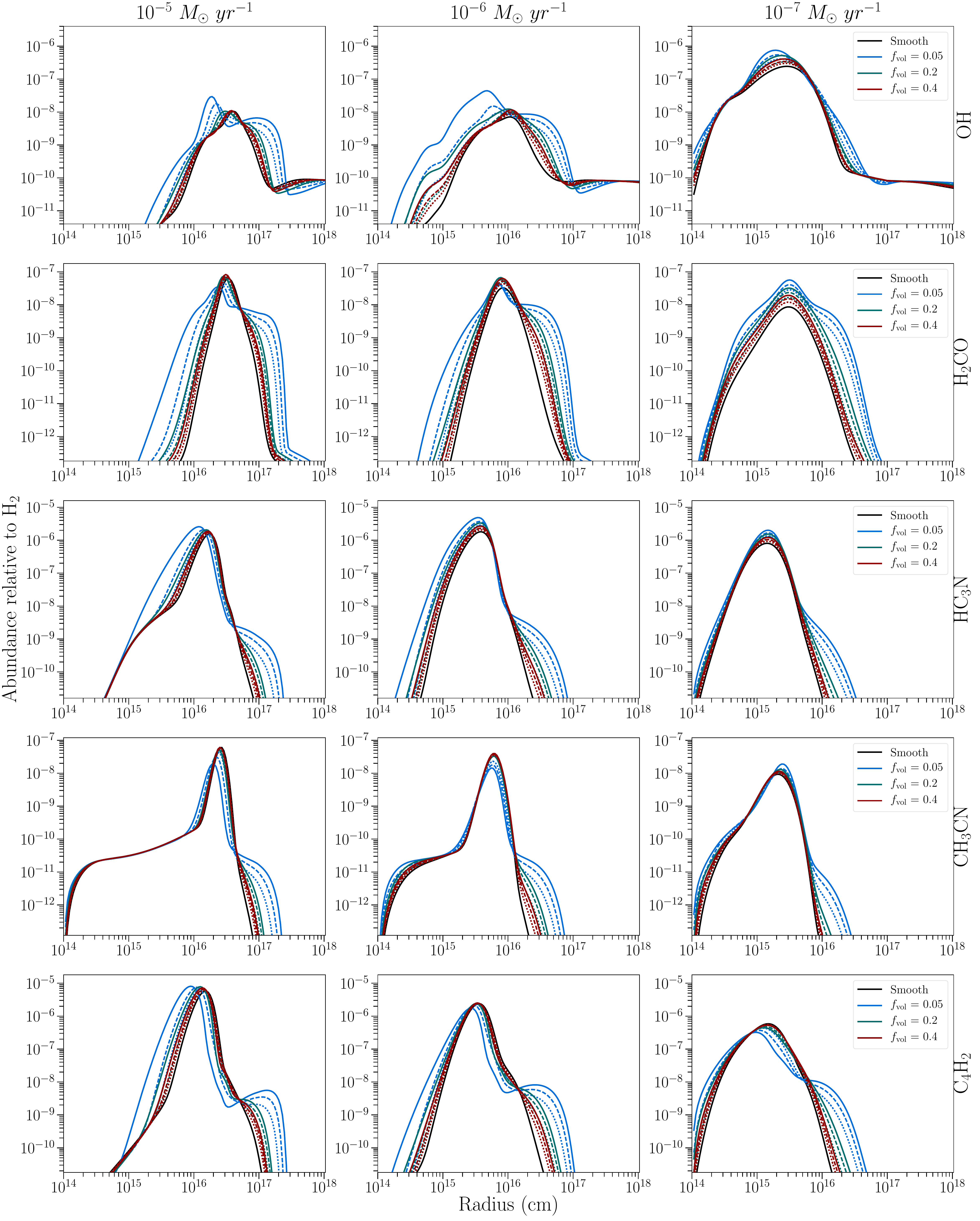}
\caption{Abundance of OH, H$_2$CO, HC$_3$N, CH$_3$CN, and C$_4$H$_2$ relative to H$_2$ throughout a two-component C-rich outflow with different mass-loss rates $\dot{M}$ and clump volume filling factors $f_\mathrm{vol}$. The characteristic size of the clumps at the stellar radius is $l_* = 10^{13}$ cm.
{
Blue lines: porosity length $h_* = 2 \times 10^{14}$ cm.
Green lines: $h_* = 5 \times 10^{13}$ cm.
Red lines: $h_* = 2.5 \times 10^{13}$ cm.
}
Solid black line: calculated abundance for a smooth, uniform outflow. 
Solid coloured line: density contrast between the inter-clump and smooth outflow $f_\mathrm{ic} = 0.1$. Dashed coloured line: $f_\mathrm{ic} = 0.3$. Dotted coloured line: $f_\mathrm{ic} = 0.5$.
{Note that the models with $f_\mathrm{vol} = 0.4$ (red) have the same porosity length as the one-component outflows with $f_\mathrm{vol} = 0.2,\ l_* = 5 \times 10^{12}$ cm and $f_\mathrm{vol} = 0.4,\ l_* = 1 \times 10^{13}$ cm.}
For reference, $1\ R_* = 5 \times 10^{13}$ cm.
}
\label{fig:App-TwoComp-Crich}
\end{figure*}

{
Figs. \ref{fig:App-OneComp-Crich} and \ref{fig:App-TwoComp-Crich} show the abundance profiles of OH, H$_2$CO, HC$_3$N, CH$_3$CN, and C$_4$H$_2$ for respectively a one-component and two-component C-rich outflow.
The corresponding column densities are listed in Tables \ref{table:App-CD-1C-C} and \ref{table:App-CD-2C-C}.
The effects of clumpiness and porosity on the chemical formation and destruction pathways of these molecules are similar to those presented in Sect. \ref{subsect:Results-Crich}.

OH, H$_2$CO, HC$_3$N are expected to have a low inner wind abundance assuming TE \citep{Agundez2010}. 
Similar to the molecules presented in Sect. \ref{sect:Results}, their abundance in the inner region of the outflow increases when taking clumpiness into account, with again the largest effects seen in one-component, highly porous outflows. The abundance of OH is linked to that of H$_2$O (Sect. \ref{subsubsect:Results-Crich-H2O}). 

H$_2$CO has been detected in the outflow of IRC+10216 with an abundance of $1.3_{-0.8}^{+1.5} \times 10^{-8}$ \citep{Ford2004}.
Highly porous models with $\dot{M} = 10^{-5}$ M$_\odot$ yr$^{-1}$ have a peak value close to the observed abundance, both for a one-component and a two-component model.

HC$_3$N is located in a molecular shell around $3 \times 10^{16}$ cm in the outflow of IRC+10216 \citep{Agundez2017}. The location of its peak in abundance clearly depends on the specific clumpiness of the outflow.

The observed abundance profile of CH$_3$CN in the outflow of IRC+10216 has a peculiar shape, which cannot be reproduced by a smooth outflow model or clumpy outflow model (using the formalism of \citeauthor{Agundez2010}~\citeyear{Agundez2010}). Its abundance peaks around $10^{16}$ cm at $5.0 \times 10^{-8}$ \citep{Agundez2015}. Our one-component and two-component highly porous models have peak abundances within an order of magnitude of the observed value, with the peak also occurring around $10^{16}$ cm for a highly porous outflow. The one-component outflows have wider distribution compared to the two-component outflows.

C$_4$H$_2$ has recently been detected in the outflow of IRC+10216 with a column density of $(2.4 \pm 1.5) \times 10^{16}$ cm$^{-2}$ \citep{Fonfria2017}. From the analysis of \citet{Fonfria2017}, it is suggested that the photodissociation occurs closer to the central star, due to e.g. a clumpy outer envelope.
Our models with $\dot{M} = 10^{-5}$ M$_\odot$ yr$^{-1}$ yield a column density close to the observed value for highly porous one-component outflows.
}

\begin{sidewaystable}[!ht]
    \caption{Column density [cm$^{-2}]$ of OH, H$_2$CO, HC$_3$N, CH$_3$CN, and C$_4$H$_2$ in a smooth C-rich outflow with different mass-loss rates, together with column density ratios relative to the smooth outflow for specific one-component outflows. The corresponding abundance profiles are shown in Fig. \ref{fig:App-OneComp-Crich}. 
{Note that the models with $f_\mathrm{vol} = 0.2,\ l_* = 5 \times 10^{12}$ cm and $f_\mathrm{vol} = 0.4,\ l_* = 1 \times 10^{13}$ cm have the same porosity length $h_* = 2.5 \times 10^{13}$ cm.}
    }
    \centering
    	\resizebox{0.95\textwidth}{!}{%
    \begin{tabular}{c c c c c c c c c c c c c c c c c  }
    \hline \hline 
    \noalign{\smallskip}
    	$\dot{M}$ & Species & & {OH} & &  & {H$_2$CO} & &  & {HC$_3$N} & &  & {CH$_3$CN} & &  & {C$_4$H$_2$} & \\
    \cmidrule(lr){1-1} \cmidrule(lr){2-2} \cmidrule(lr){3-5} \cmidrule(lr){6-8}  \cmidrule(lr){9-11}  \cmidrule(lr){12-14} \cmidrule(lr){15-17}  
    	\noalign{\smallskip}
{\multirow{5}{*}{\rotatebox[origin=r]{90}{$10^{-5}\ \mathrm{M}_\odot\ \mathrm{yr}^{-1}$}}} &    Smooth & \multicolumn{3}{c}{ 2.7e+11  cm$^{-2}$} &  \multicolumn{3}{c}{ 1.2e+07  cm$^{-2}$} &  \multicolumn{3}{c}{ 5.6e+12  cm$^{-2}$} &  \multicolumn{3}{c}{ 1.2e+11  cm$^{-2}$} &  \multicolumn{3}{c}{ 2.5e+10  cm$^{-2}$} \\
    \cmidrule(lr){2-2} \cmidrule(lr){3-5} \cmidrule(lr){6-8}  \cmidrule(lr){9-11}  \cmidrule(lr){12-14} \cmidrule(lr){15-17}  
	& $f_\mathrm{vol}$ & 0.05 & 0.2 & 0.4 & 0.05 & 0.2 & 0.4 & 0.05 & 0.2 & 0.4 & 0.05 & 0.2 & 0.4 & 0.05 & 0.2 & 0.4 \\	
    \cmidrule(lr){2-2} \cmidrule(lr){3-5} \cmidrule(lr){6-8}  \cmidrule(lr){9-11}  \cmidrule(lr){12-14} \cmidrule(lr){15-17}  
& $l_* = 5 \times 10^{12}$ cm & 3.6e+00 & 1.4e+00 & 1.1e+00 & 1.6e+00 & 1.5e+00 & 1.5e+00 & 3.5e+00 & 1.6e+00 & 1.3e+00 & 5.5e-01 & 9.3e-01 & 1.1e+00 & 3.1e+00 & 1.8e+00 & 1.4e+00 \\
& $l_* = 1 \times 10^{13}$ cm & 7.6e+00 & 1.6e+00 & 1.2e+00 & 2.0e+00 & 1.6e+00 & 1.5e+00 & 7.9e+00 & 2.0e+00 & 1.5e+00 & 5.3e-01 & 9.6e-01 & 1.1e+00 & 4.1e+00 & 2.2e+00 & 1.6e+00 \\
& $l_* = 5 \times 10^{13}$ cm & 1.5e+02 & 1.2e+01 & 2.3e+00 & 5.4e+00 & 2.5e+00 & 2.0e+00 & 5.1e+01 & 1.0e+01 & 4.1e+00 & 3.4e-01 & 1.1e+00 & 1.3e+00 & 2.7e+00 & 4.2e+00 & 3.1e+00 \\
	\noalign{\smallskip}
	\hline
    \noalign{\smallskip}
{\multirow{5}{*}{\rotatebox[origin=r]{90}{$10^{-6}\ \mathrm{M}_\odot\ \mathrm{yr}^{-1}$}}} &    Smooth & \multicolumn{3}{c}{ 8.1e+10  cm$^{-2}$} &  \multicolumn{3}{c}{ 1.8e+06  cm$^{-2}$} &  \multicolumn{3}{c}{ 5.0e+11  cm$^{-2}$} &  \multicolumn{3}{c}{ 3.3e+10  cm$^{-2}$} &  \multicolumn{3}{c}{ 3.4e+09  cm$^{-2}$} \\
    \cmidrule(lr){2-2} \cmidrule(lr){3-5} \cmidrule(lr){6-8}  \cmidrule(lr){9-11}  \cmidrule(lr){12-14} \cmidrule(lr){15-17}  
	& $f_\mathrm{vol}$ & 0.05 & 0.2 & 0.4 & 0.05 & 0.2 & 0.4 & 0.05 & 0.2 & 0.4 & 0.05 & 0.2 & 0.4 & 0.05 & 0.2 & 0.4 \\	
    \cmidrule(lr){2-2} \cmidrule(lr){3-5} \cmidrule(lr){6-8}  \cmidrule(lr){9-11}  \cmidrule(lr){12-14} \cmidrule(lr){15-17}  
& $l_* = 5 \times 10^{12}$ cm & 1.2e+01 & 1.8e+00 & 1.3e+00 & 2.1e+00 & 2.3e+00 & 2.1e+00 & 3.6e+00 & 2.0e+00 & 1.6e+00 & 5.0e-01 & 1.0e+00 & 1.2e+00 & 1.1e+00 & 1.1e+00 & 1.1e+00 \\
& $l_* = 1 \times 10^{13}$ cm & 3.7e+01 & 3.6e+00 & 1.6e+00 & 2.5e+00 & 2.4e+00 & 2.1e+00 & 4.6e+00 & 2.4e+00 & 1.7e+00 & 4.9e-01 & 1.0e+00 & 1.2e+00 & 1.1e+00 & 1.1e+00 & 1.1e+00 \\
& $l_* = 5 \times 10^{13}$ cm & 2.2e+02 & 7.7e+01 & 2.3e+01 & 7.6e+00 & 3.3e+00 & 2.5e+00 & 6.8e+00 & 4.1e+00 & 2.8e+00 & 9.4e-01 & 9.1e-01 & 1.2e+00 & 8.0e-01 & 1.1e+00 & 1.1e+00 \\
	\noalign{\smallskip}
	\hline
    \noalign{\smallskip}
{\multirow{5}{*}{\rotatebox[origin=r]{90}{$10^{-7}\ \mathrm{M}_\odot\ \mathrm{yr}^{-1}$}}} &    Smooth & \multicolumn{3}{c}{ 1.1e+11  cm$^{-2}$} &  \multicolumn{3}{c}{ 3.4e+06  cm$^{-2}$} &  \multicolumn{3}{c}{ 7.2e+09  cm$^{-2}$} &  \multicolumn{3}{c}{ 7.4e+10  cm$^{-2}$} &  \multicolumn{3}{c}{ 3.7e+10  cm$^{-2}$} \\
    \cmidrule(lr){2-2} \cmidrule(lr){3-5} \cmidrule(lr){6-8}  \cmidrule(lr){9-11}  \cmidrule(lr){12-14} \cmidrule(lr){15-17}  
	& $f_\mathrm{vol}$ & 0.05 & 0.2 & 0.4 & 0.05 & 0.2 & 0.4 & 0.05 & 0.2 & 0.4 & 0.05 & 0.2 & 0.4 & 0.05 & 0.2 & 0.4 \\	
    \cmidrule(lr){2-2} \cmidrule(lr){3-5} \cmidrule(lr){6-8}  \cmidrule(lr){9-11}  \cmidrule(lr){12-14} \cmidrule(lr){15-17}  
& $l_* = 5 \times 10^{12}$ cm & 2.0e+00 & 1.6e+00 & 1.4e+00 & 6.9e+00 & 4.0e+00 & 2.4e+00 & 2.4e+00 & 1.9e+00 & 1.5e+00 & 1.3e+00 & 1.2e+00 & 1.2e+00 & 8.9e-01 & 1.0e+00 & 1.1e+00 \\
& $l_* = 1 \times 10^{13}$ cm & 2.7e+00 & 2.0e+00 & 1.6e+00 & 8.7e+00 & 4.4e+00 & 2.6e+00 & 2.3e+00 & 1.9e+00 & 1.5e+00 & 1.6e+00 & 1.3e+00 & 1.2e+00 & 1.0e+00 & 1.1e+00 & 1.1e+00 \\
& $l_* = 5 \times 10^{13}$ cm & 5.0e+00 & 4.4e+00 & 3.3e+00 & 1.7e+01 & 7.8e+00 & 4.3e+00 & 1.9e+00 & 1.7e+00 & 1.5e+00 & 3.4e+00 & 2.0e+00 & 1.7e+00 & 1.4e+00 & 1.4e+00 & 1.3e+00 \\
    	\hline \hline
    \end{tabular}%
    }
    \label{table:App-CD-1C-C}
\end{sidewaystable}

\begin{sidewaystable}[!hb]
    \caption{Column density [cm$^{-2}]$ of OH, H$_2$CO, HC$_3$N, CH$_3$CN, and C$_4$H$_2$ in a smooth C-rich outflow with different mass-loss rates, together with column density ratios relative to the smooth outflow for specific two-component outflows. The corresponding abundance profiles are shown in Fig. \ref{fig:App-TwoComp-Crich}. 
{Note that models with $f_\mathrm{vol} = 0.4$ have the same porosity length $h_* = 2.5 \times 10^{13}$ cm as the one-component models with $f_\mathrm{vol} = 0.2,\ l_* = 5 \times 10^{12}$ cm and $f_\mathrm{vol} = 0.4,\ l_* = 1 \times 10^{13}$ cm.}
    }
    \centering
    	\resizebox{0.95\textwidth}{!}{%
    \begin{tabular}{c c c c c c c c c c c c c c c c c  }
    \hline \hline 
    \noalign{\smallskip}
    	$\dot{M}$ & Species & & {OH} & &  & {H$_2$CO} & &  & {HC$_3$N} & &  & {CH$_3$CN} & &  & {C$_4$H$_2$} & \\
    \cmidrule(lr){1-1} \cmidrule(lr){2-2} \cmidrule(lr){3-5} \cmidrule(lr){6-8}  \cmidrule(lr){9-11}  \cmidrule(lr){12-14} \cmidrule(lr){15-17}  
    	\noalign{\smallskip}
{\multirow{5}{*}{\rotatebox[origin=r]{90}{$10^{-5}\ \mathrm{M}_\odot\ \mathrm{yr}^{-1}$}}} &    Smooth & \multicolumn{3}{c}{ 2.7e+11  cm$^{-2}$} &  \multicolumn{3}{c}{ 1.2e+07  cm$^{-2}$} &  \multicolumn{3}{c}{ 5.6e+12  cm$^{-2}$} &  \multicolumn{3}{c}{ 1.2e+11  cm$^{-2}$} &  \multicolumn{3}{c}{ 2.5e+10  cm$^{-2}$} \\
    \cmidrule(lr){2-2} \cmidrule(lr){3-5} \cmidrule(lr){6-8}  \cmidrule(lr){9-11}  \cmidrule(lr){12-14} \cmidrule(lr){15-17}  
	& $f_\mathrm{vol}$ & 0.05 & 0.2 & 0.4 & 0.05 & 0.2 & 0.4 & 0.05 & 0.2 & 0.4 & 0.05 & 0.2 & 0.4 & 0.05 & 0.2 & 0.4 \\	
    \cmidrule(lr){2-2} \cmidrule(lr){3-5} \cmidrule(lr){6-8}  \cmidrule(lr){9-11}  \cmidrule(lr){12-14} \cmidrule(lr){15-17}  
& $f_\mathrm{ic} = 0.1$ & 3.9e+00 & 1.4e+00 & 1.2e+00 & 1.6e+00 & 1.5e+00 & 1.4e+00 & 3.8e+00 & 1.7e+00 & 1.4e+00 & 5.8e-01 & 9.2e-01 & 1.0e+00 & 3.0e+00 & 1.8e+00 & 1.5e+00 \\
& $f_\mathrm{ic} = 0.3$ & 2.3e+00 & 1.3e+00 & 1.2e+00 & 1.2e+00 & 1.3e+00 & 1.3e+00 & 2.0e+00 & 1.4e+00 & 1.3e+00 & 7.3e-01 & 9.4e-01 & 1.1e+00 & 2.0e+00 & 1.5e+00 & 1.4e+00 \\
& $f_\mathrm{ic} = 0.5$ & 1.5e+00 & 1.2e+00 & 1.2e+00 & 1.1e+00 & 1.2e+00 & 1.2e+00 & 1.5e+00 & 1.3e+00 & 1.3e+00 & 8.5e-01 & 1.0e+00 & 1.1e+00 & 1.5e+00 & 1.3e+00 & 1.3e+00 \\
	\noalign{\smallskip}
	\hline
    \noalign{\smallskip}
{\multirow{5}{*}{\rotatebox[origin=r]{90}{$10^{-6}\ \mathrm{M}_\odot\ \mathrm{yr}^{-1}$}}} &    Smooth & \multicolumn{3}{c}{ 8.1e+10  cm$^{-2}$} &  \multicolumn{3}{c}{ 1.8e+06  cm$^{-2}$} &  \multicolumn{3}{c}{ 5.0e+11  cm$^{-2}$} &  \multicolumn{3}{c}{ 3.3e+10  cm$^{-2}$} &  \multicolumn{3}{c}{ 3.4e+09  cm$^{-2}$} \\
    \cmidrule(lr){2-2} \cmidrule(lr){3-5} \cmidrule(lr){6-8}  \cmidrule(lr){9-11}  \cmidrule(lr){12-14} \cmidrule(lr){15-17}  
	& $f_\mathrm{vol}$ & 0.05 & 0.2 & 0.4 & 0.05 & 0.2 & 0.4 & 0.05 & 0.2 & 0.4 & 0.05 & 0.2 & 0.4 & 0.05 & 0.2 & 0.4 \\	
    \cmidrule(lr){2-2} \cmidrule(lr){3-5} \cmidrule(lr){6-8}  \cmidrule(lr){9-11}  \cmidrule(lr){12-14} \cmidrule(lr){15-17}  
& $f_\mathrm{ic} = 0.1$ & 1.2e+01 & 2.2e+00 & 1.5e+00 & 2.0e+00 & 2.2e+00 & 1.9e+00 & 3.6e+00 & 2.1e+00 & 1.6e+00 & 5.1e-01 & 1.0e+00 & 1.2e+00 & 1.1e+00 & 1.1e+00 & 1.1e+00 \\
& $f_\mathrm{ic} = 0.3$ & 2.9e+00 & 1.5e+00 & 1.3e+00 & 1.7e+00 & 1.8e+00 & 1.6e+00 & 2.4e+00 & 1.6e+00 & 1.4e+00 & 6.2e-01 & 1.0e+00 & 1.2e+00 & 1.1e+00 & 1.1e+00 & 1.1e+00 \\
& $f_\mathrm{ic} = 0.5$ & 1.6e+00 & 1.3e+00 & 1.3e+00 & 1.4e+00 & 1.6e+00 & 1.4e+00 & 1.7e+00 & 1.4e+00 & 1.3e+00 & 7.7e-01 & 1.1e+00 & 1.2e+00 & 1.1e+00 & 1.1e+00 & 1.2e+00 \\
	\noalign{\smallskip}
	\hline
    \noalign{\smallskip}
{\multirow{5}{*}{\rotatebox[origin=r]{90}{$10^{-7}\ \mathrm{M}_\odot\ \mathrm{yr}^{-1}$}}} &    Smooth & \multicolumn{3}{c}{ 1.1e+11  cm$^{-2}$} &  \multicolumn{3}{c}{ 3.4e+06  cm$^{-2}$} &  \multicolumn{3}{c}{ 7.2e+09  cm$^{-2}$} &  \multicolumn{3}{c}{ 7.4e+10  cm$^{-2}$} &  \multicolumn{3}{c}{ 3.7e+10  cm$^{-2}$} \\
    \cmidrule(lr){2-2} \cmidrule(lr){3-5} \cmidrule(lr){6-8}  \cmidrule(lr){9-11}  \cmidrule(lr){12-14} \cmidrule(lr){15-17}  
	& $f_\mathrm{vol}$ & 0.05 & 0.2 & 0.4 & 0.05 & 0.2 & 0.4 & 0.05 & 0.2 & 0.4 & 0.05 & 0.2 & 0.4 & 0.05 & 0.2 & 0.4 \\	
    \cmidrule(lr){2-2} \cmidrule(lr){3-5} \cmidrule(lr){6-8}  \cmidrule(lr){9-11}  \cmidrule(lr){12-14} \cmidrule(lr){15-17}  
& $f_\mathrm{ic} = 0.1$ & 2.5e+00 & 1.8e+00 & 1.5e+00 & 6.8e+00 & 3.7e+00 & 2.3e+00 & 2.1e+00 & 1.7e+00 & 1.4e+00 & 1.4e+00 & 1.2e+00 & 1.2e+00 & 9.3e-01 & 1.0e+00 & 1.1e+00 \\
& $f_\mathrm{ic} = 0.3$ & 2.0e+00 & 1.6e+00 & 1.4e+00 & 4.5e+00 & 2.7e+00 & 1.8e+00 & 1.8e+00 & 1.5e+00 & 1.3e+00 & 1.2e+00 & 1.2e+00 & 1.2e+00 & 9.2e-01 & 1.0e+00 & 1.1e+00 \\
& $f_\mathrm{ic} = 0.5$ & 1.6e+00 & 1.4e+00 & 1.4e+00 & 3.1e+00 & 1.9e+00 & 1.5e+00 & 1.5e+00 & 1.3e+00 & 1.3e+00 & 1.1e+00 & 1.2e+00 & 1.2e+00 & 9.5e-01 & 1.1e+00 & 1.2e+00 \\
    	\hline \hline
    \end{tabular}%
    }
    \label{table:App-CD-2C-C}
\end{sidewaystable}

\section{Predicability of models}        \label{sect:App:Predictions}

{
We have selected so far undetected species whose peak abundance and number density change considerably when assuming a clumpy outflow.
These species are selected by requiring that a clumpy outflow causes both a peak fractional abundance larger than $1 \times 10^{-9}$ and an increase in column density relative to the smooth outflow by at least two orders of magnitude.
The abundance profiles and column densities of the selected molecules are considerably changed in a clumpy outflow, making them potentially detectable.

As in Sect. \ref{sect:Results} and Appendix \ref{sect:App:AddMolecules}, highly porous one-component outflows show the largest increase in both abundance in the inner region and column density.

}

\subsection{O-rich outflows}        \label{subsect:App:Predictions:Orich}

\begin{figure*}[ht]
\centering
\includegraphics[width=0.96\textwidth]{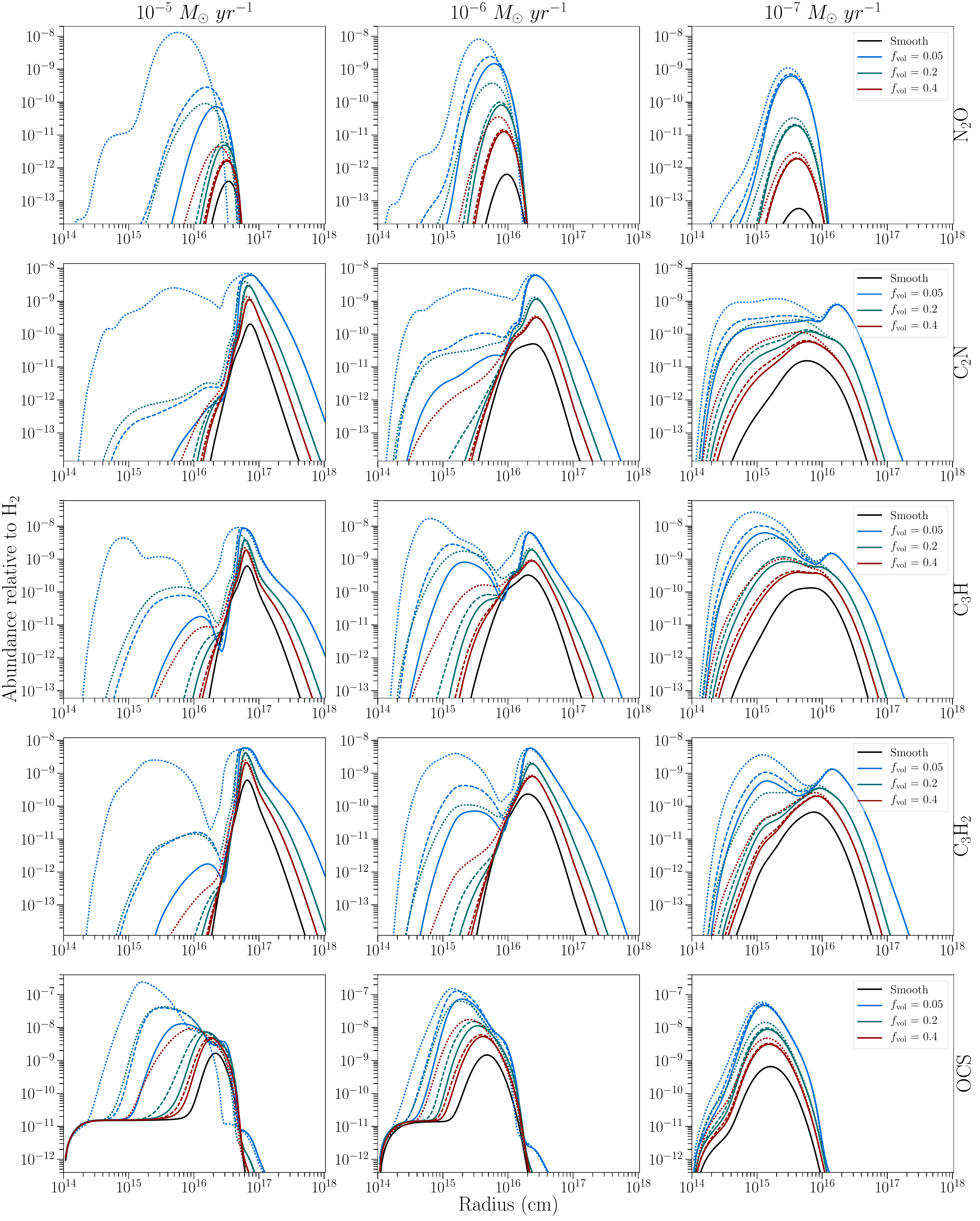}
\caption{Abundance of N$_2$O, C$_2$N, C$_3$H, C$_3$H$_2$ and OCS relative to H$_2$ throughout a one-component O-rich outflow with different mass-loss rates $\dot{M}$ and clump volume filling factors $f_\mathrm{vol}$.
Solid black line: calculated abundance for a smooth, uniform outflow. 
Solid coloured line: characteristic clump scale $l_* = 5 \times 10^{12}$ cm, {porosity length $h_* = 1 \times 10^{14}, 2.5 \times 10^{13}, 1.25 \times 10^{13}$ cm for $f_\mathrm{vol} = 0.05, 0.2, 0.4$, respectively.} 
Dashed coloured line: $l_* = 10^{13}$ cm, {$h_* = 2 \times 10^{14}, 5 \times 10^{13}, 2.5 \times 10^{13}$ cm for $f_\mathrm{vol} = 0.05, 0.2, 0.4$, respectively.}
Dotted coloured line: $l_* = 5 \times 10^{13}$ cm, {$h_* = 1 \times 10^{15}, 2.5 \times 10^{14}, 1.25 \times 10^{14}$ cm for $f_\mathrm{vol} = 0.05, 0.2, 0.4$, respectively.}
{
Note that models with $f_\mathrm{vol} = 0.2,\ l_* = 5 \times 10^{12}$ cm (green, solid) and $f_\mathrm{vol} = 0.4,\ l_* =  1 \times 10^{13}$ cm (red, dashed) have the same porosity length $h_* = 2.5 \times 10^{13}$ cm.
}
For reference, $1\ R_* = 5 \times 10^{13}$ cm.
}
\label{fig:App-OneComp-Orich-Pred}
\end{figure*}

\begin{figure*}[!hb]
\centering
\includegraphics[width=0.96\textwidth]{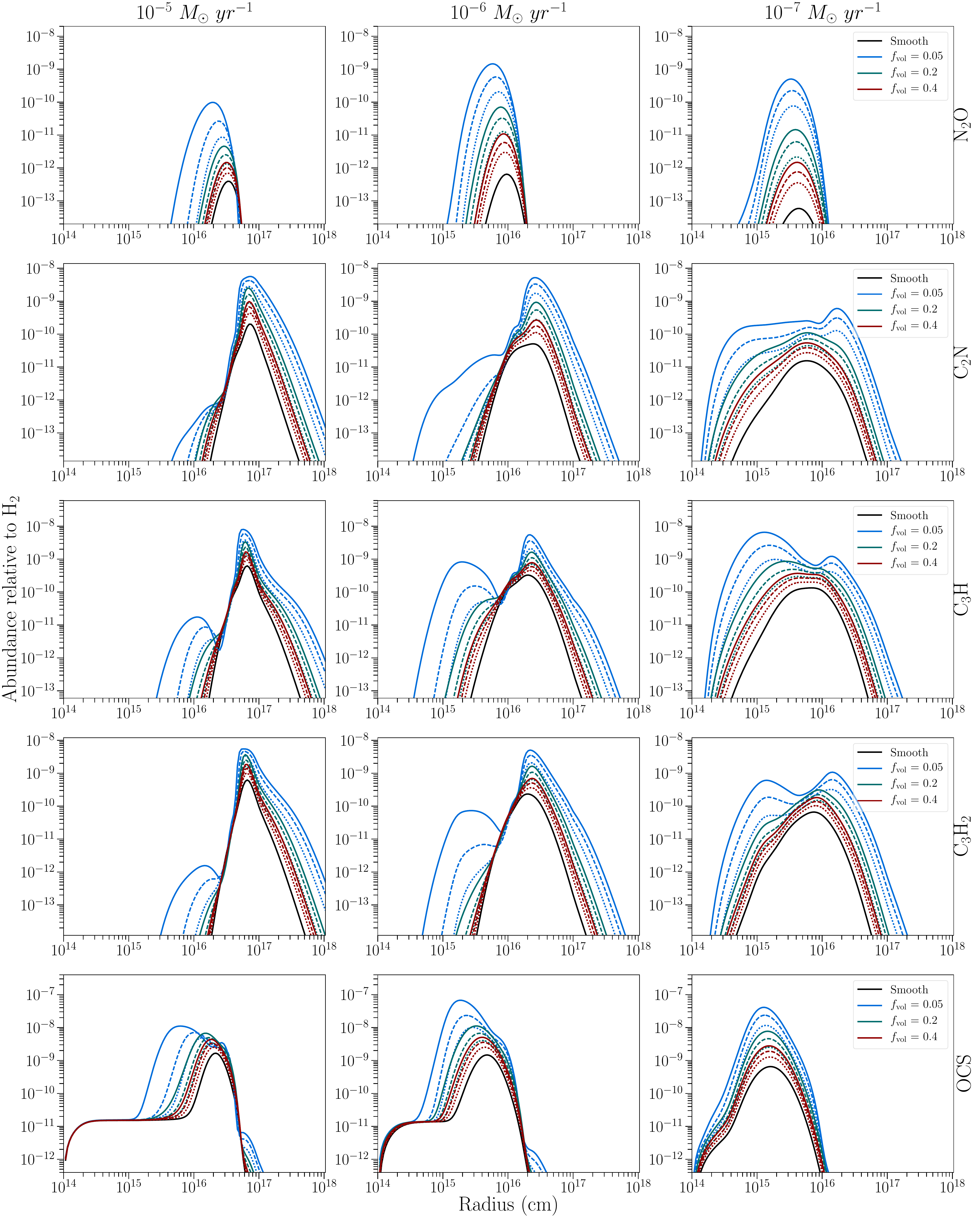}
\caption{Abundance of N$_2$O, C$_2$N, C$_3$H, C$_3$H$_2$ and OCS relative to H$_2$ throughout a two-component O-rich outflow with different mass-loss rates $\dot{M}$ and clump volume filling factors $f_\mathrm{vol}$. The characteristic size of the clumps at the stellar radius is $l_* = 10^{13}$ cm.
{
Blue lines: porosity length $h_* = 2 \times 10^{14}$ cm.
Green lines: $h_* = 5 \times 10^{13}$ cm.
Red lines: $h_* = 2.5 \times 10^{13}$ cm.
}
Solid black line: calculated abundance for a smooth, uniform outflow. 
Solid coloured line: density contrast between the inter-clump and smooth outflow $f_\mathrm{ic} = 0.1$. Dashed coloured line: $f_\mathrm{ic} = 0.3$. Dotted coloured line: $f_\mathrm{ic} = 0.5$.
{Note that the models with $f_\mathrm{vol} = 0.4$ (red) have the same porosity length as the one-component outflows with $f_\mathrm{vol} = 0.2,\ l_* = 5 \times 10^{12}$ cm and $f_\mathrm{vol} = 0.4,\ l_* = 1 \times 10^{13}$ cm.}
For reference, $1\ R_* = 5 \times 10^{13}$ cm.
}
\label{fig:App-TwoComp-Orich-Pred}
\end{figure*}

Figs. \ref{fig:App-OneComp-Orich-Pred} and \ref{fig:App-TwoComp-Orich-Pred} show the abundance profiles of N$_2$O, C$_2$N, C$_3$H, C$_3$H$_2$ and OCS for respectively a one-component and two-component O-rich outflow.
The corresponding column densities are listed in Tables \ref{table:App-CD-1C-O-Pred} and \ref{table:App-CD-2C-O-Pred}.

\begin{sidewaystable}[!ht]
    \caption{Column density [cm$^{-2}]$ of N$_2$O, C$_2$N, C$_3$H, C$_3$H$_2$ and OCS in a smooth O-rich outflow with different mass-loss rates, together with column density ratios relative to the smooth outflow for specific one-component outflows. The corresponding abundance profiles are shown in Fig. \ref{fig:App-OneComp-Crich-Pred}. 
{Note that the models with $f_\mathrm{vol} = 0.2,\ l_* = 5 \times 10^{12}$ cm and $f_\mathrm{vol} = 0.4,\ l_* = 1 \times 10^{13}$ cm have the same porosity length $h_* = 2.5 \times 10^{13}$ cm.}
    }
    \centering
    	\resizebox{0.95\textwidth}{!}{%
    \begin{tabular}{c c c c c c c c c c c c c c c c c  }
    \hline \hline 
    \noalign{\smallskip}
    	$\dot{M}$ & Species & & {N$_2$O} & &  & {C$_2$N} & &  & {C$_3$H} & &  & {C$_3$H$_2$} & &  & {OCS} & \\
    \cmidrule(lr){1-1} \cmidrule(lr){2-2} \cmidrule(lr){3-5} \cmidrule(lr){6-8}  \cmidrule(lr){9-11}  \cmidrule(lr){12-14} \cmidrule(lr){15-17}  
    	\noalign{\smallskip}
{\multirow{5}{*}{\rotatebox[origin=r]{90}{$10^{-5}\ \mathrm{M}_\odot\ \mathrm{yr}^{-1}$}}} &    Smooth & \multicolumn{3}{c}{ 4.8e+07  cm$^{-2}$} &  \multicolumn{3}{c}{ 1.0e+10  cm$^{-2}$} &  \multicolumn{3}{c}{ 3.6e+10  cm$^{-2}$} &  \multicolumn{3}{c}{ 3.0e+10  cm$^{-2}$} &  \multicolumn{3}{c}{ 9.1e+11  cm$^{-2}$} \\
    \cmidrule(lr){2-2} \cmidrule(lr){3-5} \cmidrule(lr){6-8}  \cmidrule(lr){9-11}  \cmidrule(lr){12-14} \cmidrule(lr){15-17}  
	& $f_\mathrm{vol}$ & 0.05 & 0.2 & 0.4 & 0.05 & 0.2 & 0.4 & 0.05 & 0.2 & 0.4 & 0.05 & 0.2 & 0.4 & 0.05 & 0.2 & 0.4 \\	
    \cmidrule(lr){2-2} \cmidrule(lr){3-5} \cmidrule(lr){6-8}  \cmidrule(lr){9-11}  \cmidrule(lr){12-14} \cmidrule(lr){15-17}  
& $l_* = 5 \times 10^{12}$ cm & 4.3e+02 & 1.5e+01 & 4.5e+00 & 4.2e+01 & 1.3e+01 & 4.8e+00 & 1.5e+01 & 4.6e+00 & 2.4e+00 & 1.3e+01 & 5.7e+00 & 2.9e+00 & 2.0e+01 & 3.3e+00 & 1.9e+00 \\
& $l_* = 1 \times 10^{13}$ cm & 3.4e+03 & 2.2e+01 & 5.2e+00 & 4.8e+01 & 1.4e+01 & 5.0e+00 & 2.3e+01 & 5.0e+00 & 2.5e+00 & 1.5e+01 & 6.1e+00 & 3.0e+00 & 1.2e+02 & 4.6e+00 & 2.2e+00 \\
& $l_* = 5 \times 10^{13}$ cm & 4.4e+05 & 1.2e+03 & 2.3e+01 & 1.1e+03 & 2.3e+01 & 6.8e+00 & 1.2e+03 & 2.0e+01 & 3.5e+00 & 4.4e+02 & 1.1e+01 & 4.0e+00 & 9.9e+02 & 1.3e+02 & 1.2e+01 \\
	\noalign{\smallskip}
	\hline
    \noalign{\smallskip}
{\multirow{5}{*}{\rotatebox[origin=r]{90}{$10^{-6}\ \mathrm{M}_\odot\ \mathrm{yr}^{-1}$}}} &    Smooth & \multicolumn{3}{c}{ 3.0e+07  cm$^{-2}$} &  \multicolumn{3}{c}{ 1.9e+09  cm$^{-2}$} &  \multicolumn{3}{c}{ 1.3e+10  cm$^{-2}$} &  \multicolumn{3}{c}{ 6.9e+09  cm$^{-2}$} &  \multicolumn{3}{c}{ 2.3e+11  cm$^{-2}$} \\
    \cmidrule(lr){2-2} \cmidrule(lr){3-5} \cmidrule(lr){6-8}  \cmidrule(lr){9-11}  \cmidrule(lr){12-14} \cmidrule(lr){15-17}  
	& $f_\mathrm{vol}$ & 0.05 & 0.2 & 0.4 & 0.05 & 0.2 & 0.4 & 0.05 & 0.2 & 0.4 & 0.05 & 0.2 & 0.4 & 0.05 & 0.2 & 0.4 \\	
    \cmidrule(lr){2-2} \cmidrule(lr){3-5} \cmidrule(lr){6-8}  \cmidrule(lr){9-11}  \cmidrule(lr){12-14} \cmidrule(lr){15-17}  
& $l_* = 5 \times 10^{12}$ cm & 4.6e+03 & 1.6e+02 & 2.2e+01 & 5.6e+01 & 1.0e+01 & 3.7e+00 & 3.3e+01 & 3.8e+00 & 2.2e+00 & 1.7e+01 & 4.9e+00 & 2.5e+00 & 1.0e+02 & 1.0e+01 & 3.9e+00 \\
& $l_* = 1 \times 10^{13}$ cm & 9.5e+03 & 2.2e+02 & 2.6e+01 & 8.7e+01 & 1.0e+01 & 3.7e+00 & 1.4e+02 & 4.3e+00 & 2.3e+00 & 4.1e+01 & 5.1e+00 & 2.6e+00 & 2.1e+02 & 1.6e+01 & 4.9e+00 \\
& $l_* = 5 \times 10^{13}$ cm & 4.7e+04 & 1.4e+03 & 9.3e+01 & 1.1e+03 & 2.9e+01 & 5.1e+00 & 1.5e+03 & 7.7e+01 & 5.6e+00 & 4.4e+02 & 1.3e+01 & 3.0e+00 & 3.0e+02 & 1.1e+02 & 2.4e+01 \\
	\noalign{\smallskip}
	\hline
    \noalign{\smallskip}
{\multirow{5}{*}{\rotatebox[origin=r]{90}{$10^{-7}\ \mathrm{M}_\odot\ \mathrm{yr}^{-1}$}}} &    Smooth & \multicolumn{3}{c}{ 5.0e+05  cm$^{-2}$} &  \multicolumn{3}{c}{ 3.4e+08  cm$^{-2}$} &  \multicolumn{3}{c}{ 3.1e+09  cm$^{-2}$} &  \multicolumn{3}{c}{ 8.9e+08  cm$^{-2}$} &  \multicolumn{3}{c}{ 2.7e+10  cm$^{-2}$} \\
    \cmidrule(lr){2-2} \cmidrule(lr){3-5} \cmidrule(lr){6-8}  \cmidrule(lr){9-11}  \cmidrule(lr){12-14} \cmidrule(lr){15-17}  
	& $f_\mathrm{vol}$ & 0.05 & 0.2 & 0.4 & 0.05 & 0.2 & 0.4 & 0.05 & 0.2 & 0.4 & 0.05 & 0.2 & 0.4 & 0.05 & 0.2 & 0.4 \\	
    \cmidrule(lr){2-2} \cmidrule(lr){3-5} \cmidrule(lr){6-8}  \cmidrule(lr){9-11}  \cmidrule(lr){12-14} \cmidrule(lr){15-17}  
& $l_* = 5 \times 10^{12}$ cm & 1.3e+04 & 3.4e+02 & 3.2e+01 & 9.7e+01 & 1.3e+01 & 4.7e+00 & 1.6e+02 & 1.4e+01 & 4.4e+00 & 4.2e+01 & 5.4e+00 & 2.9e+00 & 7.0e+01 & 1.3e+01 & 4.6e+00 \\
& $l_* = 1 \times 10^{13}$ cm & 1.5e+04 & 3.8e+02 & 3.4e+01 & 1.9e+02 & 2.0e+01 & 6.2e+00 & 3.0e+02 & 2.1e+01 & 5.5e+00 & 7.8e+01 & 6.6e+00 & 3.2e+00 & 8.0e+01 & 1.5e+01 & 5.0e+00 \\
& $l_* = 5 \times 10^{13}$ cm & 2.6e+04 & 6.9e+02 & 5.7e+01 & 7.1e+02 & 1.1e+02 & 2.5e+01 & 1.1e+03 & 1.0e+02 & 1.9e+01 & 3.1e+02 & 2.2e+01 & 6.1e+00 & 1.0e+02 & 2.4e+01 & 8.0e+00 \\
    	\hline \hline
    \end{tabular}%
    }
    \label{table:App-CD-1C-O-Pred}
\end{sidewaystable}

\begin{sidewaystable}[!hb]
    \caption{Column density [cm$^{-2}]$ of N$_2$O, C$_2$N, C$_3$H, C$_3$H$_2$ and OCS in a smooth O-rich outflow with different mass-loss rates, together with column density ratios relative to the smooth outflow for specific two-component outflows. The corresponding abundance profiles are shown in Fig. \ref{fig:App-TwoComp-Crich-Pred}. 
{Note that models with $f_\mathrm{vol} = 0.4$ have the same porosity length $h_* = 2.5 \times 10^{13}$ cm as the one-component models with $f_\mathrm{vol} = 0.2,\ l_* = 5 \times 10^{12}$ cm and $f_\mathrm{vol} = 0.4,\ l_* = 1 \times 10^{13}$ cm.}
    }
    \centering
    	\resizebox{0.95\textwidth}{!}{%
    \begin{tabular}{c c c c c c c c c c c c c c c c c  }
    \hline \hline 
    \noalign{\smallskip}
    	$\dot{M}$ & Species & & {N$_2$O} & &  & {C$_2$N} & &  & {C$_3$H} & &  & {C$_3$H$_2$} & &  & {OCS} & \\
    \cmidrule(lr){1-1} \cmidrule(lr){2-2} \cmidrule(lr){3-5} \cmidrule(lr){6-8}  \cmidrule(lr){9-11}  \cmidrule(lr){12-14} \cmidrule(lr){15-17}  
    	\noalign{\smallskip}
{\multirow{5}{*}{\rotatebox[origin=r]{90}{$10^{-5}\ \mathrm{M}_\odot\ \mathrm{yr}^{-1}$}}} &    Smooth & \multicolumn{3}{c}{ 4.8e+07  cm$^{-2}$} &  \multicolumn{3}{c}{ 1.0e+10  cm$^{-2}$} &  \multicolumn{3}{c}{ 3.6e+10  cm$^{-2}$} &  \multicolumn{3}{c}{ 3.0e+10  cm$^{-2}$} &  \multicolumn{3}{c}{ 9.1e+11  cm$^{-2}$} \\
    \cmidrule(lr){2-2} \cmidrule(lr){3-5} \cmidrule(lr){6-8}  \cmidrule(lr){9-11}  \cmidrule(lr){12-14} \cmidrule(lr){15-17}  
	& $f_\mathrm{vol}$ & 0.05 & 0.2 & 0.4 & 0.05 & 0.2 & 0.4 & 0.05 & 0.2 & 0.4 & 0.05 & 0.2 & 0.4 & 0.05 & 0.2 & 0.4 \\	
    \cmidrule(lr){2-2} \cmidrule(lr){3-5} \cmidrule(lr){6-8}  \cmidrule(lr){9-11}  \cmidrule(lr){12-14} \cmidrule(lr){15-17}  
& $f_\mathrm{ic} = 0.1$ & 6.7e+02 & 1.5e+01 & 4.2e+00 & 4.0e+01 & 1.1e+01 & 4.2e+00 & 1.4e+01 & 4.1e+00 & 2.2e+00 & 1.2e+01 & 5.1e+00 & 2.6e+00 & 1.6e+01 & 3.5e+00 & 1.9e+00 \\
& $f_\mathrm{ic} = 0.3$ & 1.2e+02 & 7.6e+00 & 2.8e+00 & 2.6e+01 & 7.0e+00 & 2.9e+00 & 8.7e+00 & 2.9e+00 & 1.8e+00 & 8.6e+00 & 3.5e+00 & 2.0e+00 & 5.4e+00 & 2.3e+00 & 1.6e+00 \\
& $f_\mathrm{ic} = 0.5$ & 3.0e+01 & 3.8e+00 & 1.9e+00 & 1.5e+01 & 4.0e+00 & 2.0e+00 & 5.0e+00 & 2.0e+00 & 1.5e+00 & 5.4e+00 & 2.3e+00 & 1.6e+00 & 2.9e+00 & 1.7e+00 & 1.4e+00 \\
	\noalign{\smallskip}
	\hline
    \noalign{\smallskip}
{\multirow{5}{*}{\rotatebox[origin=r]{90}{$10^{-6}\ \mathrm{M}_\odot\ \mathrm{yr}^{-1}$}}} &    Smooth & \multicolumn{3}{c}{ 3.0e+07  cm$^{-2}$} &  \multicolumn{3}{c}{ 1.9e+09  cm$^{-2}$} &  \multicolumn{3}{c}{ 1.3e+10  cm$^{-2}$} &  \multicolumn{3}{c}{ 6.9e+09  cm$^{-2}$} &  \multicolumn{3}{c}{ 2.3e+11  cm$^{-2}$} \\
    \cmidrule(lr){2-2} \cmidrule(lr){3-5} \cmidrule(lr){6-8}  \cmidrule(lr){9-11}  \cmidrule(lr){12-14} \cmidrule(lr){15-17}  
	& $f_\mathrm{vol}$ & 0.05 & 0.2 & 0.4 & 0.05 & 0.2 & 0.4 & 0.05 & 0.2 & 0.4 & 0.05 & 0.2 & 0.4 & 0.05 & 0.2 & 0.4 \\	
    \cmidrule(lr){2-2} \cmidrule(lr){3-5} \cmidrule(lr){6-8}  \cmidrule(lr){9-11}  \cmidrule(lr){12-14} \cmidrule(lr){15-17}  
& $f_\mathrm{ic} = 0.1$ & 5.0e+03 & 1.5e+02 & 1.9e+01 & 4.9e+01 & 8.3e+00 & 3.2e+00 & 3.3e+01 & 3.5e+00 & 2.0e+00 & 1.6e+01 & 4.3e+00 & 2.3e+00 & 9.7e+01 & 1.1e+01 & 3.9e+00 \\
& $f_\mathrm{ic} = 0.3$ & 1.6e+03 & 6.2e+01 & 1.0e+01 & 2.8e+01 & 5.2e+00 & 2.3e+00 & 7.9e+00 & 2.4e+00 & 1.6e+00 & 8.6e+00 & 3.0e+00 & 1.8e+00 & 2.8e+01 & 5.6e+00 & 2.6e+00 \\
& $f_\mathrm{ic} = 0.5$ & 4.7e+02 & 2.3e+01 & 4.9e+00 & 1.4e+01 & 3.0e+00 & 1.7e+00 & 3.7e+00 & 1.8e+00 & 1.4e+00 & 5.1e+00 & 2.0e+00 & 1.5e+00 & 1.0e+01 & 3.1e+00 & 1.9e+00 \\
	\noalign{\smallskip}
	\hline
    \noalign{\smallskip}
{\multirow{5}{*}{\rotatebox[origin=r]{90}{$10^{-7}\ \mathrm{M}_\odot\ \mathrm{yr}^{-1}$}}} &    Smooth & \multicolumn{3}{c}{ 5.0e+05  cm$^{-2}$} &  \multicolumn{3}{c}{ 3.4e+08  cm$^{-2}$} &  \multicolumn{3}{c}{ 3.1e+09  cm$^{-2}$} &  \multicolumn{3}{c}{ 8.9e+08  cm$^{-2}$} &  \multicolumn{3}{c}{ 2.7e+10  cm$^{-2}$} \\
    \cmidrule(lr){2-2} \cmidrule(lr){3-5} \cmidrule(lr){6-8}  \cmidrule(lr){9-11}  \cmidrule(lr){12-14} \cmidrule(lr){15-17}  
	& $f_\mathrm{vol}$ & 0.05 & 0.2 & 0.4 & 0.05 & 0.2 & 0.4 & 0.05 & 0.2 & 0.4 & 0.05 & 0.2 & 0.4 & 0.05 & 0.2 & 0.4 \\	
    \cmidrule(lr){2-2} \cmidrule(lr){3-5} \cmidrule(lr){6-8}  \cmidrule(lr){9-11}  \cmidrule(lr){12-14} \cmidrule(lr){15-17}  
& $f_\mathrm{ic} = 0.1$ & 1.0e+04 & 2.6e+02 & 2.5e+01 & 1.1e+02 & 1.4e+01 & 4.9e+00 & 1.8e+02 & 1.4e+01 & 4.4e+00 & 4.4e+01 & 5.2e+00 & 2.7e+00 & 6.2e+01 & 1.2e+01 & 4.2e+00 \\
& $f_\mathrm{ic} = 0.3$ & 4.4e+03 & 1.1e+02 & 1.3e+01 & 3.8e+01 & 7.1e+00 & 3.1e+00 & 6.1e+01 & 7.0e+00 & 2.8e+00 & 1.5e+01 & 3.3e+00 & 2.0e+00 & 3.5e+01 & 6.8e+00 & 2.9e+00 \\
& $f_\mathrm{ic} = 0.5$ & 1.4e+03 & 3.7e+01 & 6.0e+00 & 1.4e+01 & 3.7e+00 & 2.1e+00 & 2.0e+01 & 3.5e+00 & 1.9e+00 & 5.7e+00 & 2.2e+00 & 1.6e+00 & 1.7e+01 & 3.8e+00 & 2.0e+00 \\
    	\hline \hline
    \end{tabular}%
    }
    \label{table:App-CD-2C-O-Pred}
\end{sidewaystable}

\subsection{C-rich outflows}       \label{subsect:App:Predictions:Crich}

\begin{figure*}[!ht]
\centering
\includegraphics[width=0.96\textwidth]{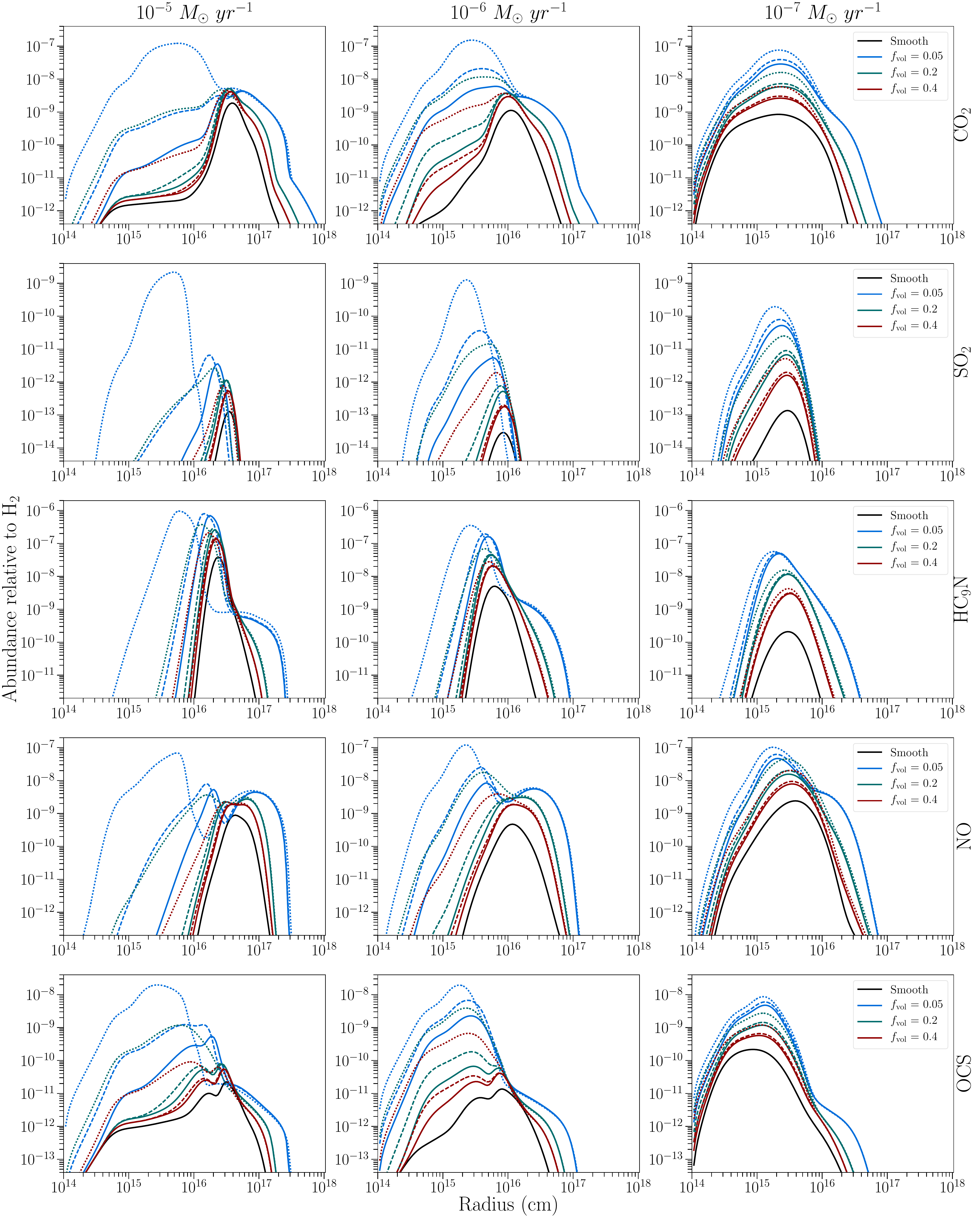}
\caption{Abundance of CO$_2$, SO$_2$, HC$_9$N, NO, and OCS relative to H$_2$ throughout a one-component C-rich outflow with different mass-loss rates $\dot{M}$ and clump volume filling factors $f_\mathrm{vol}$.
Solid black line: calculated abundance for a smooth, uniform outflow. 
Solid coloured line: characteristic clump scale $l_* = 5 \times 10^{12}$ cm, {porosity length $h_* = 1 \times 10^{14}, 2.5 \times 10^{13}, 1.25 \times 10^{13}$ cm for $f_\mathrm{vol} = 0.05, 0.2, 0.4$, respectively.} 
Dashed coloured line: $l_* = 10^{13}$ cm, {$h_* = 2 \times 10^{14}, 5 \times 10^{13}, 2.5 \times 10^{13}$ cm for $f_\mathrm{vol} = 0.05, 0.2, 0.4$, respectively.}
Dotted coloured line: $l_* = 5 \times 10^{13}$ cm, {$h_* = 1 \times 10^{15}, 2.5 \times 10^{14}, 1.25 \times 10^{14}$ cm for $f_\mathrm{vol} = 0.05, 0.2, 0.4$, respectively.}
{
Note that models with $f_\mathrm{vol} = 0.2,\ l_* = 5 \times 10^{12}$ cm (green, solid) and $f_\mathrm{vol} = 0.4,\ l_* =  1 \times 10^{13}$ cm (red, dashed) have the same porosity length $h_* = 2.5 \times 10^{13}$ cm.
}
For reference, $1\ R_* = 5 \times 10^{13}$ cm.
}
\label{fig:App-OneComp-Crich-Pred}
\end{figure*}

\begin{figure*}[!hb]
\centering
\includegraphics[width=0.96\textwidth]{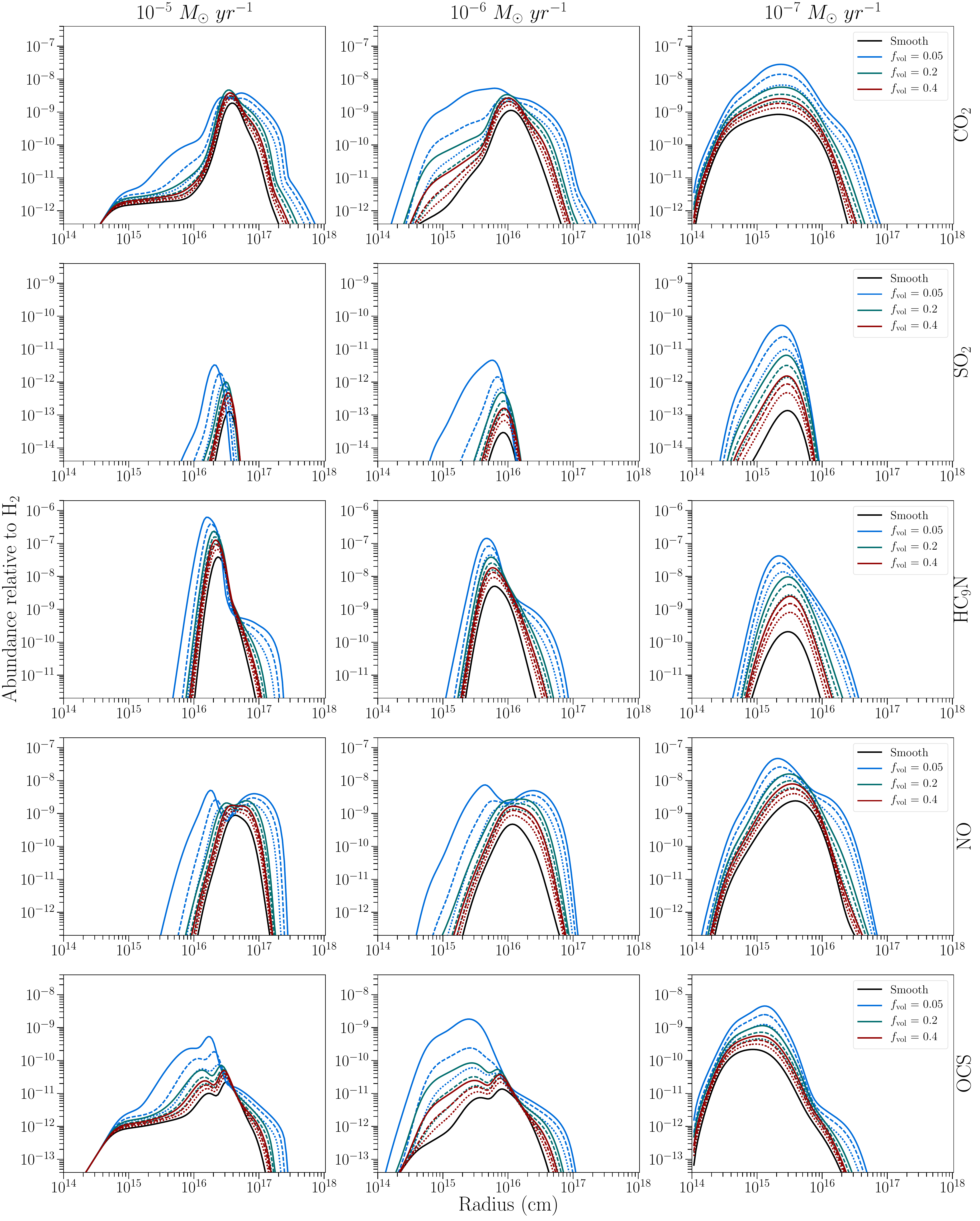}
\caption{Abundance of CO$_2$, SO$_2$, HC$_9$N, NO, and OCS relative to H$_2$ throughout a two-component C-rich outflow with different mass-loss rates $\dot{M}$ and clump volume filling factors $f_\mathrm{vol}$. The characteristic size of the clumps at the stellar radius is $l_* = 10^{13}$ cm.
{
Blue lines: porosity length $h_* = 2 \times 10^{14}$ cm.
Green lines: $h_* = 5 \times 10^{13}$ cm.
Red lines: $h_* = 2.5 \times 10^{13}$ cm.
}
Solid black line: calculated abundance for a smooth, uniform outflow. 
Solid coloured line: density contrast between the inter-clump and smooth outflow $f_\mathrm{ic} = 0.1$. Dashed coloured line: $f_\mathrm{ic} = 0.3$. Dotted coloured line: $f_\mathrm{ic} = 0.5$.
{Note that the models with $f_\mathrm{vol} = 0.4$ (red) have the same porosity length as the one-component outflows with $f_\mathrm{vol} = 0.2,\ l_* = 5 \times 10^{12}$ cm and $f_\mathrm{vol} = 0.4,\ l_* = 1 \times 10^{13}$ cm.}
For reference, $1\ R_* = 5 \times 10^{13}$ cm.
}
\label{fig:App-TwoComp-Crich-Pred}
\end{figure*}

{
Figs. \ref{fig:App-OneComp-Crich-Pred} and \ref{fig:App-TwoComp-Crich-Pred} show the abundance profiles of CO$_2$, SO$_2$, HC$_9$N, NO and OCS for respectively a one-component and two-component O-rich outflow.
The corresponding column densities are listed in Tables \ref{table:App-CD-1C-C-Pred} and \ref{table:App-CD-2C-C-Pred}.

}

\begin{sidewaystable}[!ht]
    \caption{Column density [cm$^{-2}]$ of CO$_2$, SO$_2$, HC$_9$N, NO, and OCS in a smooth C-rich outflow with different mass-loss rates, together with column density ratios relative to the smooth outflow for specific one-component outflows. The corresponding abundance profiles are shown in Fig. \ref{fig:App-OneComp-Crich-Pred}. 
{Note that the models with $f_\mathrm{vol} = 0.2,\ l_* = 5 \times 10^{12}$ cm and $f_\mathrm{vol} = 0.4,\ l_* = 1 \times 10^{13}$ cm have the same porosity length $h_* = 2.5 \times 10^{13}$ cm.}
    }
    \centering
    	\resizebox{0.95\textwidth}{!}{%
    \begin{tabular}{c c c c c c c c c c c c c c c c c  }
    \hline \hline 
    \noalign{\smallskip}
    	$\dot{M}$ & Species & & {CO$_2$} & &  & {SO$_2$} & &  & {HC$_9$N} & &  & {NO} & &  & {OCS} & \\
    \cmidrule(lr){1-1} \cmidrule(lr){2-2} \cmidrule(lr){3-5} \cmidrule(lr){6-8}  \cmidrule(lr){9-11}  \cmidrule(lr){12-14} \cmidrule(lr){15-17}  
    	\noalign{\smallskip}
{\multirow{5}{*}{\rotatebox[origin=r]{90}{$10^{-5}\ \mathrm{M}_\odot\ \mathrm{yr}^{-1}$}}} &    Smooth & \multicolumn{3}{c}{ 2.7e+11  cm$^{-2}$} &  \multicolumn{3}{c}{ 1.2e+07  cm$^{-2}$} &  \multicolumn{3}{c}{ 5.6e+12  cm$^{-2}$} &  \multicolumn{3}{c}{ 1.2e+11  cm$^{-2}$} &  \multicolumn{3}{c}{ 2.5e+10  cm$^{-2}$} \\
    \cmidrule(lr){2-2} \cmidrule(lr){3-5} \cmidrule(lr){6-8}  \cmidrule(lr){9-11}  \cmidrule(lr){12-14} \cmidrule(lr){15-17}  
	& $f_\mathrm{vol}$ & 0.05 & 0.2 & 0.4 & 0.05 & 0.2 & 0.4 & 0.05 & 0.2 & 0.4 & 0.05 & 0.2 & 0.4 & 0.05 & 0.2 & 0.4 \\	
    \cmidrule(lr){2-2} \cmidrule(lr){3-5} \cmidrule(lr){6-8}  \cmidrule(lr){9-11}  \cmidrule(lr){12-14} \cmidrule(lr){15-17}  
& $l_* = 5 \times 10^{12}$ cm & 5.2e+00 & 3.1e+00 & 2.2e+00 & 4.4e+01 & 8.8e+00 & 4.0e+00 & 2.7e+01 & 7.4e+00 & 3.7e+00 & 1.4e+01 & 4.8e+00 & 2.9e+00 & 2.3e+01 & 2.8e+00 & 1.7e+00 \\
& $l_* = 1 \times 10^{13}$ cm & 2.4e+01 & 3.3e+00 & 2.4e+00 & 1.7e+02 & 1.0e+01 & 4.3e+00 & 4.4e+01 & 9.1e+00 & 4.3e+00 & 3.7e+01 & 5.4e+00 & 3.1e+00 & 1.8e+02 & 3.8e+00 & 1.9e+00 \\
& $l_* = 5 \times 10^{13}$ cm & 1.8e+03 & 3.0e+01 & 4.3e+00 & 3.7e+05 & 1.1e+02 & 8.7e+00 & 1.7e+02 & 2.6e+01 & 9.6e+00 & 1.5e+03 & 2.4e+01 & 5.3e+00 & 4.5e+03 & 2.1e+02 & 1.2e+01 \\
	\noalign{\smallskip}
	\hline
    \noalign{\smallskip}
{\multirow{5}{*}{\rotatebox[origin=r]{90}{$10^{-6}\ \mathrm{M}_\odot\ \mathrm{yr}^{-1}$}}} &    Smooth & \multicolumn{3}{c}{ 8.1e+10  cm$^{-2}$} &  \multicolumn{3}{c}{ 1.8e+06  cm$^{-2}$} &  \multicolumn{3}{c}{ 5.0e+11  cm$^{-2}$} &  \multicolumn{3}{c}{ 3.3e+10  cm$^{-2}$} &  \multicolumn{3}{c}{ 3.4e+09  cm$^{-2}$} \\
    \cmidrule(lr){2-2} \cmidrule(lr){3-5} \cmidrule(lr){6-8}  \cmidrule(lr){9-11}  \cmidrule(lr){12-14} \cmidrule(lr){15-17}  
	& $f_\mathrm{vol}$ & 0.05 & 0.2 & 0.4 & 0.05 & 0.2 & 0.4 & 0.05 & 0.2 & 0.4 & 0.05 & 0.2 & 0.4 & 0.05 & 0.2 & 0.4 \\	
    \cmidrule(lr){2-2} \cmidrule(lr){3-5} \cmidrule(lr){6-8}  \cmidrule(lr){9-11}  \cmidrule(lr){12-14} \cmidrule(lr){15-17}  
& $l_* = 5 \times 10^{12}$ cm & 4.0e+01 & 4.5e+00 & 2.8e+00 & 5.0e+02 & 1.7e+01 & 5.8e+00 & 3.5e+01 & 8.8e+00 & 4.2e+00 & 5.2e+01 & 9.3e+00 & 4.5e+00 & 3.7e+02 & 1.2e+01 & 3.7e+00 \\
& $l_* = 1 \times 10^{13}$ cm & 1.4e+02 & 6.9e+00 & 3.1e+00 & 4.9e+03 & 3.1e+01 & 6.9e+00 & 5.2e+01 & 9.9e+00 & 4.5e+00 & 1.8e+02 & 1.2e+01 & 4.8e+00 & 1.1e+03 & 3.9e+01 & 7.0e+00 \\
& $l_* = 5 \times 10^{13}$ cm & 1.0e+03 & 9.7e+01 & 1.9e+01 & 1.9e+05 & 2.1e+03 & 1.2e+02 & 2.1e+02 & 2.3e+01 & 7.4e+00 & 1.4e+03 & 1.5e+02 & 2.3e+01 & 3.8e+03 & 8.1e+02 & 1.7e+02 \\
	\noalign{\smallskip}
	\hline
    \noalign{\smallskip}
{\multirow{5}{*}{\rotatebox[origin=r]{90}{$10^{-7}\ \mathrm{M}_\odot\ \mathrm{yr}^{-1}$}}} &    Smooth & \multicolumn{3}{c}{ 1.1e+11  cm$^{-2}$} &  \multicolumn{3}{c}{ 3.4e+06  cm$^{-2}$} &  \multicolumn{3}{c}{ 7.2e+09  cm$^{-2}$} &  \multicolumn{3}{c}{ 7.4e+10  cm$^{-2}$} &  \multicolumn{3}{c}{ 3.7e+10  cm$^{-2}$} \\
    \cmidrule(lr){2-2} \cmidrule(lr){3-5} \cmidrule(lr){6-8}  \cmidrule(lr){9-11}  \cmidrule(lr){12-14} \cmidrule(lr){15-17}  
	& $f_\mathrm{vol}$ & 0.05 & 0.2 & 0.4 & 0.05 & 0.2 & 0.4 & 0.05 & 0.2 & 0.4 & 0.05 & 0.2 & 0.4 & 0.05 & 0.2 & 0.4 \\	
    \cmidrule(lr){2-2} \cmidrule(lr){3-5} \cmidrule(lr){6-8}  \cmidrule(lr){9-11}  \cmidrule(lr){12-14} \cmidrule(lr){15-17}  
& $l_* = 5 \times 10^{12}$ cm & 1.8e+01 & 4.8e+00 & 2.5e+00 & 5.1e+02 & 5.1e+01 & 1.2e+01 & 2.7e+02 & 5.1e+01 & 1.2e+01 & 2.5e+01 & 7.1e+00 & 3.4e+00 & 1.3e+01 & 4.2e+00 & 2.3e+00 \\
& $l_* = 1 \times 10^{13}$ cm & 2.4e+01 & 5.9e+00 & 2.9e+00 & 8.2e+02 & 7.4e+01 & 1.5e+01 & 3.1e+02 & 5.6e+01 & 1.3e+01 & 3.5e+01 & 9.3e+00 & 4.1e+00 & 1.6e+01 & 5.1e+00 & 2.7e+00 \\
& $l_* = 5 \times 10^{13}$ cm & 4.7e+01 & 1.3e+01 & 5.6e+00 & 2.2e+03 & 2.6e+02 & 5.0e+01 & 4.5e+02 & 8.7e+01 & 2.0e+01 & 7.0e+01 & 2.4e+01 & 9.6e+00 & 2.5e+01 & 9.7e+00 & 5.0e+00 \\
    	\hline \hline
    \end{tabular}%
    }
    \label{table:App-CD-1C-C-Pred}
\end{sidewaystable}

\begin{sidewaystable}[!hb]
    \caption{Column density [cm$^{-2}]$ of CO$_2$, SO$_2$, HC$_9$N, NO, and OCS in a smooth C-rich outflow with different mass-loss rates, together with column density ratios relative to the smooth outflow for specific two-component outflows. The corresponding abundance profiles are shown in Fig. \ref{fig:App-TwoComp-Crich-Pred}. 
{Note that models with $f_\mathrm{vol} = 0.4$ have the same porosity length $h_* = 2.5 \times 10^{13}$ cm as the one-component models with $f_\mathrm{vol} = 0.2,\ l_* = 5 \times 10^{12}$ cm and $f_\mathrm{vol} = 0.4,\ l_* = 1 \times 10^{13}$ cm.}
    }
    \centering
    	\resizebox{0.95\textwidth}{!}{%
    \begin{tabular}{c c c c c c c c c c c c c c c c c  }
    \hline \hline 
    \noalign{\smallskip}
    	$\dot{M}$ & Species & & {CO$_2$} & &  & {SO$_2$} & &  & {HC$_9$N} & &  & {NO} & &  & {OCS} & \\
    \cmidrule(lr){1-1} \cmidrule(lr){2-2} \cmidrule(lr){3-5} \cmidrule(lr){6-8}  \cmidrule(lr){9-11}  \cmidrule(lr){12-14} \cmidrule(lr){15-17}  
    	\noalign{\smallskip}
{\multirow{5}{*}{\rotatebox[origin=r]{90}{$10^{-5}\ \mathrm{M}_\odot\ \mathrm{yr}^{-1}$}}} &    Smooth & \multicolumn{3}{c}{ 2.7e+11  cm$^{-2}$} &  \multicolumn{3}{c}{ 1.2e+07  cm$^{-2}$} &  \multicolumn{3}{c}{ 5.6e+12  cm$^{-2}$} &  \multicolumn{3}{c}{ 1.2e+11  cm$^{-2}$} &  \multicolumn{3}{c}{ 2.5e+10  cm$^{-2}$} \\
    \cmidrule(lr){2-2} \cmidrule(lr){3-5} \cmidrule(lr){6-8}  \cmidrule(lr){9-11}  \cmidrule(lr){12-14} \cmidrule(lr){15-17}  
	& $f_\mathrm{vol}$ & 0.05 & 0.2 & 0.4 & 0.05 & 0.2 & 0.4 & 0.05 & 0.2 & 0.4 & 0.05 & 0.2 & 0.4 & 0.05 & 0.2 & 0.4 \\	
    \cmidrule(lr){2-2} \cmidrule(lr){3-5} \cmidrule(lr){6-8}  \cmidrule(lr){9-11}  \cmidrule(lr){12-14} \cmidrule(lr){15-17}  
& $f_\mathrm{ic} = 0.1$ & 4.5e+00 & 2.9e+00 & 2.1e+00 & 4.7e+01 & 8.2e+00 & 3.7e+00 & 2.8e+01 & 7.3e+00 & 3.6e+00 & 1.4e+01 & 4.5e+00 & 2.8e+00 & 1.7e+01 & 2.8e+00 & 1.7e+00 \\
& $f_\mathrm{ic} = 0.3$ & 3.0e+00 & 2.2e+00 & 1.7e+00 & 1.9e+01 & 5.2e+00 & 2.6e+00 & 1.4e+01 & 4.6e+00 & 2.6e+00 & 6.9e+00 & 3.2e+00 & 2.1e+00 & 5.8e+00 & 1.9e+00 & 1.4e+00 \\
& $f_\mathrm{ic} = 0.5$ & 2.2e+00 & 1.7e+00 & 1.4e+00 & 9.2e+00 & 3.1e+00 & 1.9e+00 & 6.8e+00 & 2.8e+00 & 1.9e+00 & 3.9e+00 & 2.2e+00 & 1.6e+00 & 2.8e+00 & 1.5e+00 & 1.3e+00 \\
	\noalign{\smallskip}
	\hline
    \noalign{\smallskip}
{\multirow{5}{*}{\rotatebox[origin=r]{90}{$10^{-6}\ \mathrm{M}_\odot\ \mathrm{yr}^{-1}$}}} &    Smooth & \multicolumn{3}{c}{ 8.1e+10  cm$^{-2}$} &  \multicolumn{3}{c}{ 1.8e+06  cm$^{-2}$} &  \multicolumn{3}{c}{ 5.0e+11  cm$^{-2}$} &  \multicolumn{3}{c}{ 3.3e+10  cm$^{-2}$} &  \multicolumn{3}{c}{ 3.4e+09  cm$^{-2}$} \\
    \cmidrule(lr){2-2} \cmidrule(lr){3-5} \cmidrule(lr){6-8}  \cmidrule(lr){9-11}  \cmidrule(lr){12-14} \cmidrule(lr){15-17}  
	& $f_\mathrm{vol}$ & 0.05 & 0.2 & 0.4 & 0.05 & 0.2 & 0.4 & 0.05 & 0.2 & 0.4 & 0.05 & 0.2 & 0.4 & 0.05 & 0.2 & 0.4 \\	
    \cmidrule(lr){2-2} \cmidrule(lr){3-5} \cmidrule(lr){6-8}  \cmidrule(lr){9-11}  \cmidrule(lr){12-14} \cmidrule(lr){15-17}  
& $f_\mathrm{ic} = 0.1$ & 3.7e+01 & 4.6e+00 & 2.7e+00 & 4.2e+02 & 1.7e+01 & 5.4e+00 & 3.4e+01 & 8.0e+00 & 3.8e+00 & 5.0e+01 & 8.7e+00 & 4.1e+00 & 3.3e+02 & 1.7e+01 & 4.6e+00 \\
& $f_\mathrm{ic} = 0.3$ & 6.9e+00 & 2.9e+00 & 2.0e+00 & 5.5e+01 & 8.7e+00 & 3.5e+00 & 1.7e+01 & 5.2e+00 & 2.7e+00 & 1.5e+01 & 5.6e+00 & 2.9e+00 & 4.5e+01 & 5.6e+00 & 2.6e+00 \\
& $f_\mathrm{ic} = 0.5$ & 3.0e+00 & 2.1e+00 & 1.6e+00 & 2.1e+01 & 4.6e+00 & 2.3e+00 & 8.8e+00 & 3.2e+00 & 2.0e+00 & 8.0e+00 & 3.5e+00 & 2.1e+00 & 9.4e+00 & 2.6e+00 & 1.8e+00 \\
	\noalign{\smallskip}
	\hline
    \noalign{\smallskip}
{\multirow{5}{*}{\rotatebox[origin=r]{90}{$10^{-7}\ \mathrm{M}_\odot\ \mathrm{yr}^{-1}$}}} &    Smooth & \multicolumn{3}{c}{ 1.1e+11  cm$^{-2}$} &  \multicolumn{3}{c}{ 3.4e+06  cm$^{-2}$} &  \multicolumn{3}{c}{ 7.2e+09  cm$^{-2}$} &  \multicolumn{3}{c}{ 7.4e+10  cm$^{-2}$} &  \multicolumn{3}{c}{ 3.7e+10  cm$^{-2}$} \\
    \cmidrule(lr){2-2} \cmidrule(lr){3-5} \cmidrule(lr){6-8}  \cmidrule(lr){9-11}  \cmidrule(lr){12-14} \cmidrule(lr){15-17}  
	& $f_\mathrm{vol}$ & 0.05 & 0.2 & 0.4 & 0.05 & 0.2 & 0.4 & 0.05 & 0.2 & 0.4 & 0.05 & 0.2 & 0.4 & 0.05 & 0.2 & 0.4 \\	
    \cmidrule(lr){2-2} \cmidrule(lr){3-5} \cmidrule(lr){6-8}  \cmidrule(lr){9-11}  \cmidrule(lr){12-14} \cmidrule(lr){15-17}  
& $f_\mathrm{ic} = 0.1$ & 1.8e+01 & 4.7e+00 & 2.5e+00 & 5.3e+02 & 5.2e+01 & 1.2e+01 & 2.4e+02 & 4.4e+01 & 1.0e+01 & 2.6e+01 & 7.3e+00 & 3.4e+00 & 1.2e+01 & 4.1e+00 & 2.3e+00 \\
& $f_\mathrm{ic} = 0.3$ & 9.7e+00 & 3.1e+00 & 1.9e+00 & 2.2e+02 & 2.4e+01 & 6.5e+00 & 1.4e+02 & 2.4e+01 & 6.2e+00 & 1.4e+01 & 4.4e+00 & 2.4e+00 & 7.3e+00 & 2.8e+00 & 1.8e+00 \\
& $f_\mathrm{ic} = 0.5$ & 5.0e+00 & 2.1e+00 & 1.6e+00 & 8.0e+01 & 1.0e+01 & 3.5e+00 & 6.9e+01 & 1.1e+01 & 3.5e+00 & 6.9e+00 & 2.7e+00 & 1.8e+00 & 4.1e+00 & 2.0e+00 & 1.5e+00 \\
    	\hline \hline
    \end{tabular}%
    }
    \label{table:App-CD-2C-C-Pred}
\end{sidewaystable}

\end{appendix}

\end{document}